%% file: main.tex
\PassOptionsToPackage{dvipsnames}{xcolor}
\documentclass[twocolumn]{aastex7}
\usepackage{amsmath}

\newcommand{\editone}{}

\defcitealias{Gutermuth2009}{G09}

\shorttitle{SESHAT}
\shortauthors{Crompvoets et al.}

\begin{document}

\title{Introducing SESHAT: A Tool for Object Classification from JWST Catalogs}

\correspondingauthor{B. L. Crompvoets}

\author[0000-0001-8900-5550]{B. L. Crompvoets}
\affiliation{Department of Physics and Astronomy, University of Victoria, Victoria, BC, Canada}
\affiliation{NRC Herzberg Astronomy and Astrophysics, 5071 West Saanich Road, Victoria, BC V9E 2E7, Canada}
\email[show]{bcrompvoets@uvic.ca}

\author[0000-0002-5779-8549]{H. Kirk}
\affiliation{Department of Physics and Astronomy, University of Victoria, Victoria, BC, Canada}
\affiliation{NRC Herzberg Astronomy and Astrophysics, 5071 West Saanich Road, Victoria, BC V9E 2E7, Canada}
\email{helenkirkastro@gmail.com}

\author[0000-0002-6447-899X]{R. Gutermuth}
\affiliation{Department of Astronomy, University of Massachusetts, Amherst, MA 01003, USA}
\email{rgutermu@astro.umass.edu}

\author[0000-0002-9289-2450]{J. Di Francesco}
\affiliation{Department of Physics and Astronomy, University of Victoria, Victoria, BC, Canada}
\affiliation{NRC Herzberg Astronomy and Astrophysics, 5071 West Saanich Road, Victoria, BC V9E 2E7, Canada}
\email{james.difrancesco@nrc-cnrc.gc.ca}

% \author[0000-0002-4348-6974]{H.Teimoorinia}
% \affiliation{NRC Herzberg Astronomy and Astrophysics, 5071 West Saanich Road, Victoria, BC V9E 2E7, Canada}
% \affiliation{Department of Physics and Astronomy, University of Victoria, Victoria, BC, Canada}

\begin{abstract}

JWST's exquisite data have opened the doors to new possibilities in detecting broad classes of astronomical objects, but also to new challenges in classifying those objects. 
In this work, we introduce SESHAT, the Stellar Evolutionary Stage Heuristic Assessment Tool for the identification of Young Stellar Objects, field stars (main sequence through asymptotic giant branch), brown dwarfs, white dwarfs, and galaxies, from any JWST photometry. 
This identification is done using the machine learning method XGBoost to analyze thousands of rows of synthetic photometry, modified at run-time to match the filters available in the data to be classified. 
We validate this tool on real data of both star-forming regions and cosmological fields, and find we are able to reproduce the observed classes of objects to a minimum of 85\% recall across every class, with all available data, without additional information on the ellipticity or spatial distribution of the objects.
Furthermore, this tool can be used to test the filter choices for JWST proposals by verifying whether the chosen filters are sufficient to identify the desired class of objects.
SESHAT is released as a Python package to the community for general use.
\end{abstract}
% \keywords{Classical Novae (251) --- Ultraviolet astronomy(1736) --- History of astronomy(1868) --- Interdisciplinary astronomy(804)}

\section{Introduction}

The James Webb Space Telescope (JWST) has provided exquisite data across hundreds of observations in its first four and a half years \citep[e.g.,][]{Pontoppidan2022,Green2024}. With {38} different photometric filters, there are millions of combinations possible to tune observations for each science case. Thus, every JWST observation has the potential to use an entirely unique set of filters. It is important to be able to identify accurately the population of interest within these data, be these brown dwarfs in a cosmological field, or Young Stellar Objects (YSOs) in a star-forming region. In this work, we present the Stellar Evolutionary Stage Heuristic Assessment Tool (SESHAT)\footnote{PyPI: \url{https://pypi.org/project/seshat-classifier/}; Zenodo: \dataset[doi:10.5281/zenodo.18705173]{https://doi.org/10.5281/zenodo.18705173}}: a tool for the identification of YSOs, brown dwarfs, field stars, white dwarfs, and galaxies across any JWST observation.

Traditional methods of object classification require the use of discriminating cuts in color and magnitude space to distinguish between different populations \citep{Gutermuth2008,Gutermuth2009}. This methodology is excellent for producing homogeneous samples across observations, when every observation uses the exact same suite of filters. In the context of JWST, when every observation has the potential to use its own unique suite of filters, such consistent data selection no longer applies, and new color cuts would need to be defined for every observation. 
Furthermore, traditional methods are limited to two dimensions for the definition of various cuts, and this limitation can lead to significant impurity of the assigned classification. In the era of big data where thousands of sources are being discovered in every single image, a robust method for the identification of objects is important. 

Machine learning can overcome these multidimensional challenges. For instance, it has been used quite successfully in recent years to expedite the search for and recovery of YSOs in spatially massive catalogs \citep[e.g.,][]{Cornu2021,Kuhn2021}, including the identification of multiple evolutionary stages of stars within these catalogs \citep[e.g.,][]{Marton2016}. Machine learning can appraise every dimension possible of the input data, and determine the best cuts to separate out each type of object, in a fraction of the time that a human could do the same, and still obtain a higher degree of purity in its classifications. 

This paper is presented in self-contained sections, so the reader can focus only on that section which is necessary to them. We use synthetic models to have data in every filter, and the explanation of how these data are procured and made realistic is presented in Section~\ref{sec:data}. To analyze these data in any suite of filters, we use an XGBoost machine learning method, and the details of how this algorithm is trained and examples of verification against real Spitzer and JWST data (with pre-defined classifications from the literature) are presented in Section~\ref{sec:valid}. Section~\ref{sec:tool} describes in detail the intended use of the tool built upon this algorithm, which we call SESHAT, including its inputs, outputs, and availability. We end with our conclusions in Section~\ref{sec:concl}.

\section{Data}\label{sec:data}
One of the reasons that JWST data can be so rich is the sheer variety of filters onboard. With 29 NIRCam filters and 9 MIRI filters, JWST can sample snippets of the spectra of stars never before imaged. What this means, however, is that old data cannot be used to leverage classifications for all the options available. For instance, Spitzer had four near-infrared (IR) filters and one mid-IR, and though there are filters on JWST that have similar bandpasses to these (e.g., IRAC1 and F356W, IRAC2 and F444W, IRAC3 and F560W, IRAC4 and F770W), there are many more besides. To take full advantage of JWST, it is important to be able to rely on classifications that go beyond the Spitzer-analog bands. SESHAT is built to have flexibility in what filters are used and, consequently, we can apply the same method to any well-characterized near-IR facility in the past, present, or future.

For this work, we utilize synthetic models to inform the classifications of objects in data from JWST and Spitzer/2MASS. We use the latter to compare the performance of SESHAT to previously classified datasets. In the following sections, we outline each model grid used, and, when available, the observations these were checked against. We use only colors of objects to eliminate the need for accurate distance estimations to each source. 
We obtain Vega magnitude information for each source either from the model packages themselves, or, in cases where only spectra are supplied, by convolving the spectra with the filter's spectral response function, for each filter on JWST, Spitzer, and 2MASS using the \texttt{SEDFitter} tool \citep{Robitaille2007}.

% MIRI zero-point corrections are accounted for based on Table 6 of \citet{Gordon2024}
% 

\subsection{YSO models}\label{sec:ysos}

\begin{figure*}[t]
    \centering
    \includegraphics[width=0.45\linewidth]{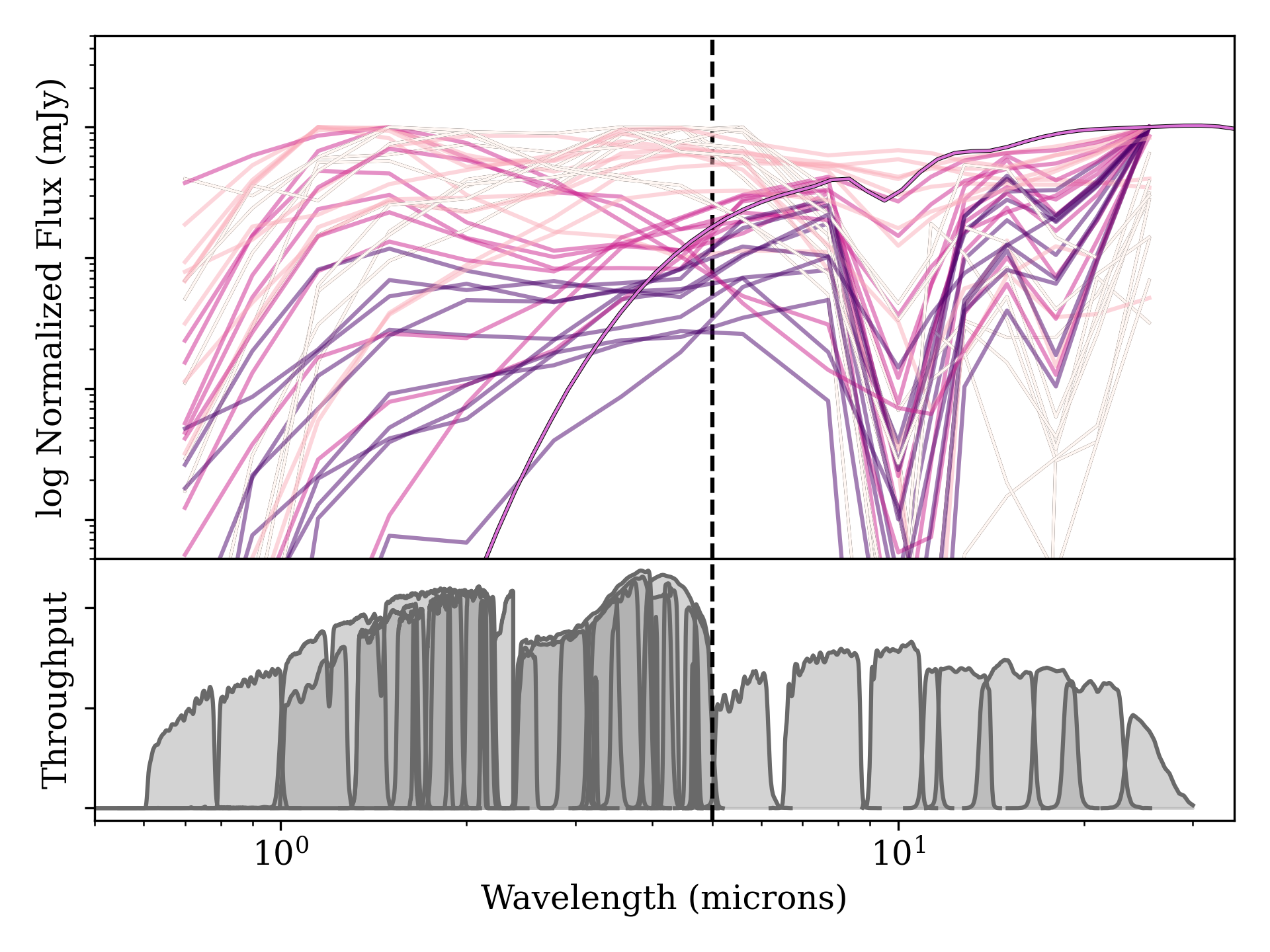}
    \includegraphics[width=0.5\linewidth]{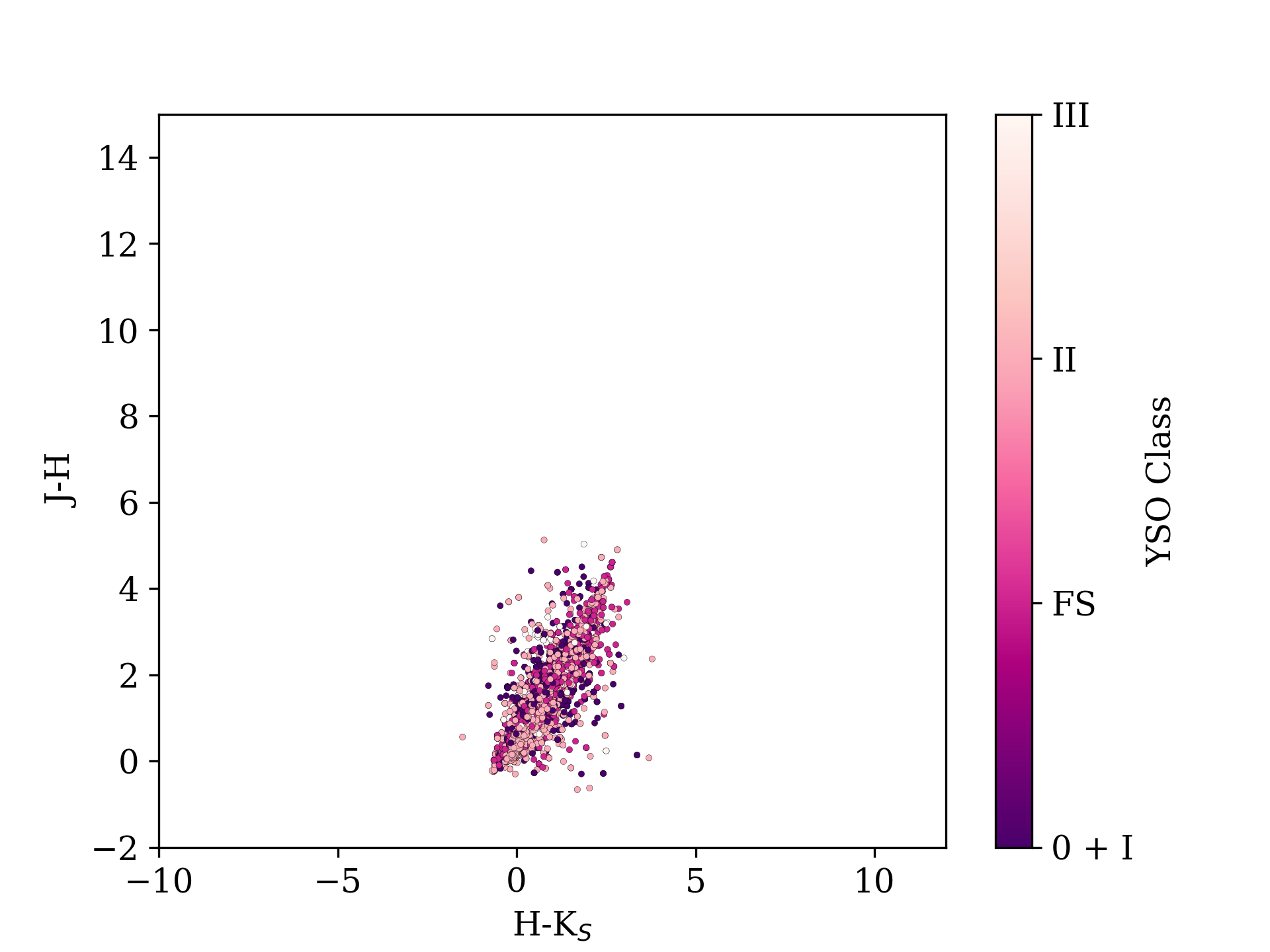}
    \caption{
    \textit{Top left}: the SED of a Class 0 YSO from \citet{Richardson2024} is shown outlined in black, \editone{on top of a sampling of other possible YSO SEDs colored by YSO Class, as defined by the spectral index calculated between IRAC2 and MIPS1, Equation~\ref{eq:alpha_col}.} \textit{Bottom left}: filter response functions for NIRCam (left of dashed line) and MIRI (right of dashed line), as indicated by light-gray filled curves. Wavelengths where the filters overlap thus have darker shading. 
    \textit{Right}: a color-color diagram of the YSOs in our training sample (without additional extinction applied), to show their distribution in color space. The x and y limits of the plot are chosen to be the same for this panel across the following figures, for ease of comparison. 
    }
    \label{fig:yso_spec}
\end{figure*}

YSOs have a wide variety of spectral energy distribution shapes influenced by many different factors (see Figure~\ref{fig:yso_spec} for an example). For this work, we use the \texttt{Hyperion} models of \citet{Richardson2024} to form the basis of our YSO set. These models vary over parameters such as age, metallicity, rotation, envelope size, envelope mass, viewing angle, disk radius, cavity angle, and more. These models follow on the radiative transfer work of Robitaille and colleagues 
\citep[][]{Robitaille2006,Robitaille2007,Robitaille2017}, updating those earlier models to include more parameters. Furthermore, they also provide a range of apertures to help with spectral fitting, determining how much of the light around the source (and surrounding disk/envelope) would actually be observed.  

To ensure that these models match realistic distributions, we use the Spitzer Extended Solar Neighborhood Archive (SESNA; R. A. Gutermuth et al. in prep) catalogs of the three nearest clouds \citep[Taurus at 145 pc; Ophiuchus at 130 pc; and Corona Australis at 150 pc; ][]{Zucker2020} as well as two more distant and active star-forming regions \citep[Orion A and Orion B at 420~pc;][]{Zucker2020}.
We fit models to these SESNA YSOs, and  use JWST filter convolutions of the model SEDs as the basis of our training set. \editone{This collection results in a sample of 7858 unique YSO models across all evolutionary stages.}

Figure~\ref{fig:yso_spec} shows the wide variations in SEDs that a YSO may have. In this case, the shift in evolutionary status is what leads to this variety, as the YSO accretes or disperses its envelope/disk. This wide shift from being far-/mid-IR dominated to near-IR dominated leads to the broad range seen in the color-color panel of Figure~\ref{fig:yso_spec}.

\subsection{Field star models}

\begin{figure*}[t]
    \centering
    \includegraphics[width=0.45\linewidth]{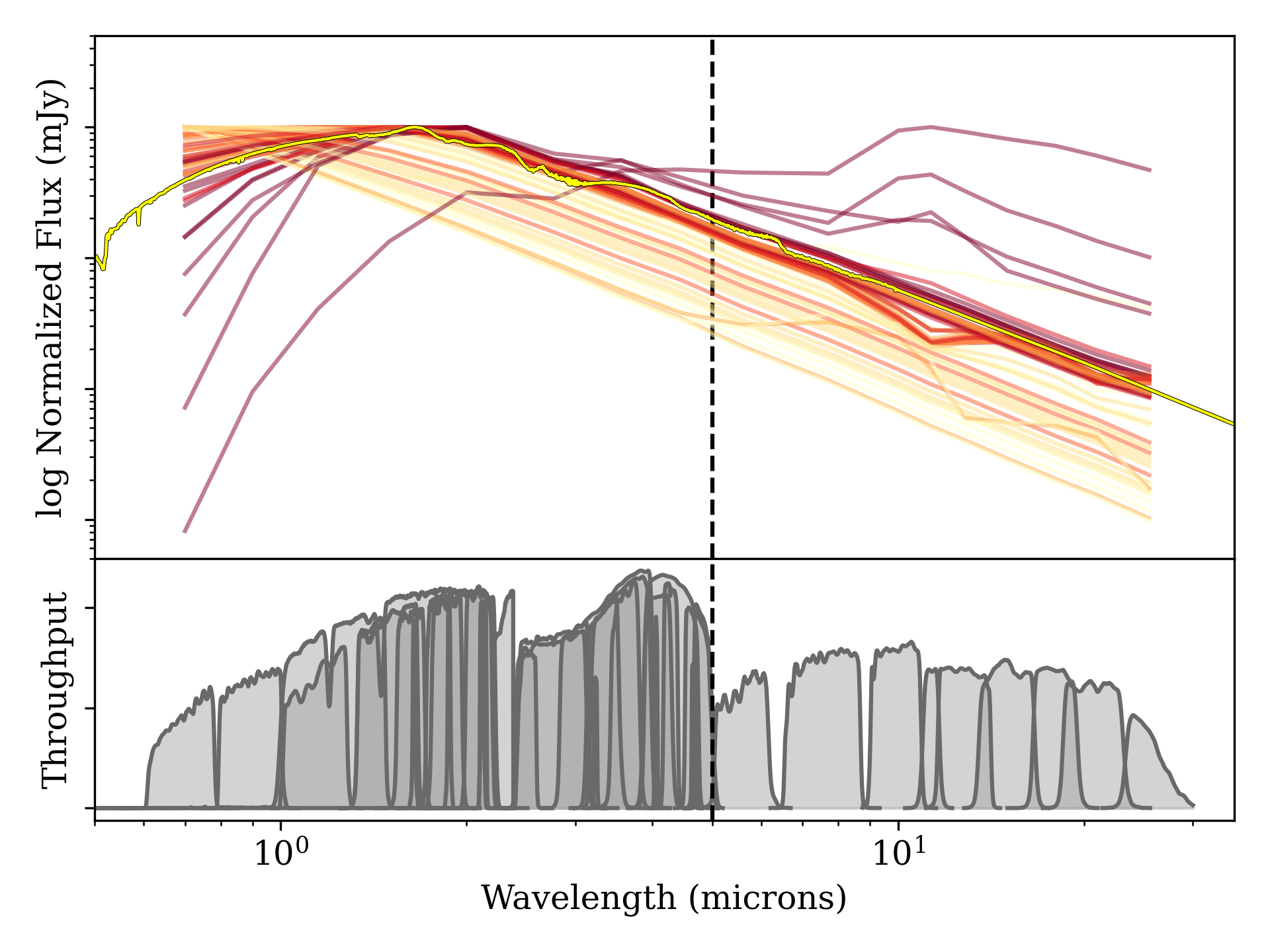}
    \includegraphics[width=0.5\linewidth]{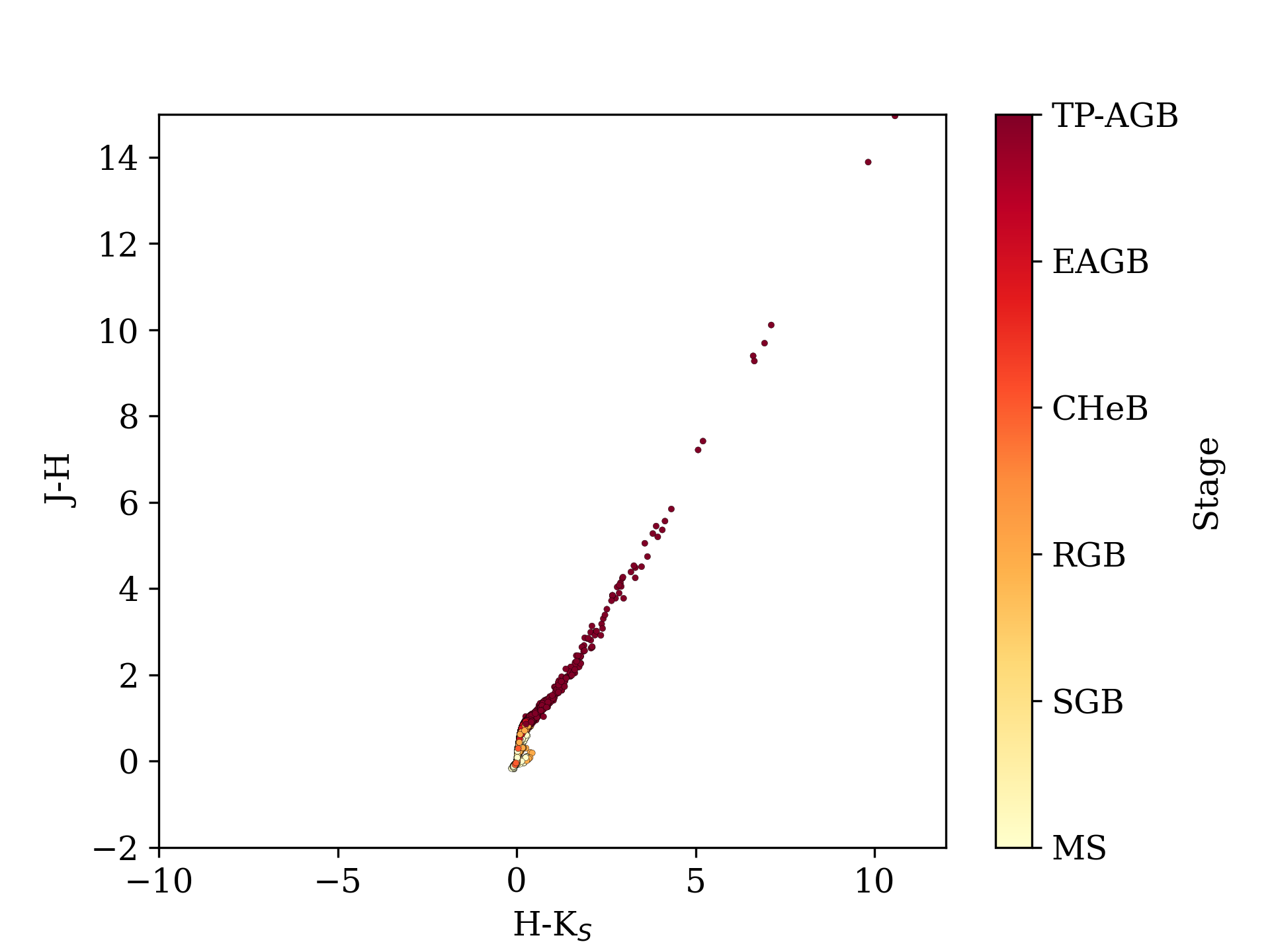}
    \caption{
    \textit{Left}: A model SED of the stellar atmosphere of a MS star from \citet{Castelli2003} is shown in yellow, outlined in black for ease in visualization, and the colored lines show ``connect-the-dots'' SEDs from PARSEC \citep{Bressan2012} CMD photometry points of JWST data from across stellar evolutionary stages (colors denoted by colorbar on far right). 
    \textit{Right}: Color-color diagram showing the field star population, with color scale indicating the evolution from MS through to AGB star. 
    }
    \label{fig:ms_spec}
\end{figure*}

Stars undergo a complex and dynamical evolution as mediated by their shifting nucleosynthesis. We use the Padova PARSEC models \citep[Version 1.2S,][]{Bressan2012} of stellar evolution to sample ages (from $\log{(t_*)} = 6.0-10.0$, in steps of $\log{(t_*)} = 0.25$) and metallicities ($[M/H]$ from $-2$ to $0.3$ in steps of $[M/H] = 0.1$) to define multiple stellar populations. The Kroupa IMF \citep{Kroupa2001}, corrected for unresolved binaries, is used to fill out the mass function. We use the PARSEC CMD tool to extract magnitudes in the medium, wide, and very-wide filters for NIRCam and MIRI. 

These models provide the populations of objects in their relative abundance at each age. Thus, when building our dataset for this work, we purposefully oversample later evolutionary stages to build representative samples that fully cover all stages\editone{, resulting in a total of 12,000 models}. 
Figure~\ref{fig:ms_spec} shows the variation of the SEDs across these stages, along with their positions in color-color space.
Many field stars are distinctive by their peak in visible wavelengths, and, for the majority of cases, a very flat tail, operating as near perfect black bodies. They are thus relatively easy to separate in color-color space, also seen in Figure~\ref{fig:ms_spec}. 
The dusty circumstellar envelopes associated with more advanced evolutionary stages (such as AGB stars) leads to the departure from this otherwise condensed relation.

\subsection{Brown dwarf models}

\begin{figure*}[t]
    \centering
    \includegraphics[width=0.45\linewidth]{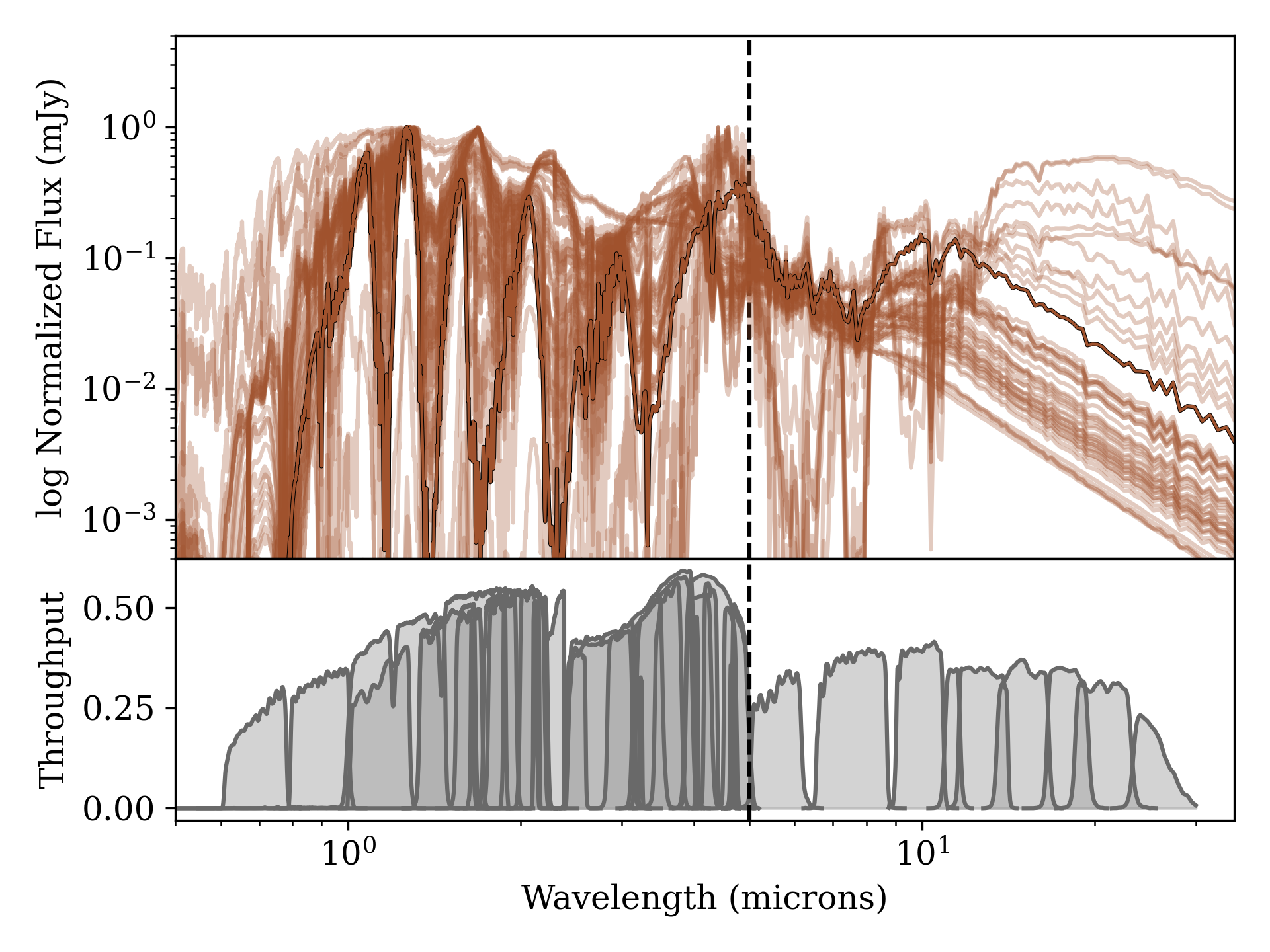}
    \includegraphics[width=0.5\linewidth]{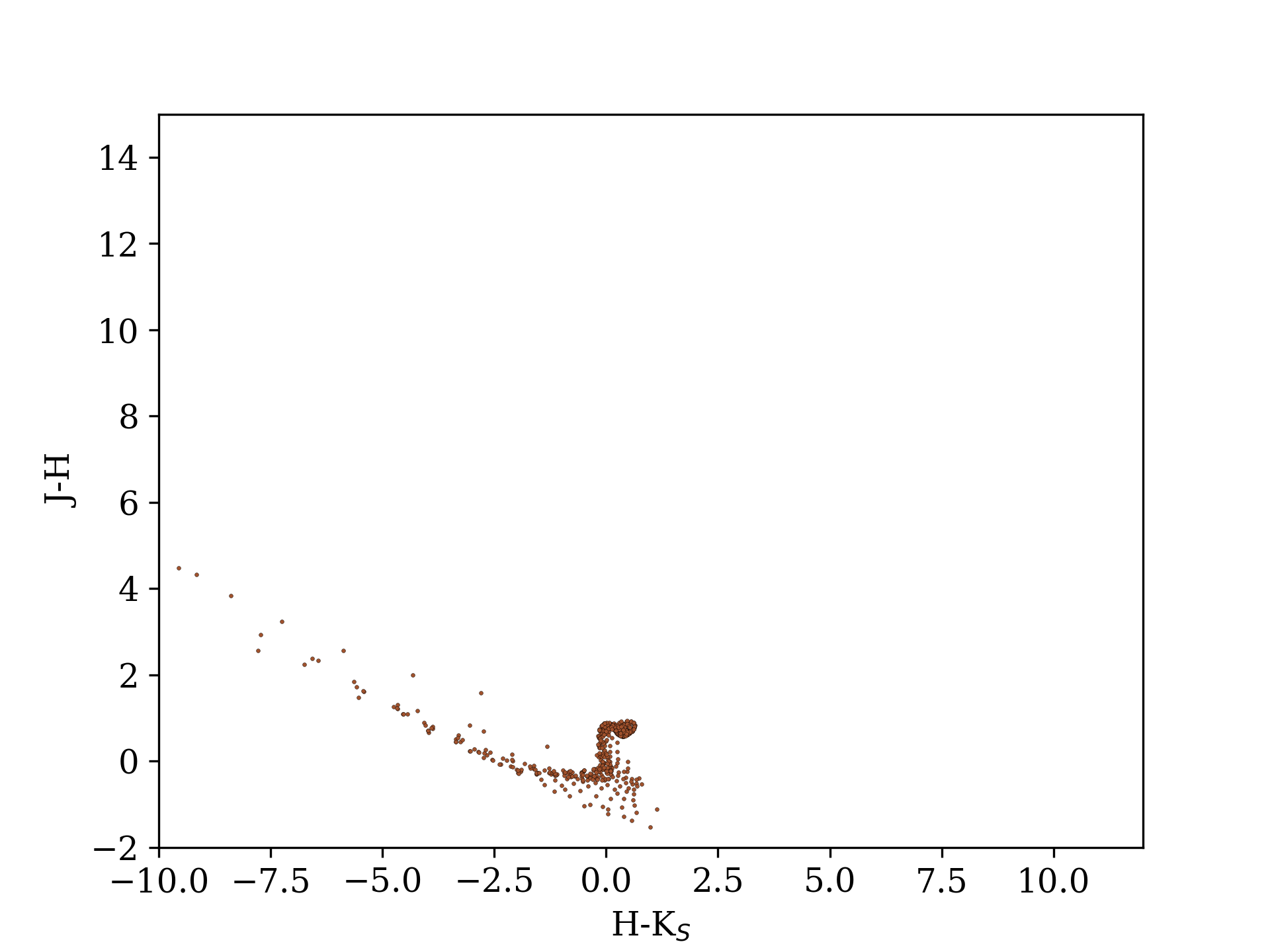}
    \caption{Similar to Figs.~\ref{fig:yso_spec} and \ref{fig:ms_spec}, now for brown dwarfs from the models of the ATMO collaboration \citep{Phillips2020}.}
    % . SED is for the model ``T900\_lg3.0\_NEQ\_strong'' from the brown dwarf models of the ATMO collaboration \citep{Phillips2020}. }
    % The model assumes chemical non-equilibrium with strong mixing, a surface gravity $\log{g} = 3.0$, and a temperature of $T_{eff}=900$ K. \textbf{MASS???} The lower panel shows all NIRCam (left of dashed line) and MIRI (right of dashed line) filter throughputs.}
    \label{fig:bd_spec}
\end{figure*}

Brown dwarf stars never reach the main sequence, and straddle the line between planets and stars \citep{Whitworth2018}. In JWST data, several searches have been performed to isolate brown dwarfs serendipitously detected in cosmological datasets, where low-metallicity brown dwarfs exist out in the halo of our Galaxy \citep[e.g.,][]{Chenbd2025}. 

For these objects, we utilize the brown dwarf evolutionary models of the ATMO team \citep{Phillips2020}. These models come from time evolution of 1D MHD simulations of brown dwarfs.
We use all available cases: assuming both chemical equilibrium and non-equilibrium, the latter with both strong and weak vertical mixing. 
This results in a sample of brown dwarfs with masses ranging from 0.001 $M_\odot$ to 0.075 $M_\odot$, with temperatures $T_{eff}=200-3000~K$ and specific gravities $\log{(g)}=2.5-5.5$ ($g$ in cgs units)\editone{, a total of 510 unique models}. The strong absorption features present in the atmospheres of brown dwarfs lead to distinctive tails in color-color space, as seen in Figure~\ref{fig:bd_spec}.

Finally, although these models have pre-determined Vega magnitudes available for all medium and wide band filters across NIRCam and MIRI, we convolve our own, and use the match between our convolved magnitudes and theirs to verify that our method is sound. We further convolve Spitzer and 2MASS fluxes, for use in our validation step. Spitzer/2MASS models are not available with the ATMO release.

\subsection{White dwarf models}

\begin{figure*}[t]
    \centering
    \includegraphics[width=0.45\linewidth]{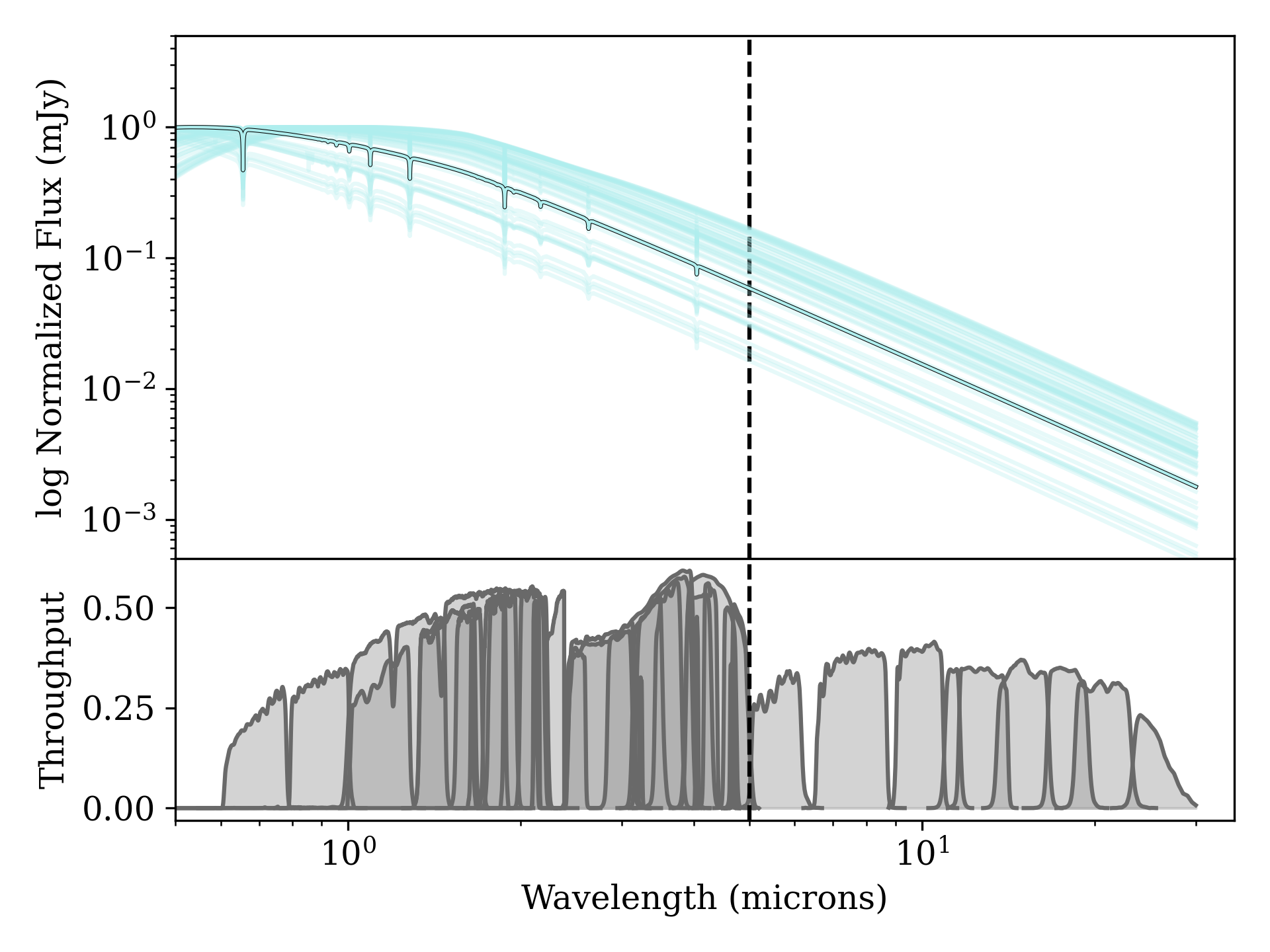}
    \includegraphics[width=0.5\linewidth]{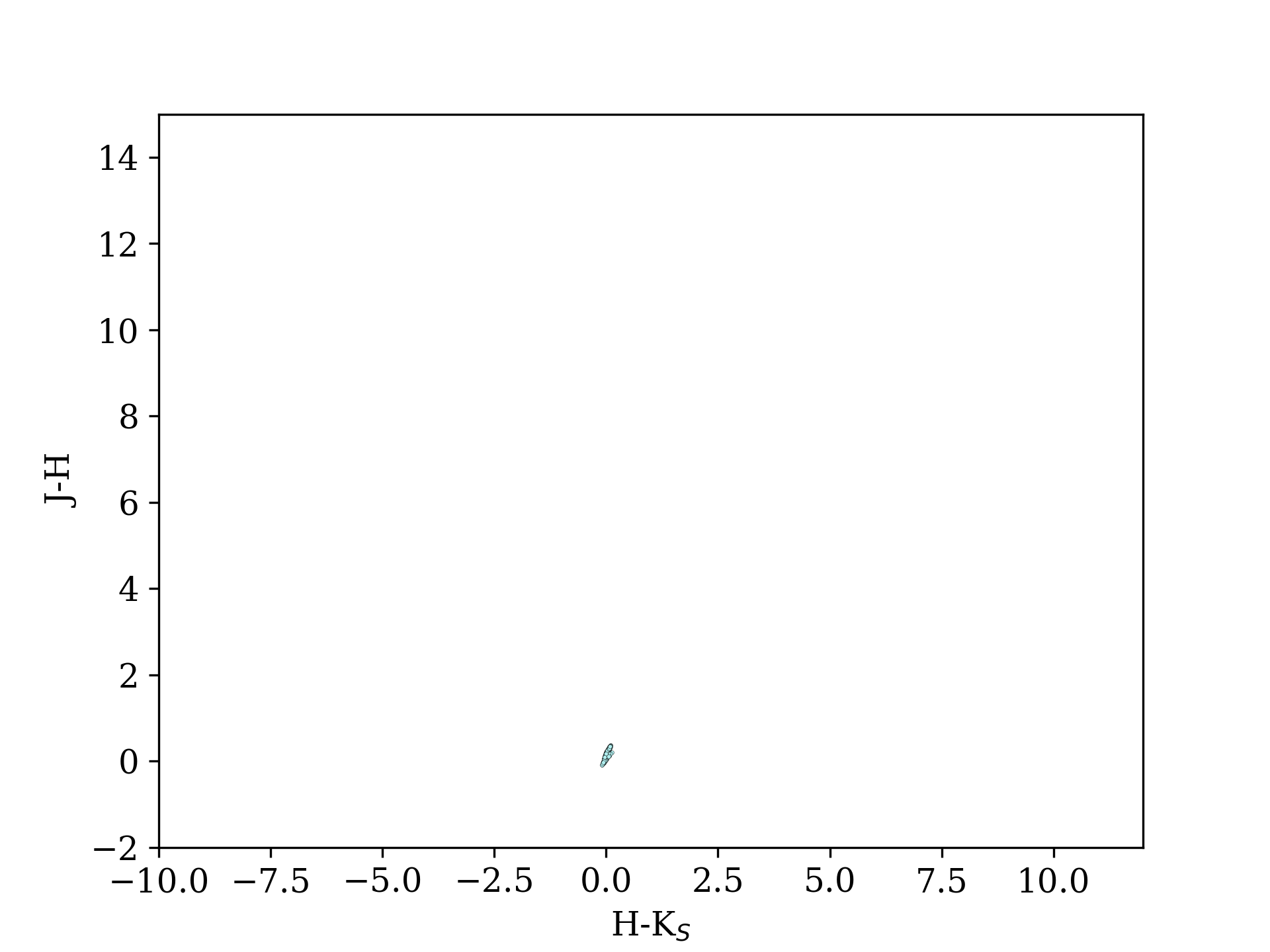}
    \caption{Similar to Figs.~\ref{fig:yso_spec}, \ref{fig:ms_spec}, and \ref{fig:bd_spec}, now for white dwarfs, using the models of \citet{Blouin2018}.}
    \label{fig:wd_spec}
\end{figure*}

White dwarfs make up 7-10\% of the local stellar population volume, as observed with Gaia \citep{Tremblay2024}. These objects have SEDs that peak in the optical, and decrease throughout the IR, but their signatures can still be observed. With JWST, these stars will only be observed as foreground stars when looking at Galactic star forming regions. For instance, assuming a limiting magnitude of 21 mags at 4.5 microns \citep[e.g.,][]{Crompvoets2024}, a typical white dwarf \citep{Kilic2006} would be detectable up to a distance of 550 pc with no extinction \citep[and the distance to the majority of nearby clouds in the Gould Belt is $\sim500$ pc;][]{Zucker2020}.

%Most white dwarf models do not extend into the near- to mid-IR due to their sharply decreasing SEDs \citep[e.g.,][]{Axelrod2023}. \citet{Blouin2018} created a set of models that continue into the MIR, and we use an extension of these in this work. 
We utilize the white dwarf models described in \citet{Blouin2018} for this work. These models were developed for JWST white dwarf searches, extending into the MIR, and
%These models 
including SEDs with absorption features for hydrogen, helium, and carbon, as well as for the variations of metals due to the absorption of exoplanets \citep{Limbach2022,Blouin2024}.
Specifically, we use models that have been fit to the nearest \editone{395} white dwarfs to ensure broad coverage of all spectral variations (Simon Blouin, private communication).
An example of a model and variations seen in white dwarf SEDs are presented in Figure~\ref{fig:wd_spec}, along with the (lack of) distribution of these models in color-color space. Most of the variations are within the optical range of the SED, and so this population has little variation in the IR, and thus little variation in the color-color distribution.

\subsection{Galaxy models}

\begin{figure*}[t]
    \centering
    \includegraphics[width=0.45\linewidth]{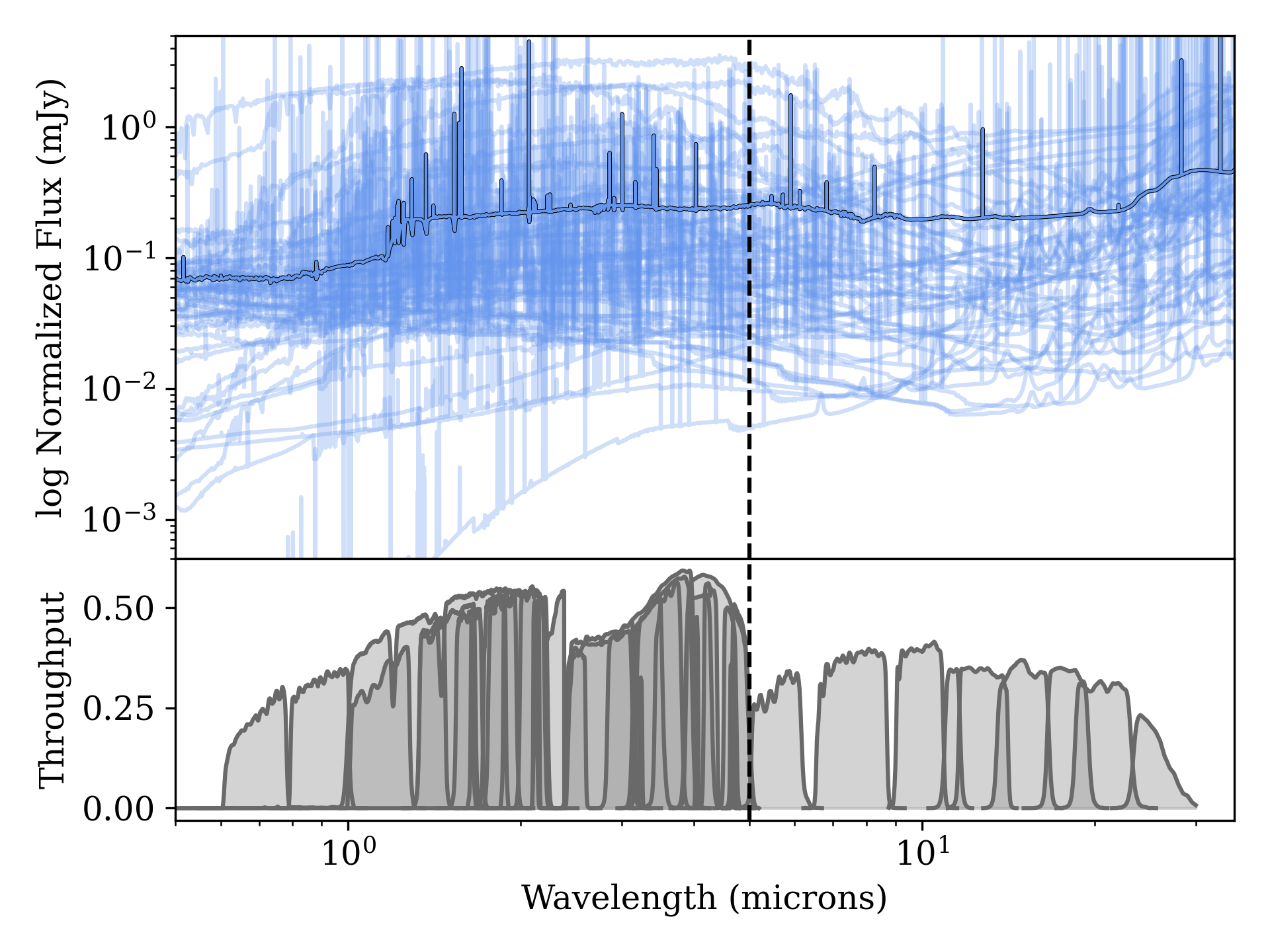}
    \includegraphics[width=0.5\linewidth]{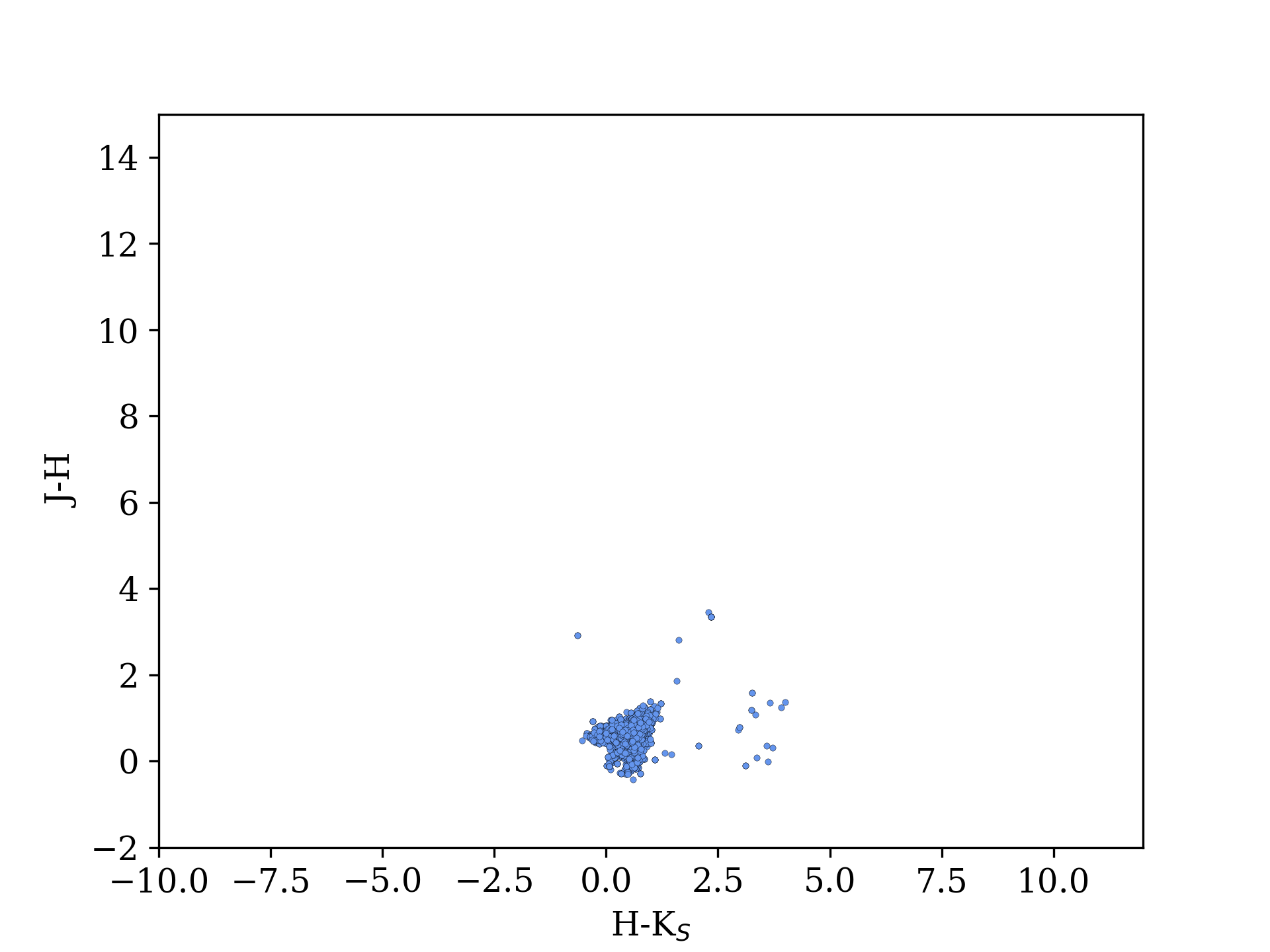}
    \caption{Similar to Figs.~\ref{fig:yso_spec}, \ref{fig:ms_spec}, \ref{fig:bd_spec}, and \ref{fig:wd_spec}, now for galaxies using CIGALE \citep{Burgarella2005,Noll2009,Boquien2019}. These galaxies are at various redshifts, with differing star formation histories, presence or absence of AGN, metallicities, and so forth, based on the parameters in \citet{Burgarella2020} and \citet{Wen2022}.}
    \label{fig:gal_spec}
\end{figure*}

JWST has surpassed expectations in its ability to image high-redshift galaxies. Galaxies, by nature, are ubiquitous throughout every field of view. They also have extreme variations across their evolution, from the star-forming main sequence, to star-bursting, those with strong AGN components, to red and dead galaxies. These variations leave distinct fingerprints in their SEDs. In our work, we have clumped these all together under the title of ``galaxy'' to keep our focus on the identification of different stellar evolutionary stages (YSO through white dwarf). Still, it is important to characterize these variations, as they cause galaxies to span a large range in color-color space, as shown in Figure~\ref{fig:gal_spec}.

To account for the myriad variations of galaxy SEDs, we utilize the models of the CIGALE group \citep{Burgarella2005,Noll2009,Boquien2019}, which account for stellar and nebular emission, dust absorption, IR re-emission by dust, star formation histories, metallicities, length of starburst, redshift, as well as AGN parameters such as optical depth, radial gradient of dust density, and polar angle effects, to name a few.  We use the parameters from Table 2 of \citet{Wen2022} for the low redshift (z=0-2) galaxies, and the parameters from ``Phase 4'' in Table 2 of \citet{Burgarella2020} for high redshift (z=2-10). 
To ensure a realistic sample of galaxies, we match the CIGALE-created galaxy models to a cross-matched catalog (Chris Willott, private communication) from two JWST surveys: the Systematic Mid-infrared Instrument Legacy Extragalactic Survey \citep[SMILES;][]{SMILESRieke2024,SMILESAlberts2024}, and the JWST Advanced Deep Extragalactic Survey (JADES; Chris Willott, private communication).
We then compare color-color diagrams to ensure that the population of JADES/SMILES data is appropriately recreated. In total, we create \editone{7815 unique} models using this method. 

Figure~\ref{fig:gal_spec} shows an example of a galaxy model SED, along with the variations one may see in such SEDs due to different processes, and the distribution of our training set of galaxies in color-color space. 
In the case of galaxies, the variations in possible environment are directly reflected in those seen across color-color space. The presence or lack of AGN, of star forming signatures, the 4000 \AA\ break, PAH signatures, etc. all contribute to a vast diversity in possible SEDs and thus wide extent across color-color space.

\subsection{Realistic Effects}

Up until this point, the data we have been discussing are entirely synthetic, without realistic effects. When taking into account realistic effects, one must consider the scenario at hand. 
For instance, when observing a star-forming region, it is important to take into account the presence of dust grains within molecular clouds, causing the dimming and reddening of light (extinction) from background sources. As well, Polycyclic Aromatic Hydrocarbon (PAH) molecules are present and excited in star-forming regions, and can also imprint their signatures on the measured starlight from background sources. Conversely, when considering a cosmological field, neither extinction nor PAH emission may be relevant, and the inclusion of those effects could be detrimental to the accurate identification of objects in such fields that do not have those signatures.

In the following sections, we explain how each realistic effect is added, with the caveat that their presence depends on the specific use case.

\subsubsection{Extinction Application}\label{sec:extinct}
Extinction, caused by the absorption of light on dust grains, is an inherently wavelength-dependent process; blue light will be preferentially absorbed or scattered and red light less so. Certain molecules also lead to specific absorption features. The level of extinction within a cloud is thus commonly defined based on a single bandpass. In this work, we describe extinction in terms of $A_V$, or visual-band extinction. 

As we will be considering the identification of objects embedded within dusty molecular clouds, it is critical to account for extinction. We thus apply extinction to each object from the above models, where the degree of extinction applied is the product of two probability distributions. After determining the extinction of a given object, we determine its effect on each wavelength/frequency band of the object's SED.

Firstly, we model the extinction of a molecular cloud as a log-normal distribution with an extended wing towards high density \citep{Burkhart2018}. In particular, we use a log-normal function with scale $m = 0$ and dispersion $\sigma =0.59$ for extinctions up to 1 mag, after which we switch to a power-law extension. The power-law is modeled as $10^{-\mathrm{ln}(A_V/\bar{A_V})}$ as determined qualitatively from the maps of \citet{Kainulainen2009}. When applying this distribution to our dataset, we assume an average visual extinction of 1.2 \citep[$\sigma$ and $\bar{A_V}$ are from the Orion B cloud;][]{Kainulainen2009}. 

Secondly, each type of object receives different extinction depending on its likelihood to be in front of or behind the cloud. For galaxies, they are assumed to be distributed uniformly behind the cloud. For field stars, these are assumed to have a bimodal distribution, either being in front of (40\%), or behind (60\%) the cloud. Since the brown dwarf and white dwarf model SEDs are faint, we anticipate that they will only be visible in front of the cloud, hence we do not apply any extinction to them. The YSOs are assumed to be within the cloud (with 10-90\% of extinction at their location being applied), and that they are only found in parts of the cloud with $A_V>3$ mag, i.e. they are embedded within some measure of cloud. We recognize that this will place some more-evolved YSOs in higher column density regions than one might anticipate if they have migrated from their birth sites, or if the part of the cloud from which they were born has begun to disperse \citep{Gupta2022}. Similarly, this assumption will place some less-evolved YSOs in lower column density regions than would be expected, since they are typically found above $A_V=5-7$ \citep[e.g.,][]{Lada2010}. The extinction model used here is chosen as the optimal for model performance, as well as being realistic. Doubling or halving the minimum $A_V$ leads to slightly worse performance in classifying real objects.

Finally, we use the extinction values from Table 5 of \citet{Wang&Chen2019} to compute the loss of flux expected in each filter by the given $A_V$ value to JWST NIRCam, 2MASS, and Spitzer IRAC filters. This reference does not provide similar conversions for longer wavelength IR bands, thus requiring us to take a different approach.
For JWST MIRI and Spitzer MIPS filters, we use the OH5 model from \citet{Ossenkopf1994} to approximate the extinction, interpolating to account for gaps in the table. We use the standard assumption that $R_V = 3.1$ and hence $A_V/A_K = 8.8$ \citep{Rieke1985} to convert from absorption coefficients to the appropriate extinction in each filter. %\editone{The OH5 model assumes an icy mantle on the dust grains, thus allowing us to further take this effect into account.}

\subsubsection{PAH models}

Polycyclic Aromatic Hydrocarbon (PAH) emission is ubiquitous throughout the universe and particularly in molecular clouds where it is associated strongly with massive YSOs that illuminate the surrounding gas. For our purposes, we are most concerned with the ways in which PAH signatures can contaminate other sources to provide excess emission. Indeed, aperture contamination by PAHs is a well-known phenomenon, and is taken into account in traditional methods of color cut separation \citep[see, e.g.,][]{Gutermuth2009}.

Regardless of the many unknowns associated with PAH emission, the progenitors of this emission have been broken down into four main species: PAH$^+$s (cationic PAHs), PAH$^0$s (neutral PAHs), PAH$^x$s (large -- $\sim 100 $ C atoms -- ionized PAHs), and eVSGs (evaporating very small grains). Observed PAH emission is further categorized by different ``classes,'' with each class having different levels of contribution from the four basic species. Class A and B are the most common, with Class A emission indicative of interstellar material illuminated by a star, be these HII regions, reflection nebulae, or the general ISM, while Class B is indicative of circumstellar material around Herbig Ae and Be stars, post-AGB stars, and planetary nebulae \citep{Peeters2002}. A third class, Class C, is much rarer and has been detected only in association with a few extreme post-AGB stars \citep{Peeters2002,Tielens2008}.

PAH emission is most clearly measured around the wavelengths of 3.30 $\mu m$, 6.20 $\mu m$, 7.70 $\mu m$, 8.60 $\mu m$, 11.30 $\mu m$, and 12.70 $\mu m$ \citep{Tielens2008}. The JWST bands thus most affected by PAH emission are: F335M, F770W, F1130W and F1280W. To account for this emission, we utilize the template models from \citet{Pilleri2012}, along with the associated 3.3~$\mu m$ emission inferred from \citet{Foschino2019}. These models include the four different species of PAHs, from which we compose Class A PAH emission spectra, since this class is the most spatially wide-spread within star forming regions, and thus is most likely to occur between a background source and the viewer. Class A PAH emission is composed primarily of the PAH$^0$ and PAH$^+$ species \citep{Peeters2002}, with minor contributions from the PAH$^x$ and eVSG species. 
As a basis for the relative flux contributions of each species, we refer to the distributions seen across real objects, as measured in Table 6 of \citet{Foschino2019}. 
Similarly, to determine what an appropriate level of flux would be for a given source's SED, we use the ratios in Table 6 of \citet{Foschino2019} as a guide.

Not every sight-line in a molecular cloud has PAH emission, and it is important to have representative samples of both cases with and without PAH emission.
As such, PAH emission is added to 50\% of all objects experiencing an extinction greater than $A_V=0.5$, to get a solid sample of all types of objects that may experience such contamination within our training set. 
Since the PAH emission originates within the molecular cloud, it may also be subject to some of the dust extinction present along the same line of sight.  To address this effect, we apply the extinction and PAH emission as three steps.
We take the extinction assigned to the source (see previous section), and choose some random value between 0 and 1. This fraction is then multiplied by the extinction to determine how much material is between the source and the PAH emission source. Then, the PAH emission is added. Finally, the remaining extinction is added to the combined signal of source object and PAH emission.

\subsubsection{Noise}

Along with the above effects, we add noise to the flux information in each band. 
Noise is defined at runtime based on the input real dataset being classified, where the noise is sampled from the distribution of errors of these data.
For each filter, these errors are used to define the standard deviation of a Gaussian with mean 0, from which a value is randomly pulled and then added to the synthetic data to generate a noisier set of magnitudes and thus colors.
In the specific case of where no real data have been supplied and the filter choice is being evaluated (see Section~\ref{sec:tool}), we leave it to the user to decide the best method of defining errors and thus noise. For example, one simple approach would be to use Gaussian errors with a mean of 0.1 mag and standard deviation of 0.02 mag.

\section{Methods}\label{sec:valid}

\subsection{XGBoost}

SESHAT is built on top of a machine learning framework. We use machine learning to be able to classify \textit{any} set of JWST data, regardless of the number of bands present. In the previous section, we defined the various SED models, which by their synthetic nature allow us to use the same set of data for all possible filter combinations, and thus to create a model with a consistent basis for the determination of class.

In this work, we use XGBoost \citep{xgboost}, a decision tree-based machine learning method for multi-class classification. This method is chosen in part for its similarity to color cuts as it makes splits at each node of the tree. XGBoost was also chosen for its ability to avoid overfitting through the use of parameters and early stopping, i.e., when a model stops being trained when it begins to perform worse on a dataset used to validate the model. Finally, XGBoost natively handles missing values without the need for imputation. 

Missing values are common in photometric datasets, either due to incomplete coverage, or saturation / obscuration of the source. The methods of imputation thus far available cannot take into account the possible variations, unless one has flags on the data to indicate what appropriate values may be. XGBoost dynamically determines optimal branch directions for missing values during training. In the case where there are no null values in the training data, XGBoost always chooses the right-side branch when it must make a decision based on missing data. It is thus important to include missing values in the training so the model can learn the optimal branch direction.

We add nulls to the dataset in three separate ways. First, we account for sources where some data are below the limiting magnitude of the observation by taking a subset of the training set, and setting the brightest filter to be between 2 mags and 4 mags brighter than the limiting magnitude of that filter, and scale the rest of the data appropriately to match. Then, for each filter we determine if it is above or below the limiting magnitude of that filter. Similarly, we take another subset of data and set the dimmest magnitude to be between 1 mags and 5 mags dimmer than the saturation limit, rescale, and set all filters with magnitude brighter than the saturation limits to null. Finally, we set random filters to null throughout the dataset to account for when data are not imaged or are otherwise lost to artifacts in the data. This occurrence happens most frequently at the edge of the field, where the offsets between images can lead to some objects having data in some, but not all, filters.

It is important to note that increasing the number of occasions of null data (i.e., decreasing the amount of usable data) leads to decreasing performance. We especially note that inclusion of MIR data (MIPS/MIRI) is essential for the recovery of most YSOs. Without these longer wavelength data, as much as half the YSOs may be misclassified as galaxies, see Appendix~\ref{app:sesna-missing-data}.
% To account for this need, we include $n$ additional subsets of 1000 sources randomly sampled across all objects, where
% $n$ is the number of filters under consideration. 
% In each subset, the data from one filter are set to null, and thus all the colors that would use that filter are similarly set to null.

\subsubsection{Training, Validation, and Test Sets}
With machine learning methods, there are three different types of datasets: training, validation, and test sets. For our training set, we use 75\% of the synthetic models previously derived. We then over-sample (resample the same dataset) to achieve 10,000 objects \editone{in each of the YSO, field star, and galaxy classes, and 5,000 in each of brown dwarf and white dwarf classes, where the latter classes are rarer in JWST data compared to the former \citep[e.g.,][]{Chenbd2025}, and as such are chosen to be less represented in the training set to moderate against overfitting to less likely classes.} \editone{We perform this over-sampling by resampling our original dataset, and then each row is given a random amount of noise as discussed previously to make each row unique. Although other methods of over-sampling exist \citep[e.g.,][]{SMOTE2002}, these methods will fill the space between different true models, in a naive fashion for the complex and overlapping populations we are considering here. Thus, we chose to keep our datasets constrained to those objects which have been either compared to or fitted to real populations (\S \ref{sec:data}), and simply add noise to the data to increase the number of unique samples whilst ensuring the oversampling does not create rows with SEDs that would match an object type other than the one intended. We note that, since we have 12,000 field stars available, we are not over-sampling to 10,000, but rather choosing 10,000 sources at random. Thus, our training set is composed of 40,000 sources in total.}

\editone{Meanwhile, the validation and testing sets use the remaining data that was not over-sampled}. The validation set is used to determine when to stop the training, and is necessary to prevent over-fitting. When possible, it is best to use a real dataset for this step. Since we are creating a method for application to any suite of filters and thus real classifications are not readily available, we validate using 50\% of the remaining synthetic data not used in the training set, i.e., 12.5\% of the total synthetic data. Finally, the test set is a set of data with no influence on the machine learning method, and is a check to see how well the model generalizes to new data. We use the remaining 12.5\% of the synthetic data as our test set.
In the following subsections we verify these methods on real data as an additional check.

The input parameters for our model are listed in Table~\ref{tab:xgb-pars}. \editone{With the exception of the objective and eval\_metric,} these parameters are  chosen to minimize over-fitting to synthetic data, based on a comparison to the SESNA catalogs that were used to constrain the YSO synthetic population. We note, however, that these parameters are set before run-time, and thus are applied to all datasets. \editone{The objective was set as multi:softprob to return a matrix of probabilities for each possible class. This allows the user to determine what the most appropriate cut in probability is for their science case, whether that is probability of X class $>90\%$ for a high purity but small sample, or $>50\%$ for a lower purity but more complete sample as done here.}

\begin{deluxetable}{cc}
\tablecaption{Parameters for XGBoost}
\label{tab:xgb-pars}
\tablehead{
\colhead{Hyperparameter} & \colhead{Value}}
    \startdata
    $\eta$ & 0.01\\
    $\gamma$ & 15\\
    subsample & 0.3\\
    max\_depth & 1 \\
    num\_class & 5 \\
    objective & multi:softprob\\
    eval\_metric & mlogloss\\
    \enddata
    \tablecomments{$\eta$ is the learning rate, $\gamma$ is set high to minimize overfitting, subsample is used to vary the objects seen by each tree to also minimize overfitting, max\_depth is kept shallow to minimize overfitting, the objective is set to return probabilities for each class rather than classes, and we use the log loss as our evaluation metric to determine where to call early stopping.}
\end{deluxetable}

%Finally, SESHAT is designed to return probabilities as well as classifications based on the most probable outcome. With machine learning methods, however, the probability can be more aptly considered the likelihood the machine will classify that source as that class. To obtain more realistic probabilities, they must be calibrated. This is done using the CalibratedCV method from sklearn. \bc{Did this, but I feel it's more confusing than helpful until I've had a good think about phrasing.}

\subsection{Generic Synthetic Test}
\begin{figure}
    \centering
    \includegraphics[width=\linewidth]{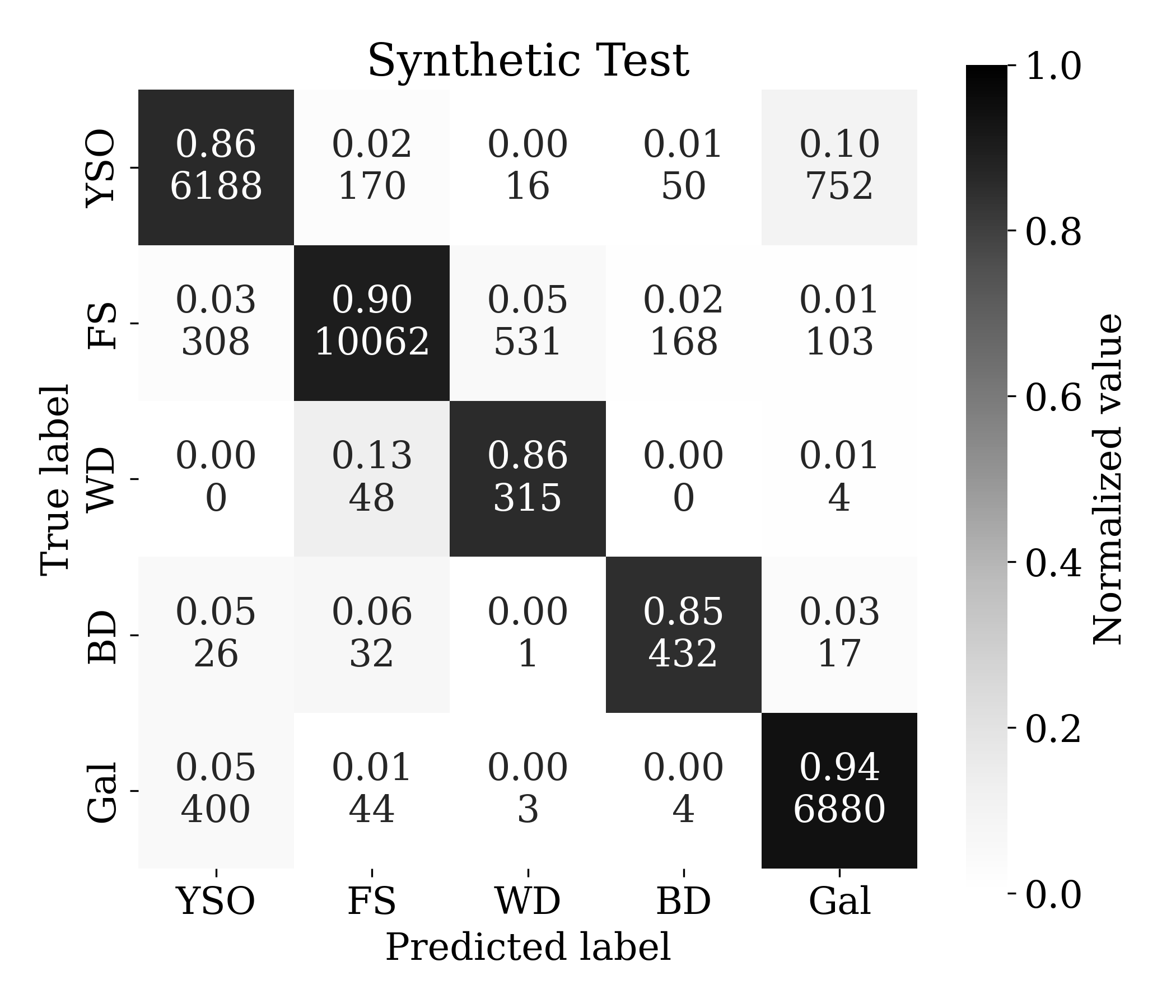}
    \caption{Confusion matrix for the test set of synthetic data, using Spitzer and 2MASS filters. In each case, we include the fraction of objects of the true label in each bin, as well as the total number. Grayscale is given by the fraction.}
    \label{fig:test_synth}
\end{figure}

The success of classification with the synthetic test set is shown in Figure~\ref{fig:test_synth}. 
Overall, we perform to a minimum of 85\% recall across all objects, when using the Spitzer IRAC and MIPS 1 bands, along with the 2MASS J, H, and Ks bands, where recall is a measure of how many objects are correctly classified vs. how many objects are of that class. Ideally, one would want to see values approaching 1 along the diagonal, indicating accurate classifications. The values in the off-diagonal boxes describe what percent of the objects of that true label have been mislabeled as the predicted label.
The exact performance will vary depending on the input filters, and so Figure~\ref{fig:test_synth} acts as an example of possible performance. For any application, the test set acts as an approximate bound on the expected contamination rates. For example, in this case we can expect that 10\% ($752/7176$ from the number of objects in the test set) of YSOs will be mislabeled as galaxies, and approximately 5\% of the galaxies will be labeled as YSOs.

\subsection{Performance with Major Contaminators}

\begin{figure}[htbp]
    \centering
    \includegraphics[width=\linewidth]{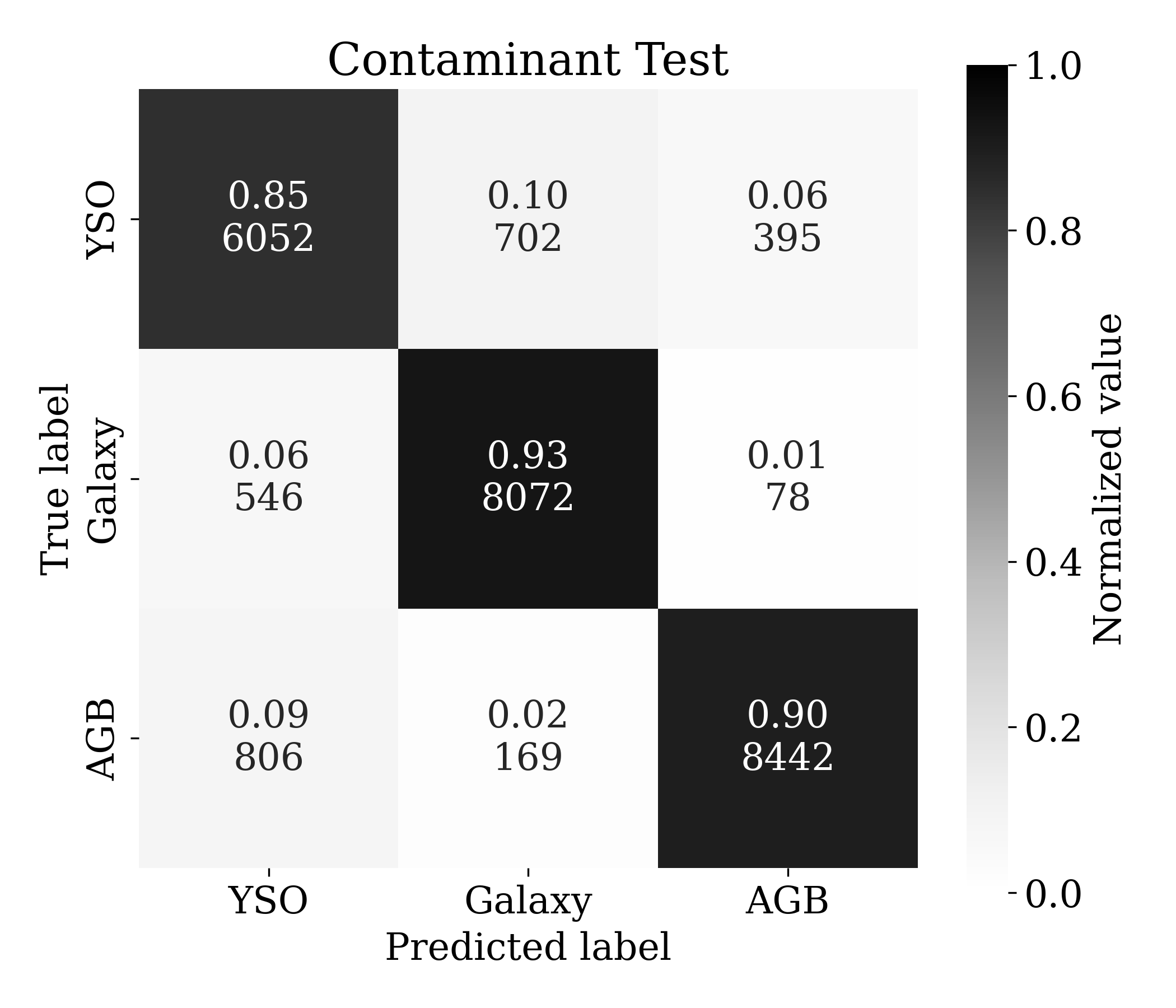}
    \caption{Confusion matrix of the test set where we sought only to identify YSOs, AGB stars, and galaxies, using Spitzer and 2MASS filters.}
    \label{fig:yso-agb-gal-test}
\end{figure}

Color cuts for YSOs are most often plagued by contamination from AGB stars \citep[e.g.,][]{Dunham2015}, and unresolved star-forming galaxies \citep[e.g.,][]{Gutermuth2008,Gutermuth2009}. AGB stars can be easily confused for YSOs, as they often have a double-peaked SED due to being surrounded by shells of circumstellar material, while AGN and unresolved star-forming galaxies also can be obscured by dust. We hence create a new synthetic catalog of only YSOs, AGB stars, and galaxies, so we may have a statistically large sample to test SESHAT on these most troublesome classes. Similar to before, we use the 2MASS and Spitzer channels for this test. The results of this test are shown in Figure~\ref{fig:yso-agb-gal-test}, where all classes are recovered to greater than 85\%.

\subsection{Performance in Star-Forming Regions Observed with Spitzer}\label{sec:sesna-sf}

Star-forming regions are some of the most difficult regions in which to classify objects, due to extinction, PAH emission, foreground and background stars, background galaxies, and, of course, YSOs. YSOs have a wide variety of spectral types, with the peak wavelength of their distributions varying rapidly with age. A YSO is generally characterized as a point source with a circumstellar dust component, be this an envelope or disk. The signatures of YSOs can be easily confused with other objects, however, as described with the previous test. In this section, we test SESHAT on Spitzer data from the Spitzer Extended Solar Neighborhood Archive (SESNA; R.A. Gutermuth et al., in prep) of star-forming regions. 

The SESNA catalogs are a rigorously tested and well-defined set of catalogs of several star-forming regions, varying in both distance from the Sun and environment (e.g., those highly irradiated by massive stars or quiescently collapsing). These data have been homogeneously analyzed to identify the following types of objects: YSOs (in their various classes of evolution), galaxies (both unresolved star-forming and AGN), field stars, and knots of shocked emission. SESNA thus allows us to test SESHAT on several different regions with homogeneously classified catalogs. 

We used fits to photometry of YSOs in the three clouds nearest to the Sun (Ophiuchus, Taurus, and  Corona Australis) plus the Orion A+B clouds, in the SESNA catalog to aid in the building of our synthetic sample of YSOs (see previous section for further discussion). As such, we exclude objects in those regions from our test set. As well, the Pipe Nebula data contain several artifacts, leading to less dependable categorization of YSOs, and suffers from high contamination rates due to its location within the Galactic plane. Thus, we also exclude its objects from our test set. Similarly, the data for Cygnus X were processed differently in comparison to the rest of the sample. Cygnus X is located at a much greater distance than the rest of the clouds, and has a very massive spatial extent. The distance measurements of this region may then be uncertain. Furthermore, Cygnus X has a higher rate of field stars (due to its distance and spatial extent), a higher rate of patches of bright nebulosity, and is shallower in the IRAC bands. UK Infrared Deep Sky Survey Galactic Plane Survey \citep[UKIDSS GPS][]{Lucas2008} rather than 2MASS data were used for Cygnus X as well. Due to the variety of these factors, we elected to also exclude Cygnus X from our test set.

\begin{figure*}[t]
    \centering
    \includegraphics[width=0.7\linewidth]{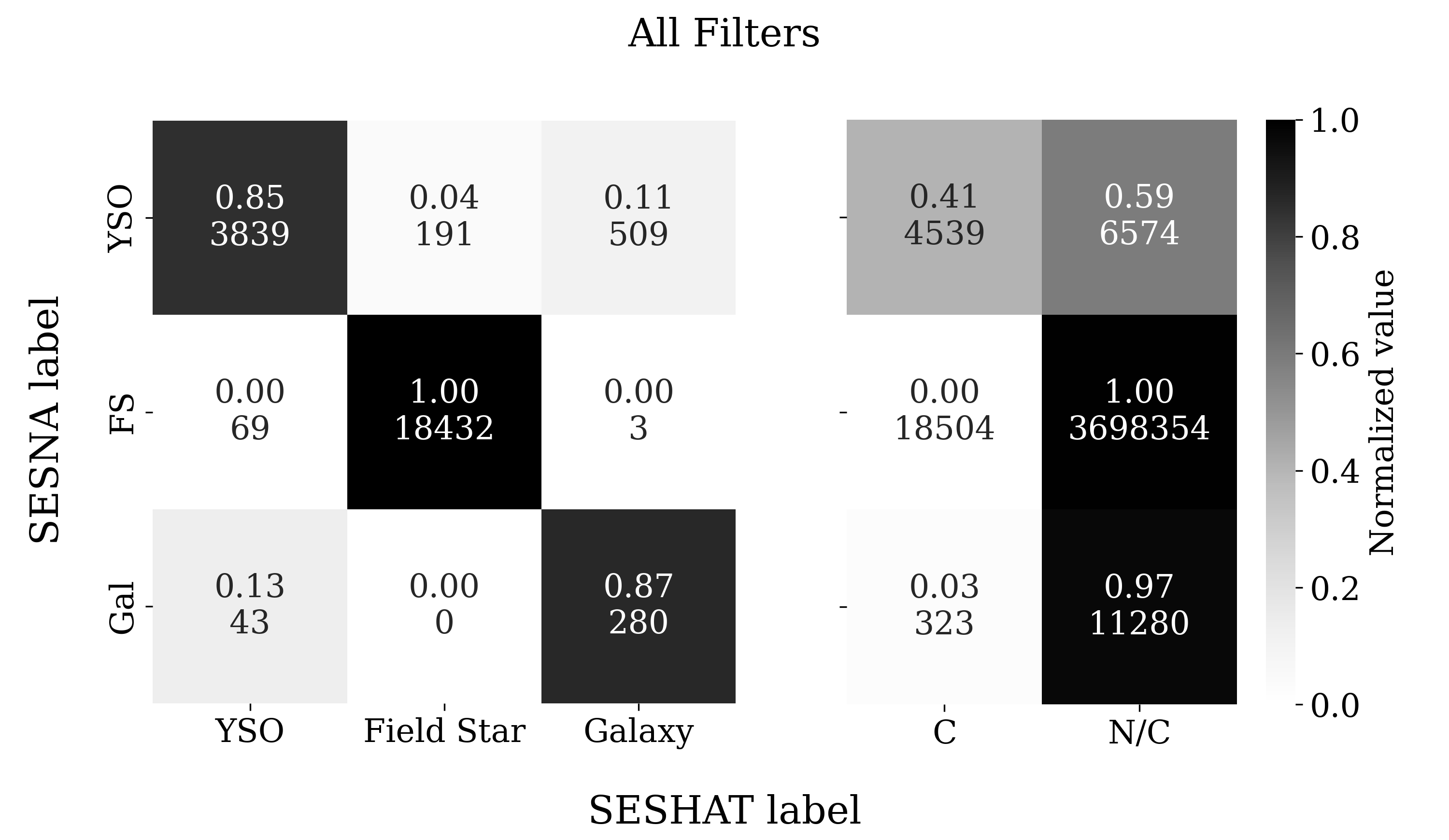}
    \includegraphics[width=0.7\linewidth]{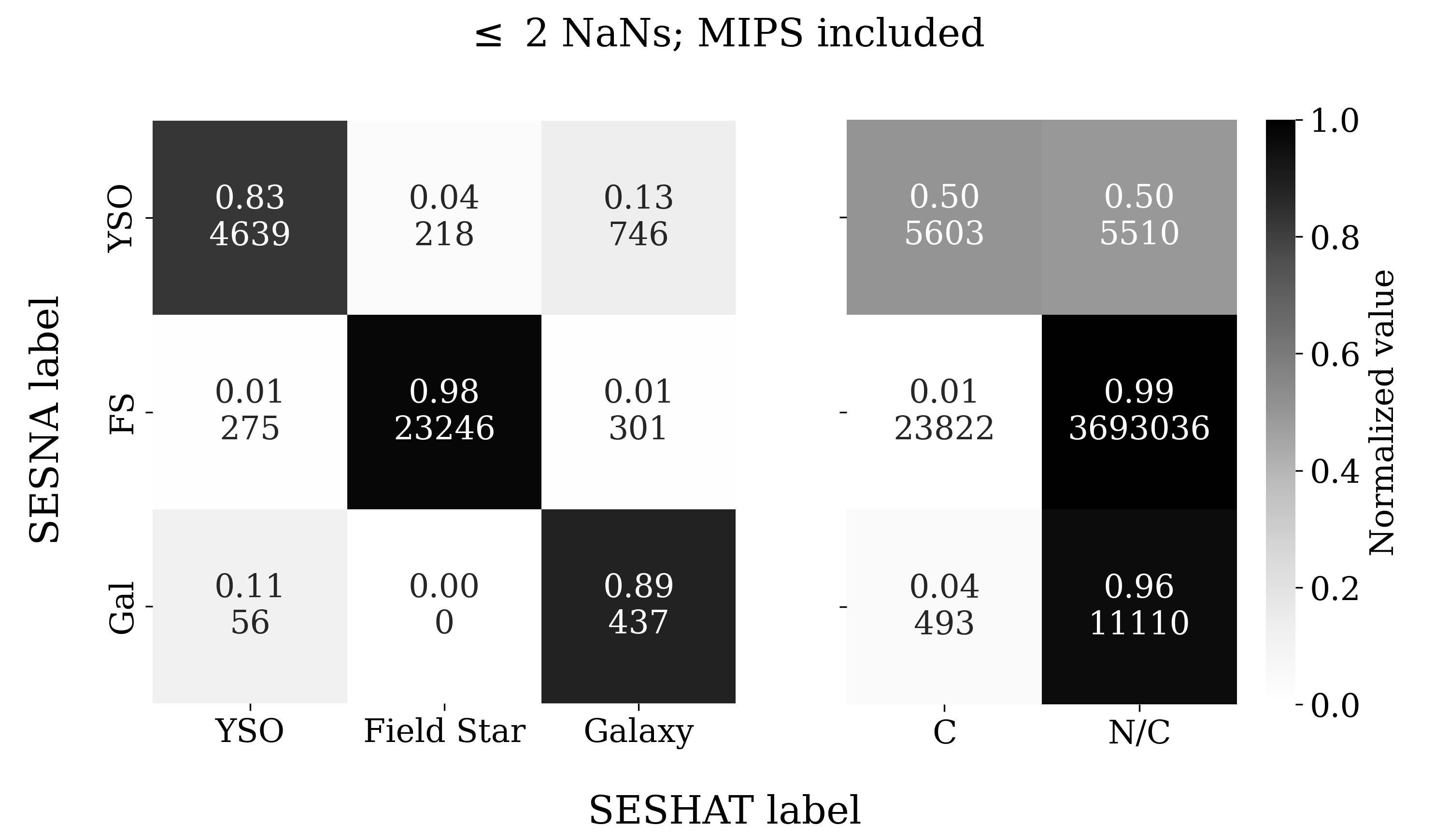}
    \includegraphics[width=0.7\linewidth]{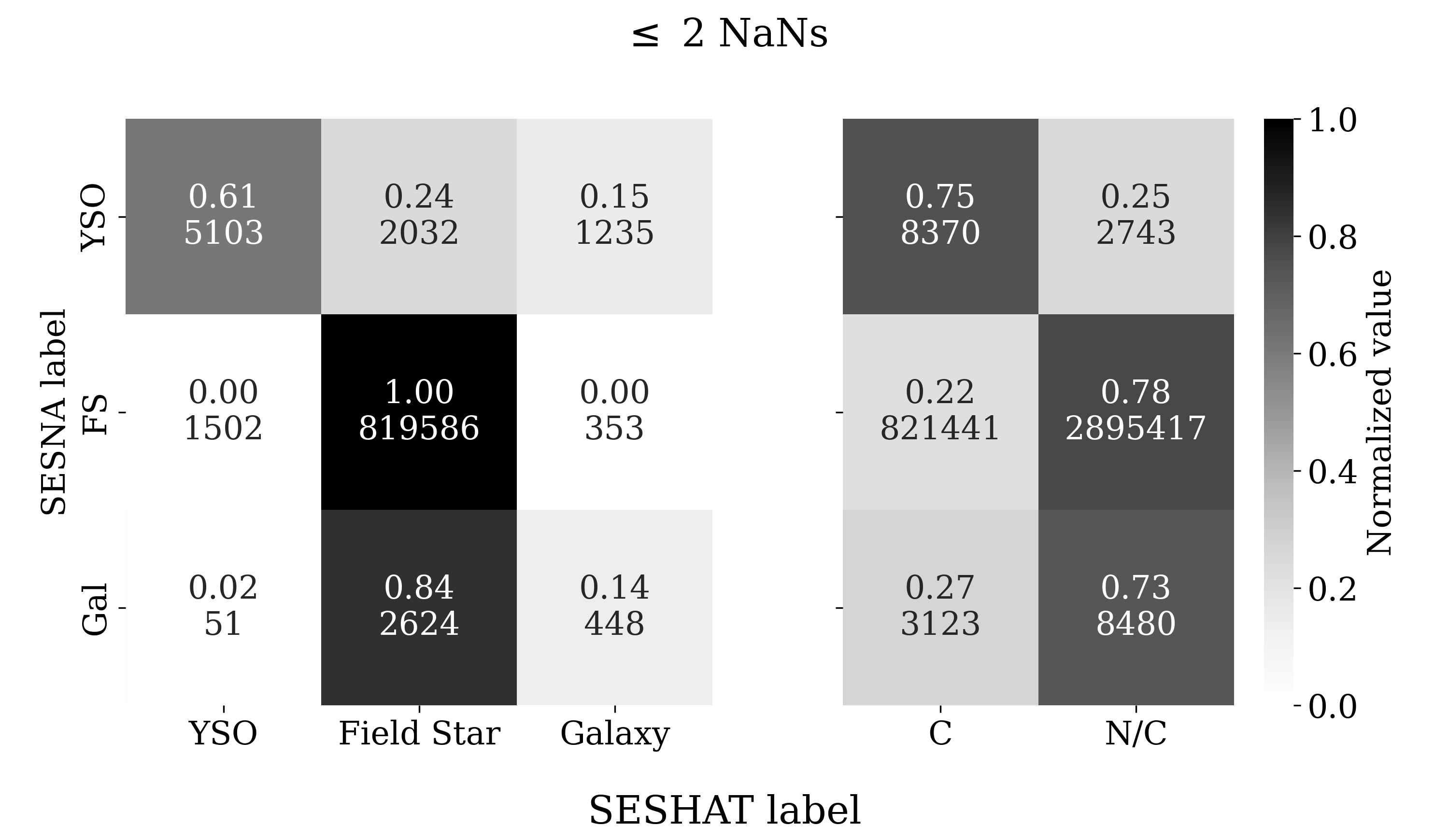}
    \caption{Confusion matrices for the test set portion of the SESNA catalog, for three different cases. In all cases, the matrix to the left shows the agreement between SESNA and SESHAT classifications, and the matrix on the right shows the number of objects classified (C) or not classified (N/C) for each case. \textit{Top}: this case is for when all filters must be present. \textit{Middle:} in this case, the MIPS filter must be present, but as many as 2 missing filters are allowed. \textit{Bottom:} in this case, two missing filters are allowed, including MIPS, to show how necessary MIR data are for classification.}
    % \includegraphics[width=\linewidth]{SESNA_Test_Sept132025.png}
    % % {Test_Apr82025_all.png}
    % \caption{The confusion matrix for the test set portion of the SESNA catalog. }
    \label{fig:sesna_cm}
\end{figure*}

The test set is thus composed of objects from 22 different star-forming regions (see Table~\ref{tab:app-sesna-deets} for a full list of regions).
As described in detail in Appendix \ref{app:sesna-missing-data}, the SESNA catalog uses archival data from multiple facilities to classify as many objects as possible. Because of this heterogeneity, a significant fraction of the catalog (99.2 \%) is missing data in one or more bands.  In early testing, we found that SESHAT's effectiveness in classification varied depending on which and how many bands of missing data were present.  Here, we present three cases to showcase this variation. Figure~\ref{fig:sesna_cm} shows the results for three different tests, with varying cuts on the filters \editone{required for classification by SESHAT}, allowing any SNR in all cases. Appendix~\ref{app:sesna-missing-data} explores the results when no constraints are applied, as well as the distribution and number of sources with missing data. In general, we find that the classifications of SESHAT in comparison to SESNA is heavily dependent on having MIPS 1 data available. \editone{In general, SESHAT will classify the entire catalog given to it, regardless of the number of filters available, thus it is up to the user's discretion to determine what a reasonable cut on the available data is for their science case.}
When \editone{only those data that have all filters available are classified}, we retrieve greater than 85\% recall across YSOs, galaxies, and field stars. Indeed, we also have a similarly good recovery rate \editone{when only those data with up to two missing filters are classified}, so long as MIPS data are available and included.
Inconsistent classifications are discussed in detail in the next section. In short, some fraction of the SESHAT ``misclassifications'' are actually correctly identifying expected contaminants in the original catalog. Similarly, there are objects that we misclassify that the color cuts from SESNA appropriately classify. Appendix~\ref{app:sesna_cm} shows the confusion matrices for each individual region, including the regions excluded from our main analysis.

\begin{figure}
    \centering
    \includegraphics[width=\columnwidth]{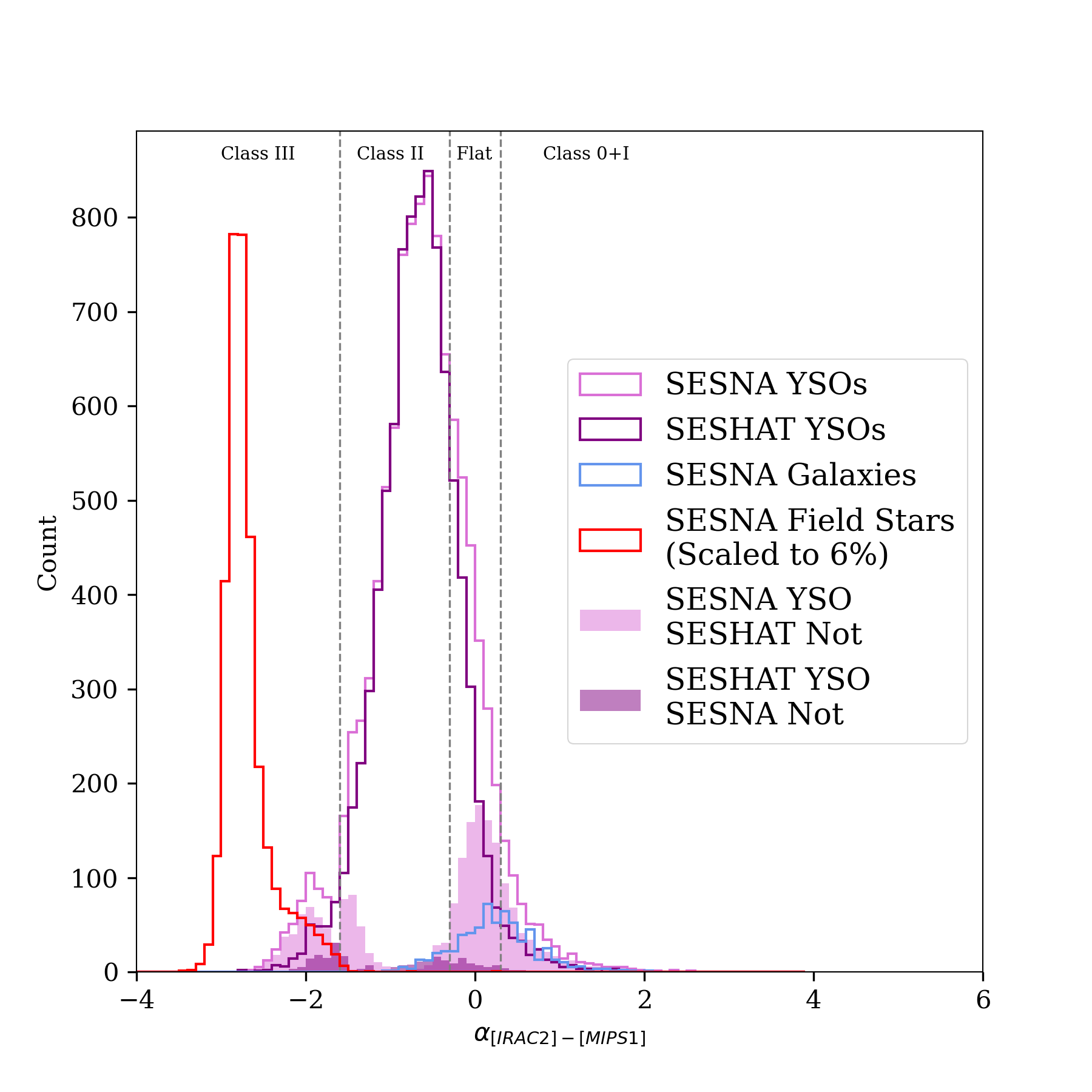}
    \caption{Recovery of the different classes of YSOs from SESNA, based on spectral index, Equation~\ref{eq:alpha_col}. The population of objects identified as YSOs by SESNA, and not by SESHAT are shown in pink, and the population identified as YSOs by SESHAT and not SESNA are shown in purple. Less evolved YSOs typically have higher spectral indices, and entirely overlap with the galaxy population (blue). Meanwhile, more evolved YSOs have lower spectral indices and overlap with the field star population (red). The field star population histogram is multiplied by 0.06, in order to see the full shape of the histogram. This figure is for those objects with data in all filters.}
    \label{fig:yso_recovery}
\end{figure}

We further check to see if there is any class of YSO that we are preferentially classifying differently than SESNA. Class, here, is defined using the spectral index, 
\begin{equation}
    \alpha = \frac{d\log{\lambda f_\lambda}}{d\log{\lambda}}
    \label{eq:alpha}
\end{equation}
which is a measure of the slope of the spectrum. Approximating this using the bands least affected by reddening, as done in \citet{Kuhn2021}, we get 
\begin{equation}
    \alpha_\mathrm{[IRAC2]-[MIPS1]} = 0.55(\mathrm{[IRAC2]-[MIPS1]}) - 2.94.
    \label{eq:alpha_col}
\end{equation}
In general, less evolved YSOs and galaxies are typically found to have higher spectral indices, while more evolved YSOs and field stars are found have lower spectral indices. Figure~\ref{fig:yso_recovery} shows the spectral index of all sources, along with demarcations for the typical class designations of YSOs. We find that, with SESHAT, we have a higher likelihood of classifying an object of very low or very high spectral index differently than SESNA, precisely where the overlaps with other classes of objects lie. That is -- SESNA YSOs of high spectral index are often labeled as galaxies with SESHAT, and SESNA YSOs of low spectral index are often labeled as field stars in SESHAT.
These results are for when all filters are available. As more data are allowed to be missing, the peaks seen here increase in height.

We note that neither the SESNA classification method nor SESHAT use spatial information. Exploring the spatial distribution of the differently classified objects could act as an independent check for the validity of classification. We recommend that the users perform similar sanity checks of any obviously strange classifications when applying SESHAT to their own data.

\subsection{Performance On Cosmological Fields}

\subsubsection{Analyzing YSO Contamination in Spitzer Bo{\"o}tes Data}

\begin{figure}
    \centering
    \includegraphics[width=\linewidth]{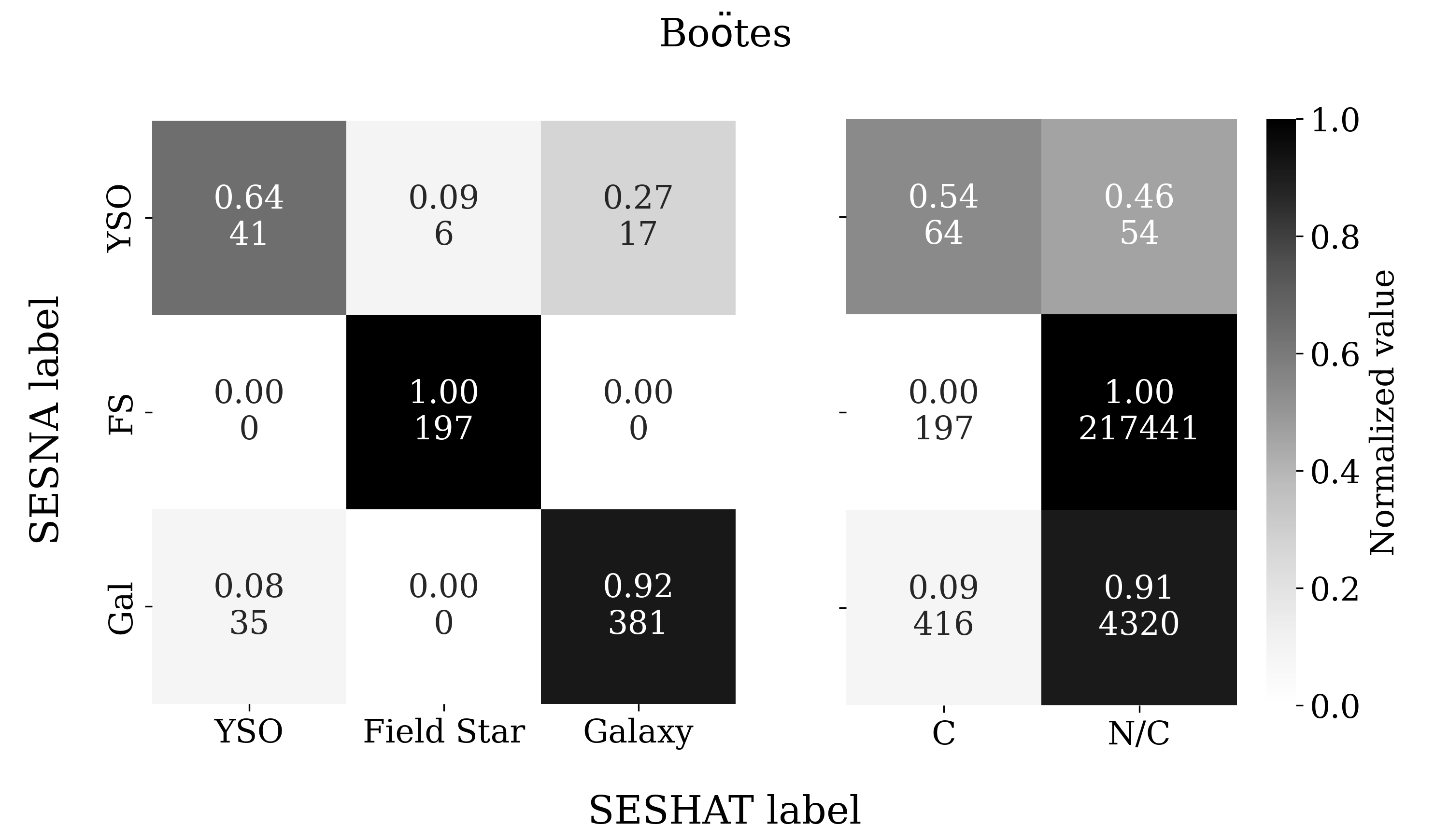}
    \caption{Confusion matrix for the Bo{\"o}tes field from SESNA.}
    \label{fig:bootes}
\end{figure}

Along with the aforementioned catalogs of star-forming clouds, the Spitzer Extended Solar Neighborhood Archive (SESNA; R.A. Gutermuth et al., in prep) data include an 8.84 square degree field of the Bo{\"o}tes cosmological field. The SESNA archive cataloged these data the same as for the star-forming regions, with classes of Field Star, Galaxy, or YSO, to determine the contamination rate of their YSO selection. This rate was determined as $9\pm1$ per square degree \citep{Gutermuth2008,Gutermuth2009,Pokhrel2020}. We re-analyze the same data with SESHAT to determine our own contamination rate of the results described in the previous section. We note, again, that the performance will vary dependent on both the filters present and the classes chosen, and thus this example is a demonstration of its performance, with the filters those of J, H, and Ks from 2MASS and IRAC 1-4 and MIPS 1 from Spitzer. \editone{We thus choose to classify only those objects with data in all filters, as before, and the results are shown in Figure~\ref{fig:bootes}.}

With these data, we are able to recover the field stars and galaxy classes to nearly the same degree as SESNA. We find that there is a trade-off in the objects classified as YSOs between the two methods, with SESHAT finding more false YSOs than SESNA.  In either case, there should be no YSOs, and yet some misclassification persists, at similar levels to what can be achieved through color cuts. 

\subsubsection{Searching for Brown Dwarfs in JWST COSMOS Data}

Along with application to Galactic regions, SESHAT can also be used for the identification of brown dwarfs in cosmological fields.
% , with a couple minor tweaks. First, such fields will be out of the Galactic plane and so it is no longer necessary to include intervening extinction $>1$ mag in $A_V$, or PAH emission. Second, YSOs can be removed from the training sample, as they are unlikely to exist in the Galactic halo. We perform these adjustments and 
We review the data of the JWST COSMOS field \citep{Cosmos2023,cosmoscat2025}, which have been previously searched for brown dwarfs using color cut criteria \citep{Chenbd2025}. We compare our results to \citet{Chenbd2025} as further validation of these methods. 

\begin{figure}[htbp]
    \centering
    \includegraphics[width=\linewidth]{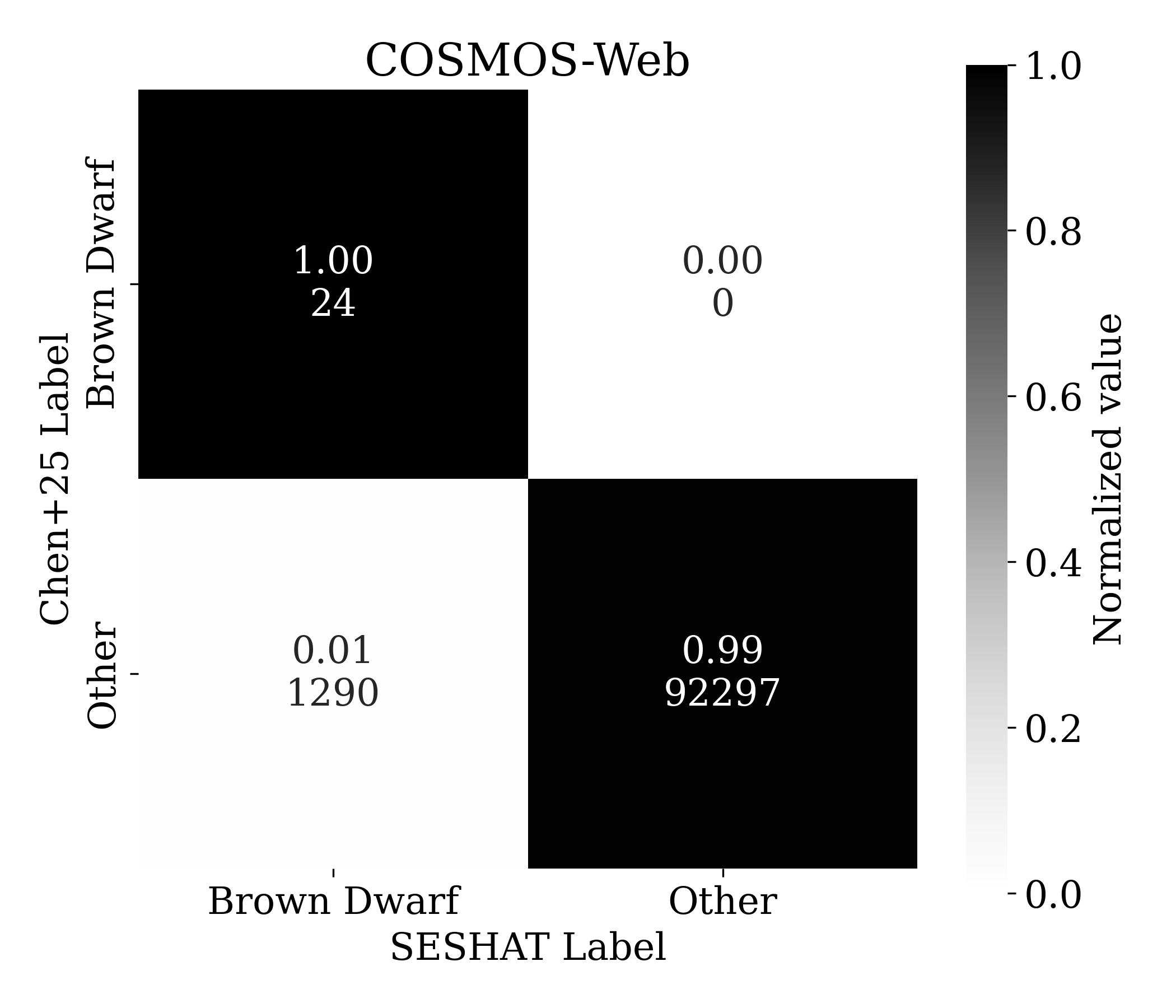}
    \caption{The confusion matrix for the COSMOSWeb field, based on the classifications of brown dwarfs in \citet{Chenbd2025}.}
    \label{fig:cos-field-cm}
\end{figure}

The COSMOSWeb data were taken in only five JWST bands: F115W, F150W, F277W, F444W, and F770W, with the MIRI F770W band having significantly less coverage than the NIRCam bands. The COSMOSWeb catalog \citep{cosmoscat2025} includes flags for the quality of the photometric detection. We use all sources that are labeled as secure, as well as those labeled inconsistent between ground and space observatories, according to their scheme. We also set to NaN all measurements whose error on their flux exceeds 10\% of the flux itself. These choices were made to eliminate all sources that were likely to be hot pixels, artifacts, or only detected in one band. We further removed all sources where the major axis was more than 10\% larger than the minor axis, to remove objects that are unambiguously galaxies, by shape alone.

\citet{Chenbd2025} used several cuts in color space as well as the ellipticity of each point source for the determination of their brown dwarf candidates. We run SESHAT on only the colors and classify sources as one of brown dwarfs, white dwarfs, field stars, or galaxies. Doing this, we retrieve 100\% of the previously identified brown dwarf candidates. 
The remaining data (previously unlabeled) are split between field stars (22,094 or 24\%), galaxies (70,203 or 75\%) and many additional brown dwarfs (1290 or 1\%). Figure~\ref{fig:cos-field-cm} shows the confusion matrix of these results, where all additional classes are lumped into ``Other'' for ease of comparison with \citet{Chenbd2025}. NOTE: these data include null values, and thus demonstrate the accuracy of classification of sources other than YSOs when MIR data are not present.
%SESHAT further identifies 7\% of the remaining sources as brown dwarf candidates, which is far more than the 120 identified by the color criteria of \citet{Chenbd2025}.
%This precludes the classification of field stars within the population. When attempting to identify galaxies, field stars, and brown dwarfs, the contamination rate drops considerably, from 7\% to 2\%, with 22\% of the unlabeled sources categorized as field stars.
To further narrow the selection of brown dwarfs, it is useful to include size and shape information. SESHAT, however, provides a first pass that can identify a much smaller population for further follow up as brown dwarfs.

\section{SESHAT Implementation and Use}\label{sec:tool}
We have produced SESHAT: Stellar Evolutionary Stage Heuristic Assessment Tool; a tool for the classification of JWST, Spitzer, or 2MASS photometric data, and have made it available to the community as a Python package (\url{https://pypi.org/project/seshat-classifier/}). This tool has two intended uses. The first, is to provide probabilities and classifications for any JWST dataset. The second, is to aid in the writing of JWST proposals by evaluating a synthetic test set on a given suite of filters.

% The use of this tool is straightforward. It takes as input items:
% \begin{enumerate}
%     \item A catalog that must be classified
%     % \item Whether or not that catalog has labels already (for comparing to other methods of classification)
%     % \item If the region under review is a cosmological field or a star-forming field
%     \item The classes the user would like searched for. These can be ``YSO'' for YSOs, ``FS'' for field stars, ``BD'' for brown dwarfs, ``WD'' for white dwarfs, or ``Gal'' for galaxies)
%     \item The filters to be tested if no catalog is provided (for testing proposals)
% \end{enumerate}

To classify a set of objects, only two pieces of information are necessary: the catalog to be classified, and the classes to be searched for. These can be ``YSO'' for YSOs, ``FS'' for field stars, ``BD'' for brown dwarfs, ``WD'' for white dwarfs, or ``Gal'' for galaxies.
The catalog must contain the Vega magnitudes and errors for each source. From this input, the color information will be extracted, the model will be updated based on the uncertainties, limiting magnitudes, and saturation magnitudes of this dataset. A copy of the catalog with the probability of being classified as each object, along with the final class (based on whatever class has the maximum probability for that object), will be returned. Optionally, a copy of the true and predicted classes and probabilities for the synthetic test set can be returned to aid in determining contamination rates.  \editone{Note that SESHAT will assign probabilities and classify each source based on the data available. It is up to the user to decide how many columns must be filled with data to return a reasonably well-classified dataset for their science case. For instance, YSO classification requires mid-IR data, while these data are unnecessary if YSOs are excluded from the analysis, e.g., in a cosmological field. The users themselves must adapt their input catalog to remove those rows without the requisite number of filters available, based on their unique science case and goal.}

% To test JWST proposals, only a set of filters and desired classes must be specified. When this is done, the synthetic test set confusion matrix will be returned. This is provided in order to verify that the chosen bands are satisfactory for the evaluation of the desired class.
% \hk{[HK curiousity question: does the user need to supply estimated uncertainties / depths for each filter as well?  It seems like the ability to accurately classify objects would be reliant on the S/N of the data obtained.]}
To test JWST proposals, one must supply the filters, the limiting and saturating magnitudes of the proposed observation, and an appropriate distribution of uncertainties for each filter. From this information, a synthetic test set will be built and the true class, predicted class, and probabilities for each class will be returned.

\subsection{Limitations}
SESHAT has two limitations that bear specifying. (1) Classifications of YSOs are limited when mid-IR data are not available. Indeed, when only near-IR data are used, the classification accuracy of real data drops significantly, and thus, at this time, we recommend against using SESHAT to classify YSOs unless mid-IR data are available. For all other classes, it performs as expected when only near-IR data are available. (2) Though SESHAT can account for missing data, the overlap of classifications with those obtained from the color cut method drops significantly with increasing fractions of missing data.

\subsection{Probabilities}

For both real and synthetic datasets, a set of probabilities for each source is returned. These probabilities have been calibrated using \texttt{CalibratedClassifierCV} from the \texttt{sklearn} libraries. The probabilities quoted are the likelihood that the source will be classified under that label\editone{, and we encourage the user to refer to the probabilities rather than the labels given as most probable class}. The benefit of using probabilities rather than strict classifications lies in their flexibility. For instance, in a cosmological field, one may wish to set a higher threshold for the classification of an object as anything other than a galaxy \editone{to encourage greater purity of classes}. Or, one may decide to apply Bayesian inference to restructure the probabilities based on prior knowledge of the region or object. Or, if the difference between the probabilities of the top two or more objects is less than a certain threshold, an ``ambiguous'' classification can be applied. We encourage users of SESHAT to perform any modifications they deem fit to these probabilities, \editone{including modifying the threshold to which an object is attributed to a given class} to best benefit their science case.

\section{Conclusions}\label{sec:concl}
In this work, we have defined SESHAT, a Python tool that is made available to the community through the package \texttt{seshat-classifier}, available on PyPI. 
This tool is capable of classifying a range of evolutionary stages of stars, from YSO through to white dwarfs, and galaxies. In the era of JWST, where every observation has the potential to use its own unique set of filters, having a method that can be applied regardless of the exact filters available can be a great aid in both expediting classifications and allowing consistent classifications across datasets.

The key uses of SESHAT are as follows:
\begin{itemize}
    \item It allows the user to process a catalog of any JWST, Spitzer, or 2MASS data, to obtain probabilities for each object to be one of YSOs, field stars, brown dwarfs, white dwarfs, or galaxies. 
    \item Optionally, a copy of the test set is can be returned, so the purity of the classifications can be estimated.
    \item JWST proposals can be tested by specifying the filter setup and observational limits, whereby the test set classifications are returned.
\end{itemize}

We note to the reader the caveat that the classifications provided by SESHAT have a maximum accuracy given by the test set. The probabilities may be the more useful output, as they can be combined with Bayesian inference to obtain stronger classifications, depending on the region. It is also important to note that classical color cut methods also suffer from lack of purity, as the overlap of objects in color-color space can sometimes simply not be distinguished regardless of the cut applied.
Finally, the performance of both classical methods as well as SESHAT will vary with what input bands are available, due to the measuring, or lack thereof, of important features. SESHAT has been tested against Galactic star-forming regions and cosmological surveys to assess its real-world performance. We quantify this as:
\begin{itemize}
    \item When testing against synthetic data of the Spitzer/2MASS bands, recall above 90\% is achieved across all classes.
    \item When testing against real Spitzer/2MASS data of star-forming regions, recall above 80\% is achieved across all classes, when MIR data are present.
    \item When testing against a cosmological set for the identification of brown dwarfs, 100\% of the previously identified brown dwarfs were recovered, and the remaining unlabeled objects were split into  field stars (24\%), white dwarfs ($<1\%$), galaxies (76\%), and additional brown dwarfs (2\%).
\end{itemize}

We expect that SESHAT will be useful to a large portion of the community, especially as more JWST data become available.

\begin{acknowledgements}

The authors thank the anonymous referee for their helpful feedback.
The authors would like to thank Drs. Simon Blouin, Hossen Teimoorinia, and Chris Willott for their helpful discussions. 
BLC acknowledges the support of an NSERC Doctoral Award held at the University of Victoria.
HK and JDiF acknowledge support from their respective NSERC Discovery Grants.
Support for RAG to participate in this work was provided by NASA ADAP award 80NSSC23K0476. RAG also acknowledges funding support from: NASA ADAP awards NNX11AD14G, NNX13AF08G, NNX15AF05G, and NNX17AF24G that supported the development of SESNA; NASA ADAP award 80NSSC19K0591; NSF AST grants 1636621, 1812747, and 2107705; NASA-USRA SOFIA grants 05-0181, 07-0225, and 08-0181; NASA-JPL/Caltech Spitzer grants 1373081, 1424329, and 1440160; NASA-JPL/Caltech Herschel grant 148938;  NASA-STScI HST grant 14181.003; and NASA-STScI JWST grants 01783.012, 01802.010, 02092.006, 05405.004, 05804.008.
This work was carried out at the University of Victoria, on the unceded lands of the Songhees and Esquimalt peoples, whom we respectfully acknowledge. 
We acknowledge as well the resources of the NASA ADS service.
\end{acknowledgements}

\appendix 
\section{Missing Data in SESNA}\label{app:sesna-missing-data}

In \S~\ref{sec:sesna-sf}, we showed the performance of SESHAT in comparison to the Spitzer Extended Solar Neighborhood Archive (SESNA) classifications of YSOs, field stars, and galaxies across 22 different star-forming regions. We found that the performance of SESHAT was heavily impacted by the availability of mid-IR data in the form of the MIPS 1 band. With the MIPS data, we were able to reach $>80\%$ recovery of YSOs, even when as many as 2 of the other filters were missing data. Without, the performance was drastically reduced.

\begin{figure}[h]
    \centering
    \includegraphics[width=0.5\linewidth]{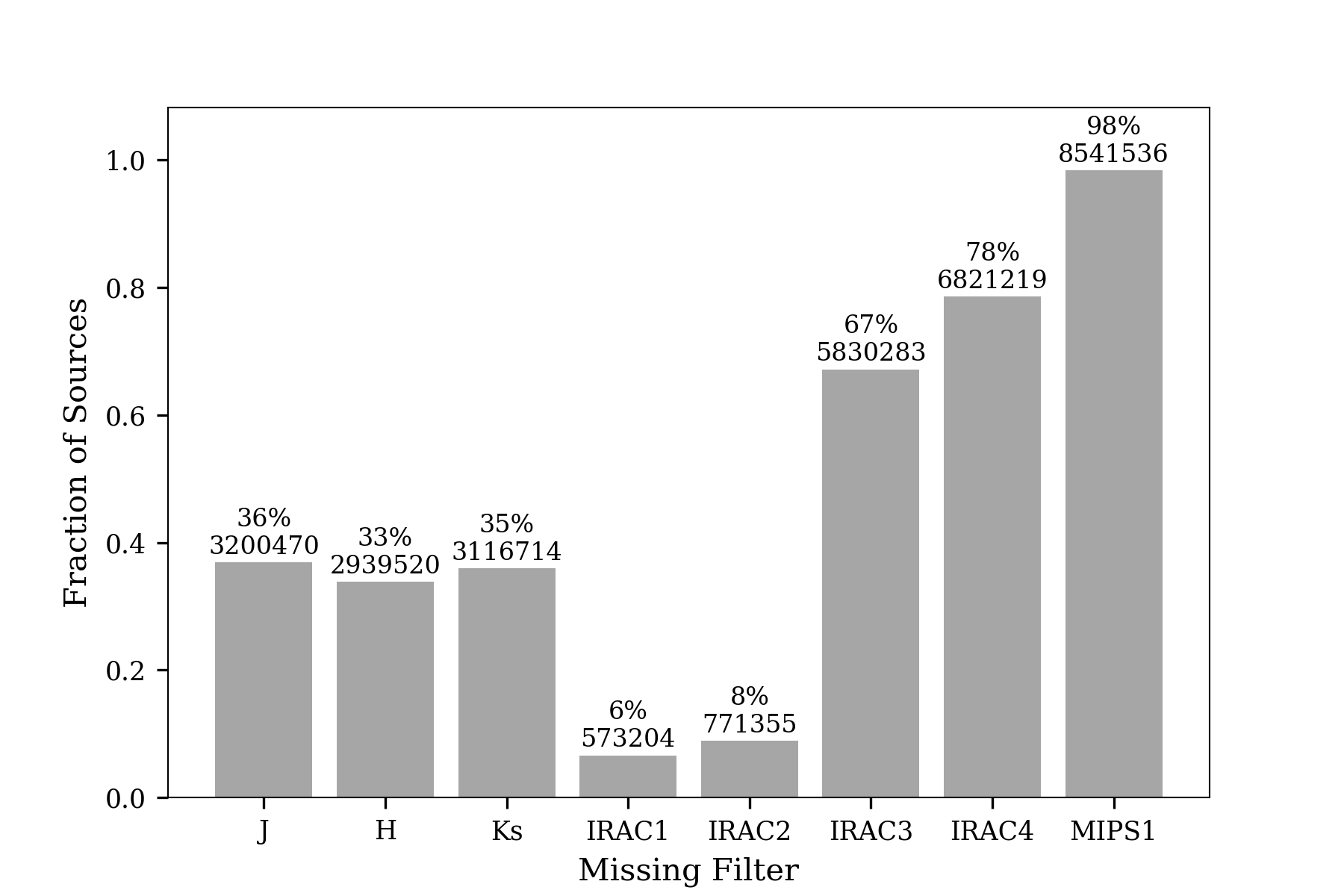}
    \caption{Histogram showing the fraction of sources with missing data in each filter. }
    \label{fig:missing_filters}
\end{figure}

\begin{figure}[h]
    \centering
    \includegraphics[width=0.5\linewidth]{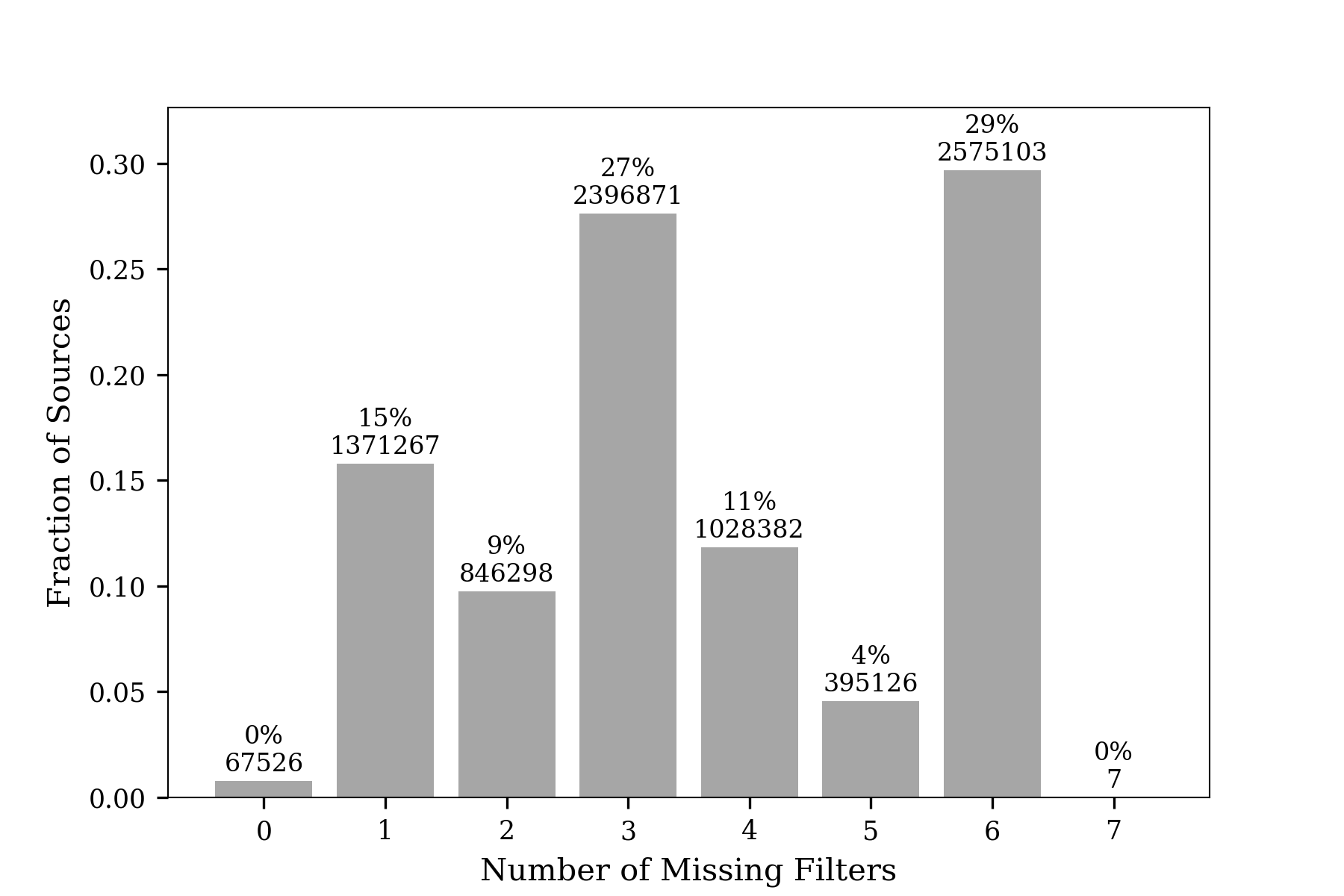}
    \caption{Histogram of the fraction of sources with X number of filters missing data.}
    \label{fig:overall_missing}
\end{figure}

Here, we briefly describe the extent of missing data in the entire SESNA catalog of 8,680,580 sources. Figure~\ref{fig:missing_filters} shows the distribution of missing filters. In general, data become sparser with increasing wavelength due to the combination of three affects: 1) the population of all regions is dominated by field stars, whose SEDs sharply decrease at longer wavelengths. Thus, the brighter completeness limits of the longer wavelengths are insufficient to detect the tails of these SEDs. 2) Background nebulosity in star-forming regions becomes increasingly brighter at longer wavelengths, and can wash out the signal from other sources. 3) Instrumental effects, such as poorer angular resolution, and increasing thermal noise. 
In general, the IRAC 1 and 2 channel arrays are far more effective at measuring photons than the IRAC 3 and 4 arrays, leading to the stark difference from the shorter to longer wavelengths \citep{Gutermuth2009}. Finally, the 2MASS filters are most often lost due to the shallow depth of field of the all-sky survey.
Altogether, this results in 99.2\% of sources missing data in at least one filter, and 46.1\% of sources missing data in at least 4 bands (half those used for classification). Figure~\ref{fig:overall_missing} shows the distribution of the fraction of missing filters across the entire sample.

\begin{figure}
    \centering
    \includegraphics[width=0.75\linewidth]{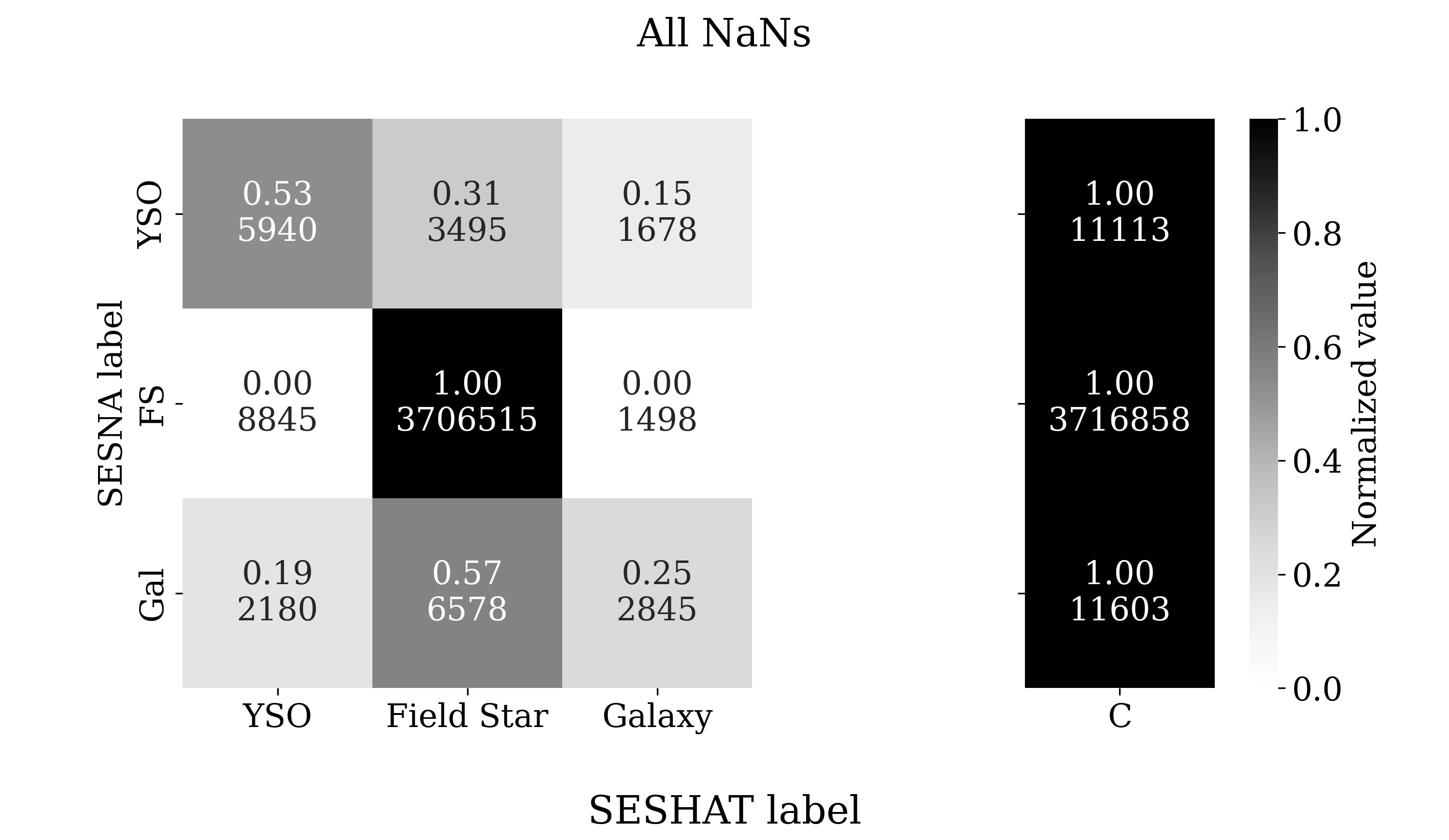}
    \caption{Confusion matrix for the SESNA test set, without any constraints on the number of missing filters allowed, at any SNR.}
    \label{fig:sesna_test-all-nans}
\end{figure}

Finally, we include here the match between SESHAT and SESNA when we classify the entire dataset, regardless of the number of filters with or without data, of any SNR. Figure~\ref{fig:sesna_test-all-nans} shows the confusion matrix for this comparison. SESHAT's classifications match those of SESNA for around 50\% of the YSOs and 25\% of the galaxies, however, a significant number of SESNA-identified field stars are labeled as YSOs, leading to a much higher contamination rate of that population. We note that some of this crossover difference in classifications may be correctly identifying YSOs and galaxies that were hidden in the field star population, but the majority are likely incorrect. Meanwhile, crossover between the YSOs and galaxies is near equivalent.

\section{SESNA Region Classifications}\label{app:sesna_cm}

% \textbf{Appendix removed to attempt arXiv upload.}
In this section, we include the confusion matrices with comparisons to the spatial distribution of objects for each region in SESNA, for objects with data available in all bands. We show all regions against the all-sky extinction map of \citet{Chiang2023}, accessed though the \texttt{dustmaps} Python package \citep{Green2018DustMaps}, as a guide to the underlying cloud structure. In all cases, the visual extinction range of each map is from the 5th to the 95th percentile of the extinction distribution in that region. Table~\ref{tab:app-sesna-deets} lists important details for each region, including whether its data were used to define the training set. 
% We calculate the expected contamination rate as $9\pm1 \times A$, where $A$ is the area (in degrees) of the SESNA map of the region.

\begin{deluxetable}{llcc}
\tablewidth{\textwidth}
\tablecaption{SESNA regions, approximate distances, area of Spitzer observations, and whether the region was used in training or not. {The confusion matrix and an image of each region are available in the online figure set associated with Figure~\ref{fig:afgl490}.} 
\label{tab:app-sesna-deets}}
    \tablehead{
    \colhead{Region}&
    \colhead{Approximate Distance (pc)} &
    \colhead{Area (deg$^2$)} &
    \colhead{Train?}
}
    \startdata
AFGL 490 & $\sim900$ \citep{Snell1984} & $0.76$ & F \\ % & Fig. ~\ref{fig:afgl490} \\
Aquila & $\sim436$ \citep{Ortiz-Leon2018} & $20.9$ & F \\ % & Fig. ~\ref{fig:aquila}\\
Auriga-California & $\sim450$ \citep{Lada2009} & $4.9$ & F \\ % & Fig. ~\ref{fig:aur-cal} \\
Cepheus Flare & $\sim350$ \citep{Szilagyi2021} & $4.7$ & F \\ % & Fig. ~\ref{fig:ceph-flare}\\
Cepheus OB3 & $\sim800$ \citep{Moreno-Corral1993} & $8.0$ & F \\ % & Fig. ~\ref{fig:ceph-ob3}\\
Chamaeleon & $\sim 190$ \citep{Zucker2020} & $9.7$ & F \\ % & Fig. ~\ref{fig:chamaeleon} \\
Corona Australis & $\sim155$ \citep{Zucker2020} & $2.0$ & T \\ % & Fig. ~\ref{fig:cra}\\
Cygnus X & $\sim 760-1660$ \citep{Zucker2020} & $40.9$ & F \\ % & Fig. ~\ref{fig:cygnusx}\\
GGD4, CB34 & $\sim1370$ \citep{Zucker2020} & $0.31$ & F \\ % & Fig. ~\ref{fig:ggd4_cb34}\\
IC 5146 & $\sim750$ \citep{Zucker2020} & $1.3$ & F \\ % & Fig. ~\ref{fig:ic5146}\\
IRAS 20050+2720 & $\sim700$ \citep{Wilking1989} & $0.39$ & F \\ % & Fig. ~\ref{fig:iras20050}\\
L988 & $\sim620$ \citep{Zucker2020} & $0.23$ & F \\ % & Fig. \ref{fig:l988}\\
Lupus & $\sim160$ \citep{Zucker2020} & $13.8$ & F \\ % & Fig. ~\ref{fig:lupus}\\
Mon OB1 & $\sim750$ \citep{Zucker2020} & $3.1$ & F \\ % & Fig. ~\ref{fig:monob1}\\
Mon R2 & $\sim850$ \citep{Zucker2020} & $8.7$ & F \\ % & Fig. ~\ref{fig:monr2}\\
Musca & $\sim140$ \citep{Bonne2020} & $4.2$ & F \\ % & Fig. ~\ref{fig:musca}\\
NGC 7129 & $\sim1150$ \citep{Straizys2014} & $0.17$ & F \\ % & Fig. ~\ref{fig:ngc7129}\\
North America Nebula & $\sim800$ \citep{Zucker2020} & $7.7$ & F \\ % & Fig. ~\ref{fig:northamerica}\\
Ophiuchus & $\sim130$ \citep{Zucker2020} & $18.2$ & T \\ % & Fig. ~\ref{fig:ophiuchus}\\
Orion A & $\sim420$ \citep{Zucker2020} & $22.8$ & T \\ % & Fig. ~\ref{fig:oriona}\\
Orion B & $\sim420$ \citep{Zucker2020} & $4.0$ & T \\ % & Fig. ~\ref{fig:orionb}\\
Perseus & $\sim280$ \citep{Zucker2020} & $17.2$ & F \\ % & Fig. ~\ref{fig:perseus}\\
Pipe & $\sim180$ \citep{Zucker2020} & $13.9$ & F \\ % & Fig. ~\ref{fig:pipe}\\
S131 & $\sim925$ \citep{Pelayo-Balarrago2023} & $0.17$ & F \\ % & Fig. ~\ref{fig:s131}\\
S140 & $\sim764$ \citep{Hirota2008} & $1.9$ & F \\ % & Fig. ~\ref{fig:s140}\\
S171 & $\sim1000$ \citep{Pandey2008} & $0.1$ & F \\ % & Fig. ~\ref{fig:s171}\\
Scorpius & $\sim130$ (Taken to match Ophiuchus) & $3.8$ & F \\ % & Fig. ~\ref{fig:scorpius}\\
Taurus & $\sim140$ \citep{Zucker2020} & $1.3$ & T \\ % & Fig. ~\ref{fig:taurus}\\
Vela D & $\sim700$ \citep{Liseau1992} & $2.2$ & F \\ % & Fig. ~\ref{fig:velad}\\
\enddata    
\end{deluxetable}

\input{fig15set}

\begin{figure*}[h]
    \centering
    \digitalasset
    \includegraphics[width=0.75\linewidth]{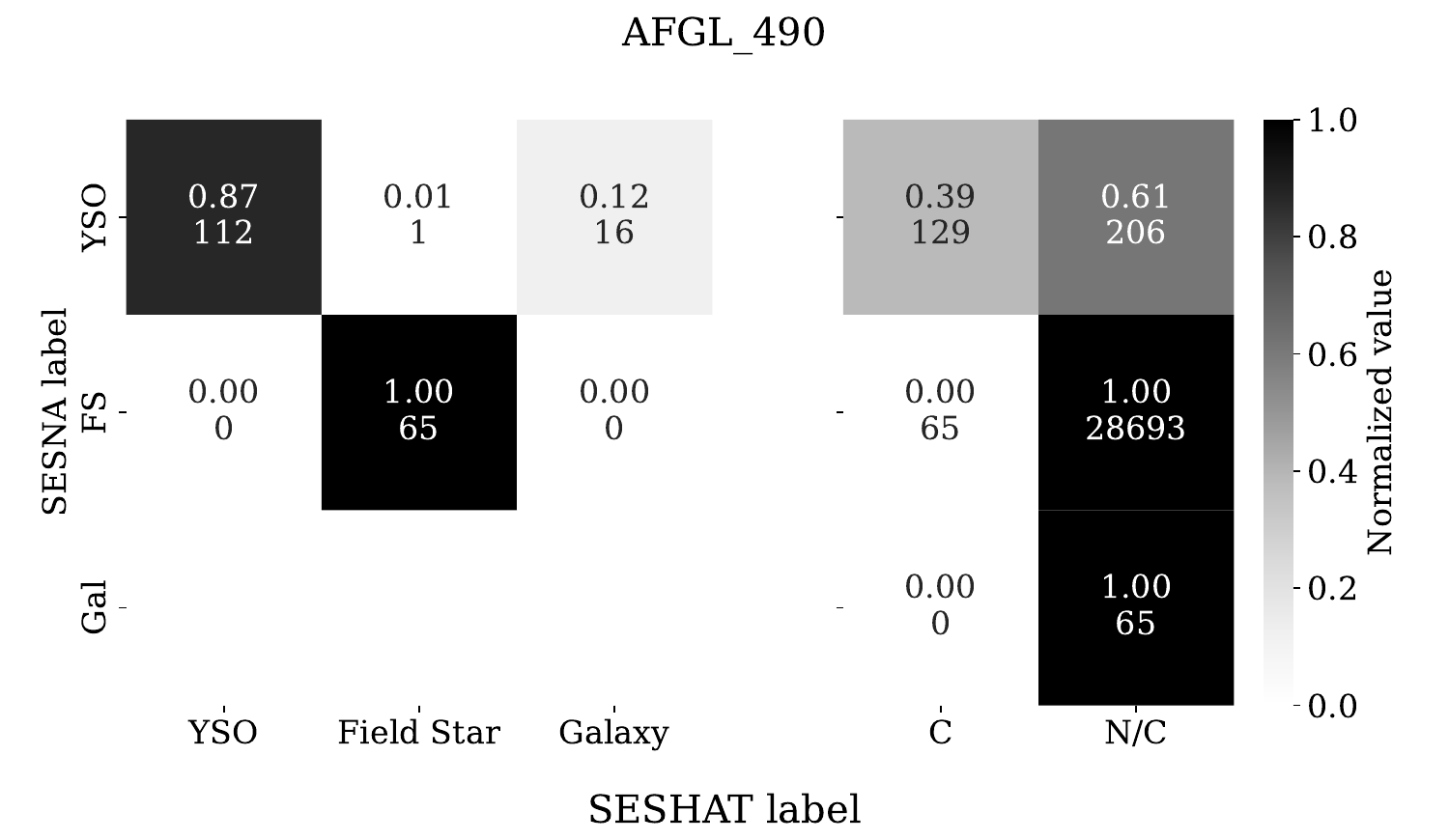}
    \includegraphics[width=0.75\linewidth]{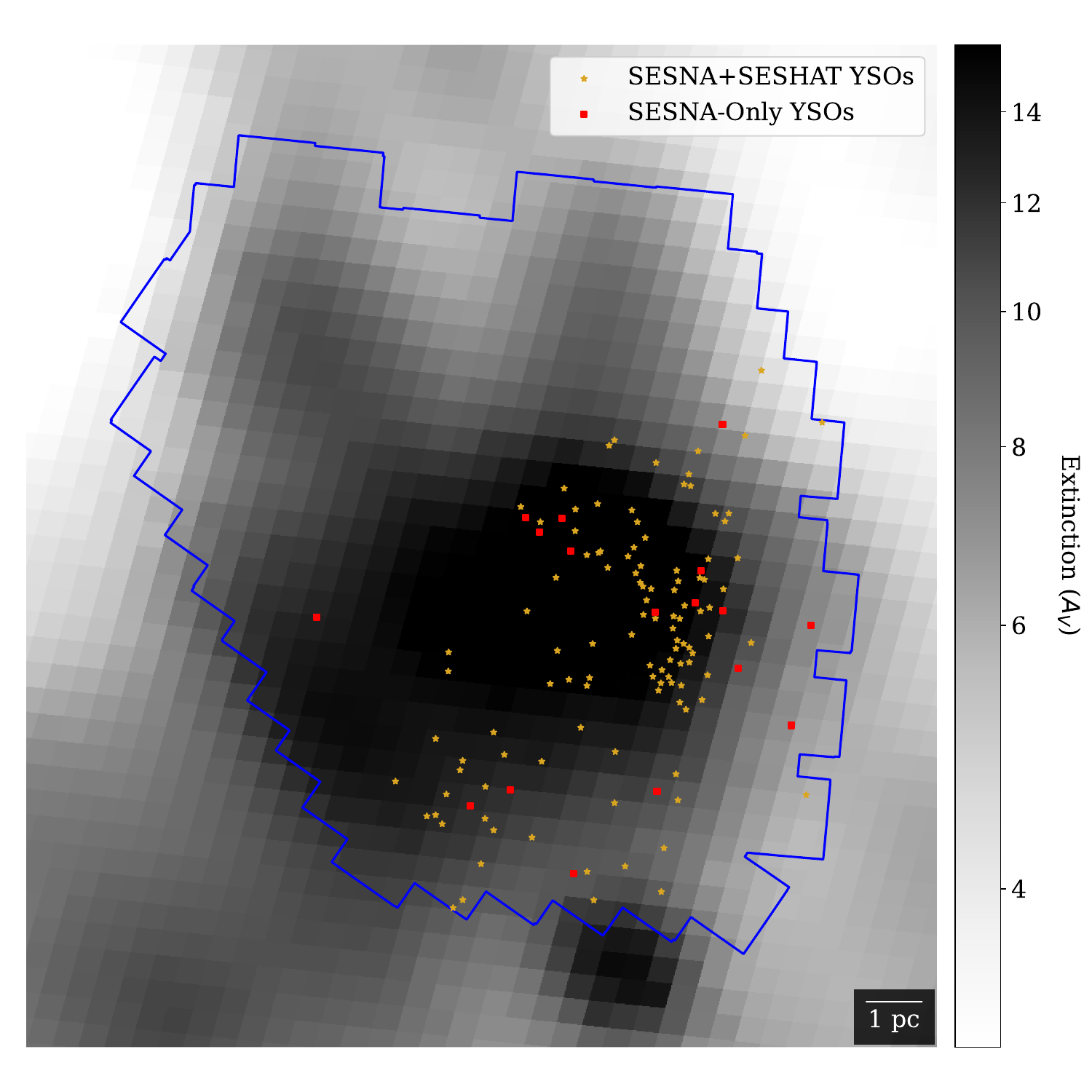}
    \caption{
    \editone{
    \textit{Top}: Confusion matrix for AFGL 490 (part of test set). \textit{Bottom}: The grayscale image is extinction from \citet{Chiang2023} with the YSOs identified in this work and SESNA labeled with gold stars, the YSOs only identified by SESHAT labeled with red circles, and the YSOs only identified in SESNA labeled with red squares.
    The complete figure set (28 figures) is available in the online journal.
    %Contours from the extinction maps of \citet{Dobashi2005}: $A_V=1$ mag in white, $A_V=2$ mag in light gray, $A_V = 5$ mag in dark gray, and $A_V=10$ mag in black.
    \label{fig:afgl490}}}
\end{figure*}

% \begin{nolinenumbers}

\bibliography{main,2024bib}{}
\bibliographystyle{aasjournal}
% \end{nolinenumbers}
\end{document}

%% file: fig15set.tex
\figsetstart
\figsetnum{15}
\figsettitle{SESNA Regions}

\figsetgrpstart
\figsetgrpnum{15.1}
\figsetgrptitle{AFGL 490}
\figsetplot{AFGL_490_cm.pdf}
\figsetplot{AFGL_490_hersch_col.pdf}
\figsetgrpnote{\textit{Top}: Confusion matrix for AFGL 490 (part of test set). \textit{Bottom}: The grayscale image is extinction from \citet{Chiang2023} with the YSOs identified in this work and SESNA labeled with gold stars, the YSOs only identified by SESHAT labeled with red circles, and the YSOs only identified in SESNA labeled with red squares. \label{fig:afgl490}}
\figsetgrpend

\figsetgrpstart
\figsetgrpnum{15.2}
\figsetgrptitle{Aquila}
\figsetplot{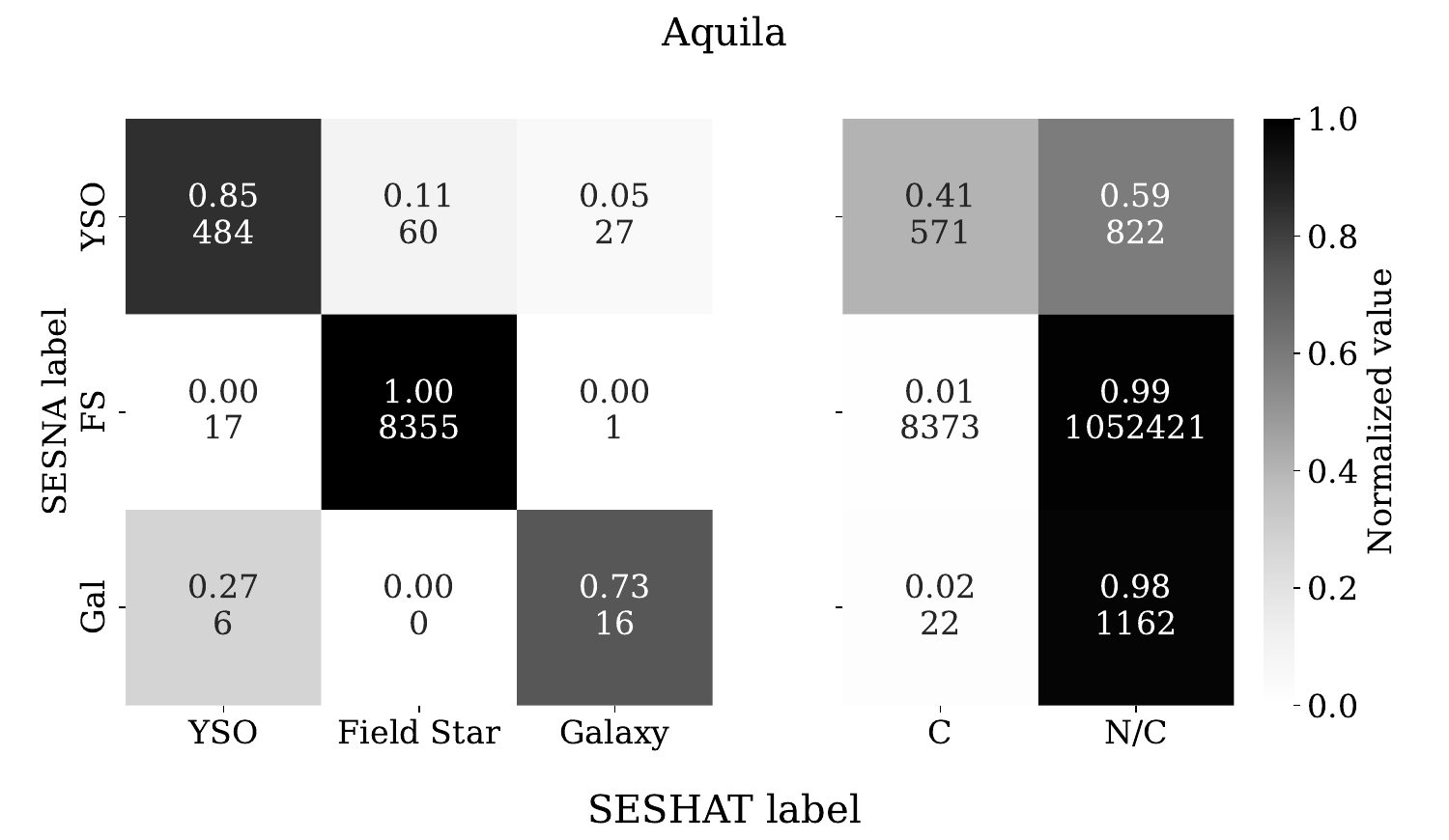}
\figsetplot{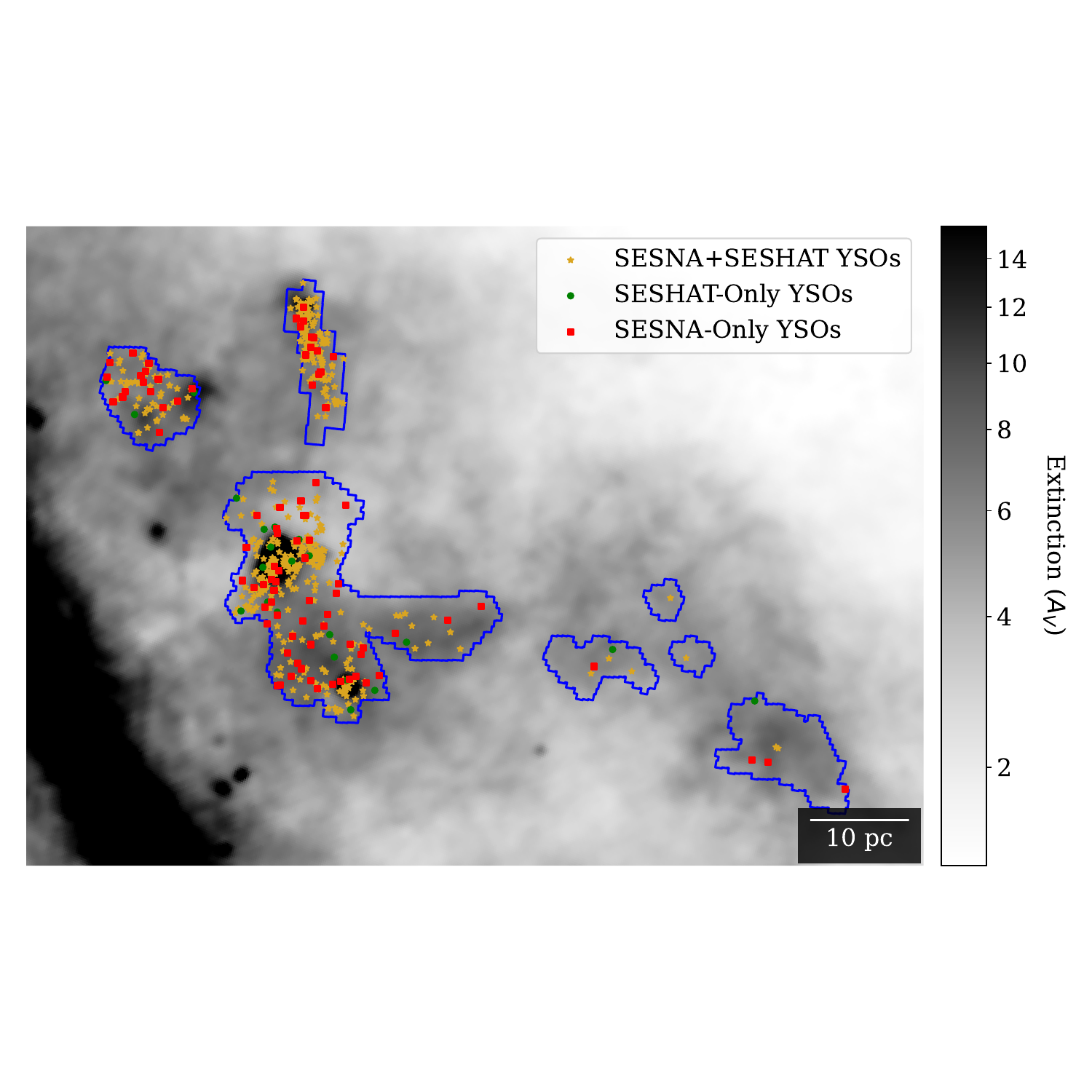}
\figsetgrpnote{The same as Figure~\ref{fig:afgl490}, but now for the Aquila star-forming region, which is part of the test set.  \label{fig:aquila}}
\figsetgrpend

\figsetgrpstart
\figsetgrpnum{15.3}
\figsetgrptitle{Auriga-California}
\figsetplot{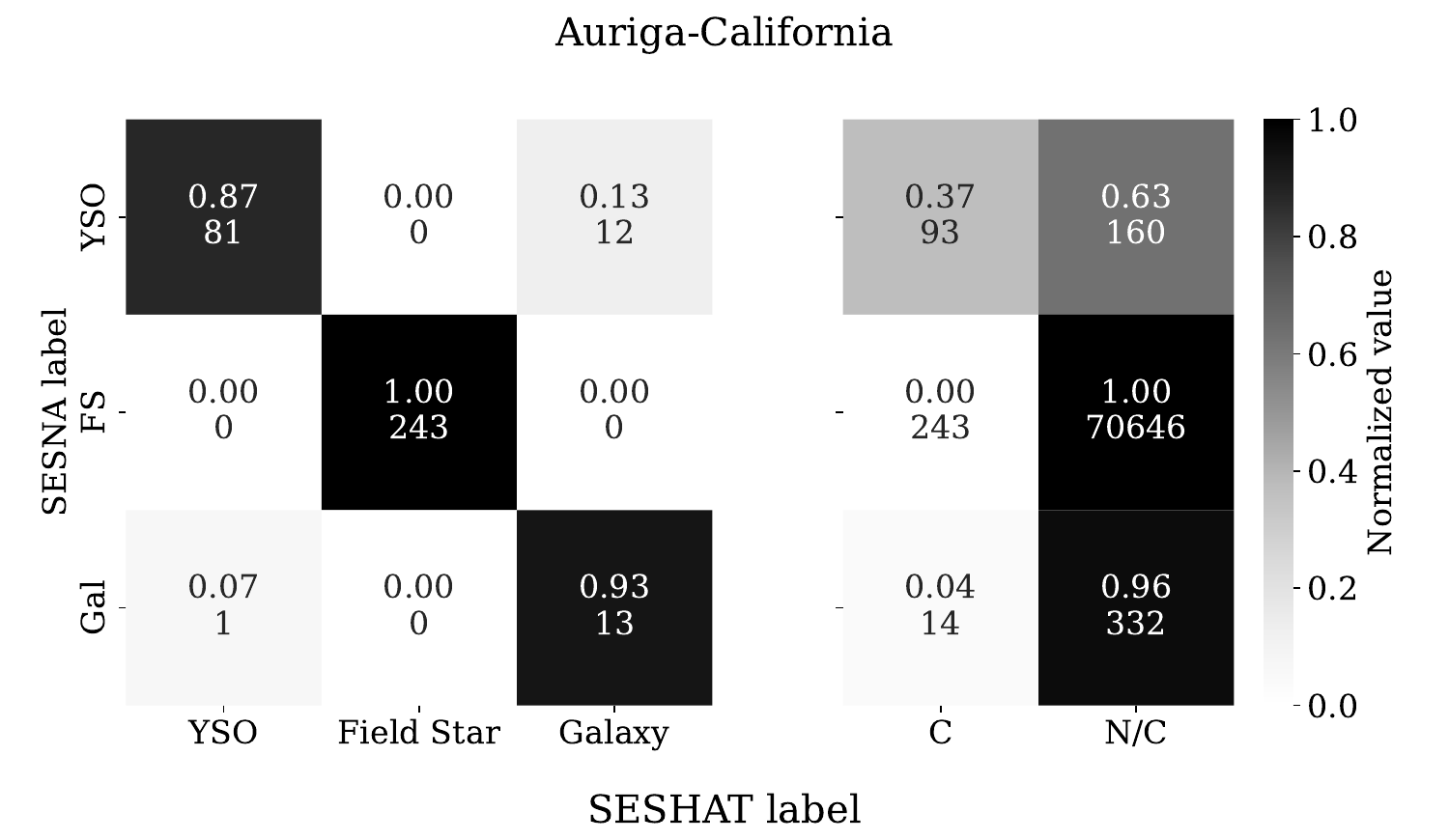}
\figsetplot{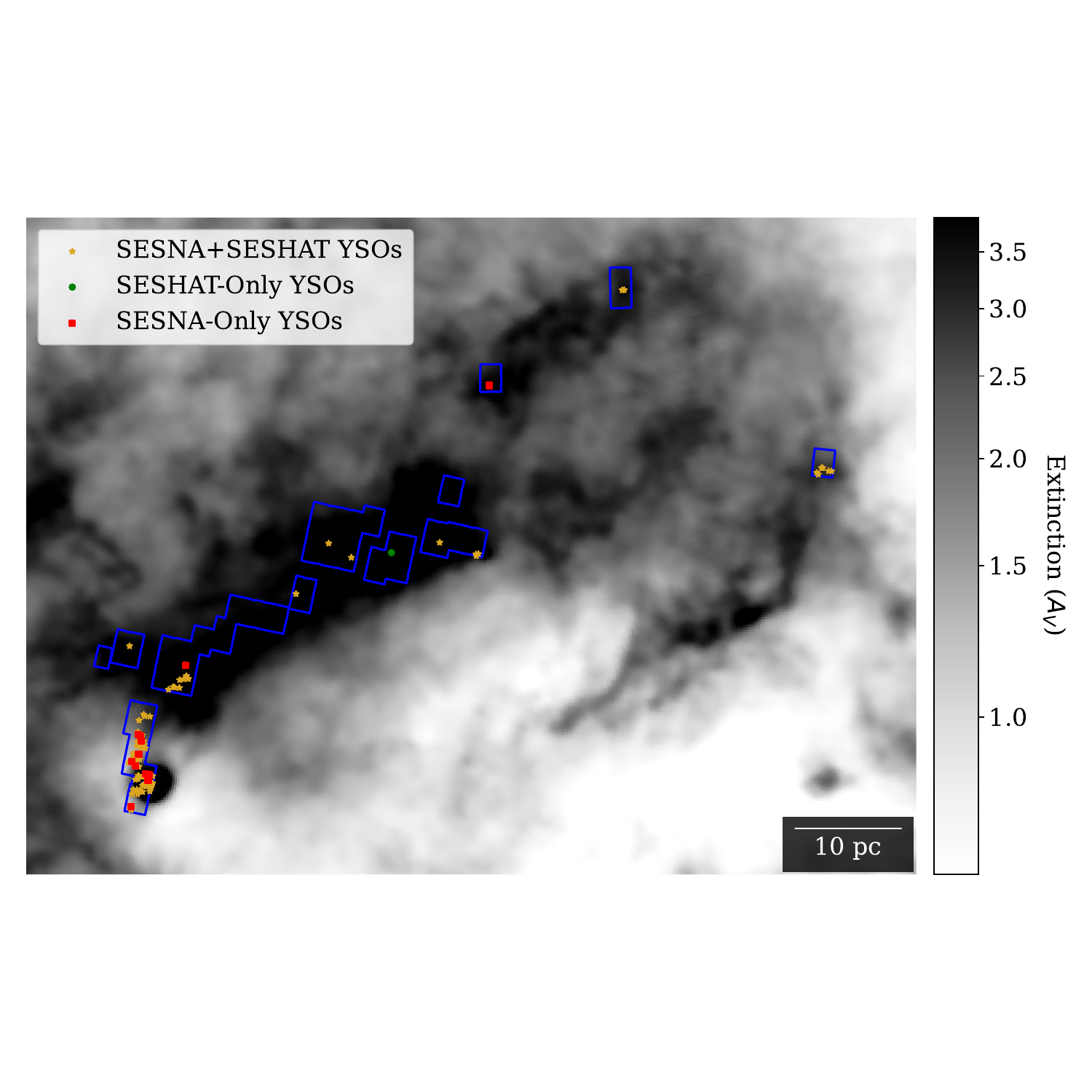}
\figsetgrpnote{The same as Figure~\ref{fig:afgl490}, but now for the Auriga-California star-forming region, which is part of the test set.  \label{fig:aur-cal}}
\figsetgrpend

\figsetgrpstart
\figsetgrpnum{15.4}
\figsetgrptitle{Cepheus Flare}
\figsetplot{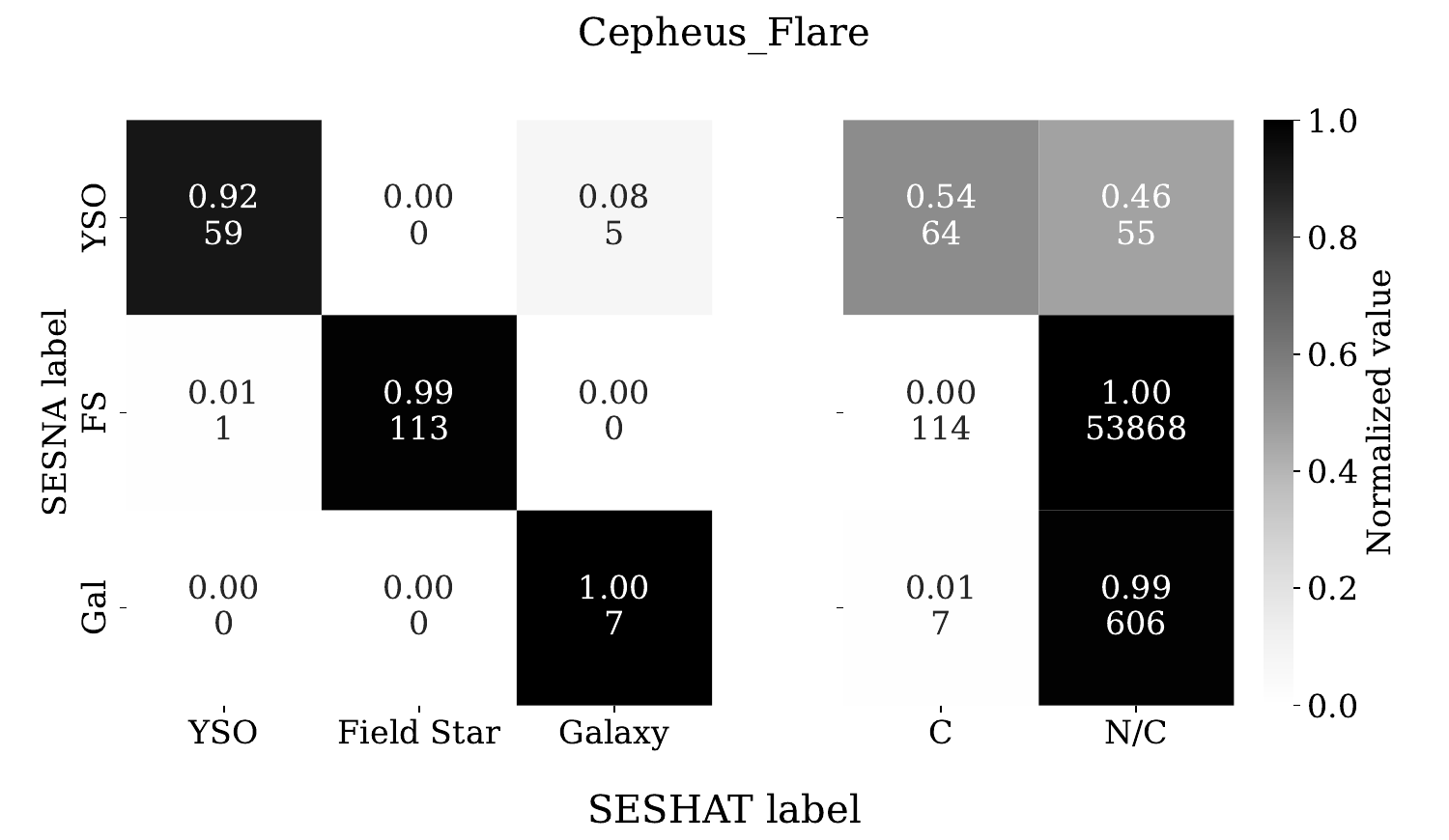}
\figsetplot{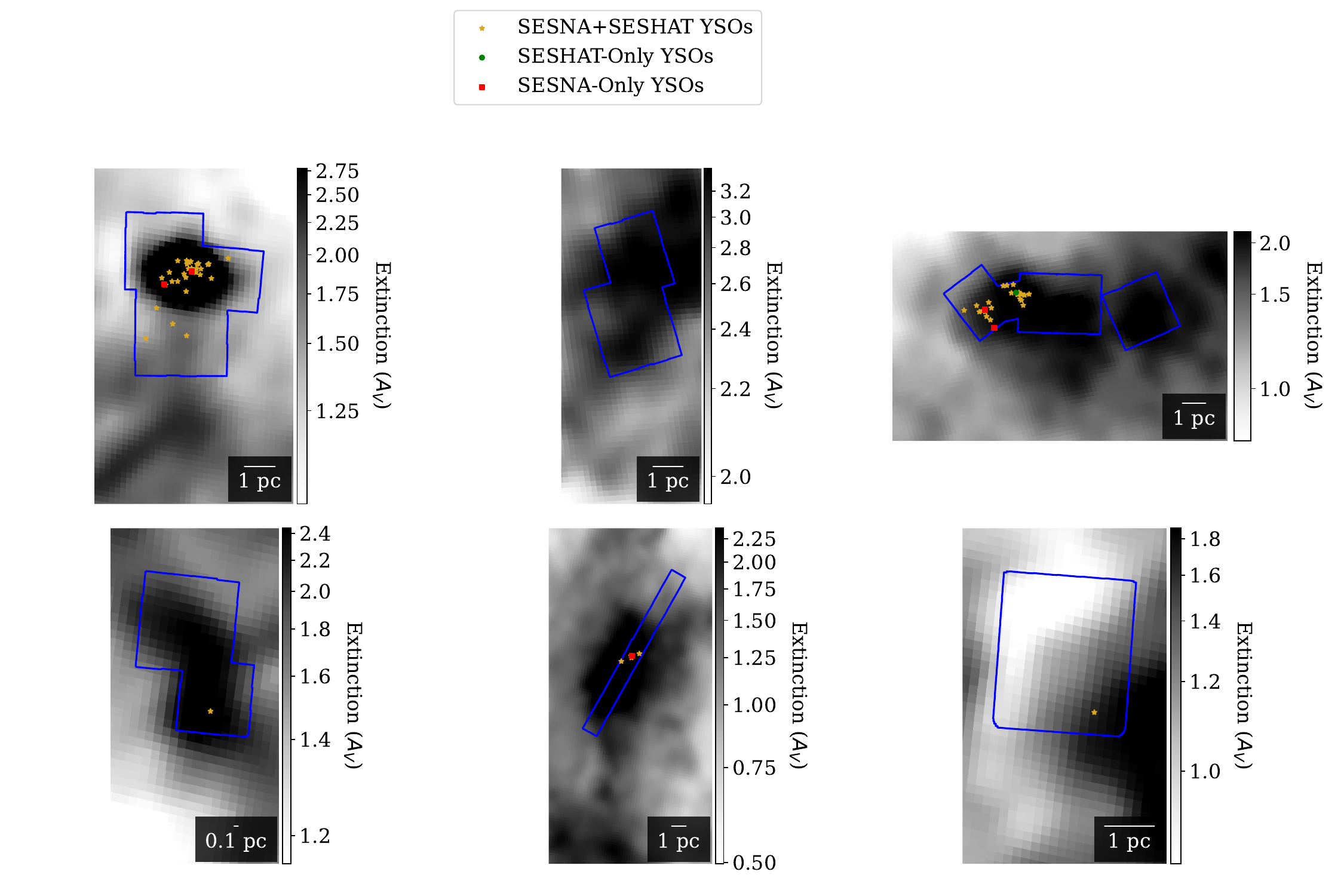}
\figsetgrpnote{The same as Figure~\ref{fig:afgl490}, but now for the Cepheus Flare region, which is part of the test set. The observations were taken with significant separation to require separate cutouts. \label{fig:ceph-flare}}
\figsetgrpend

\figsetgrpstart
\figsetgrpnum{15.5}
\figsetgrptitle{Cepheus OB3}
\figsetplot{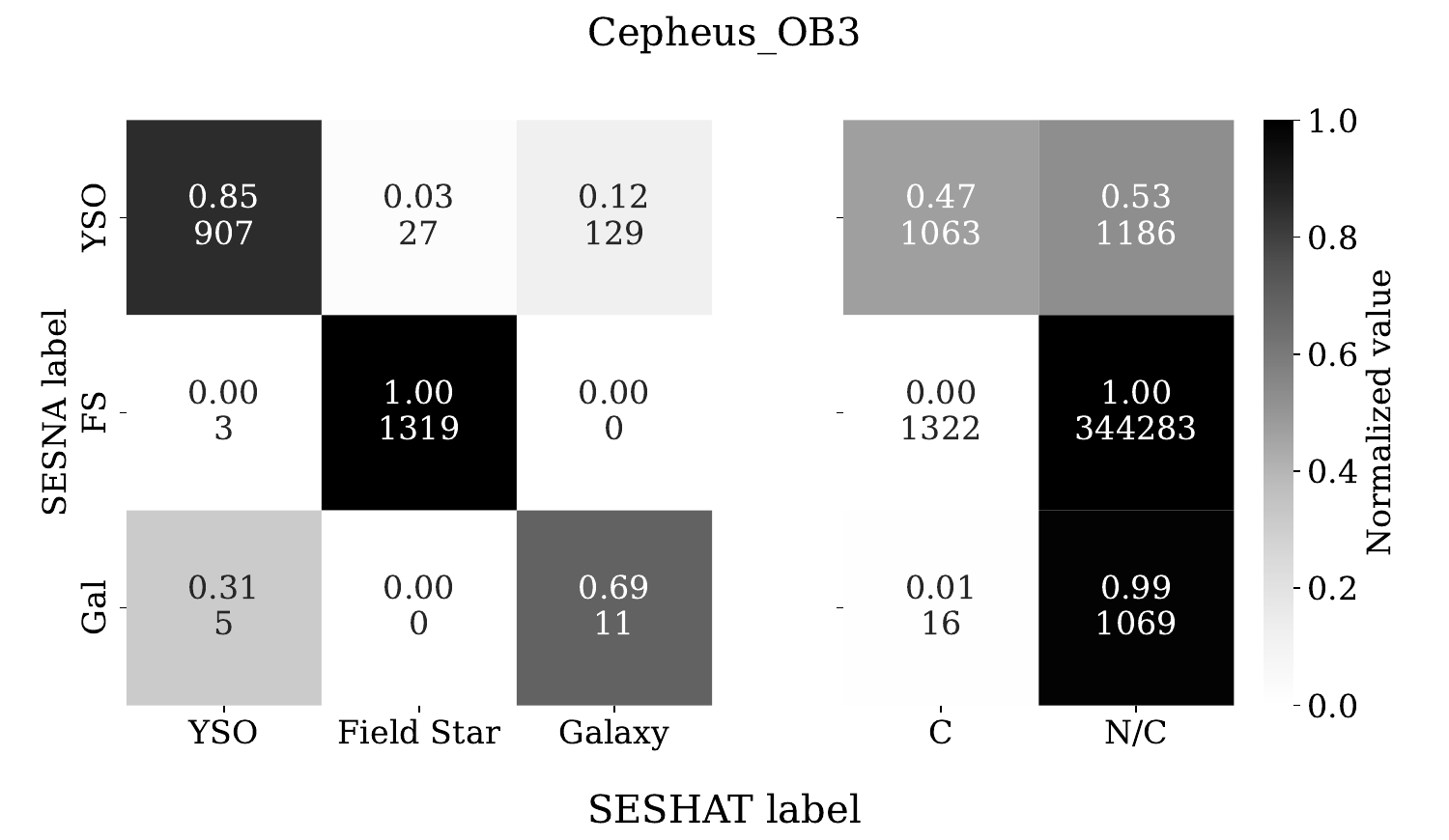}
\figsetplot{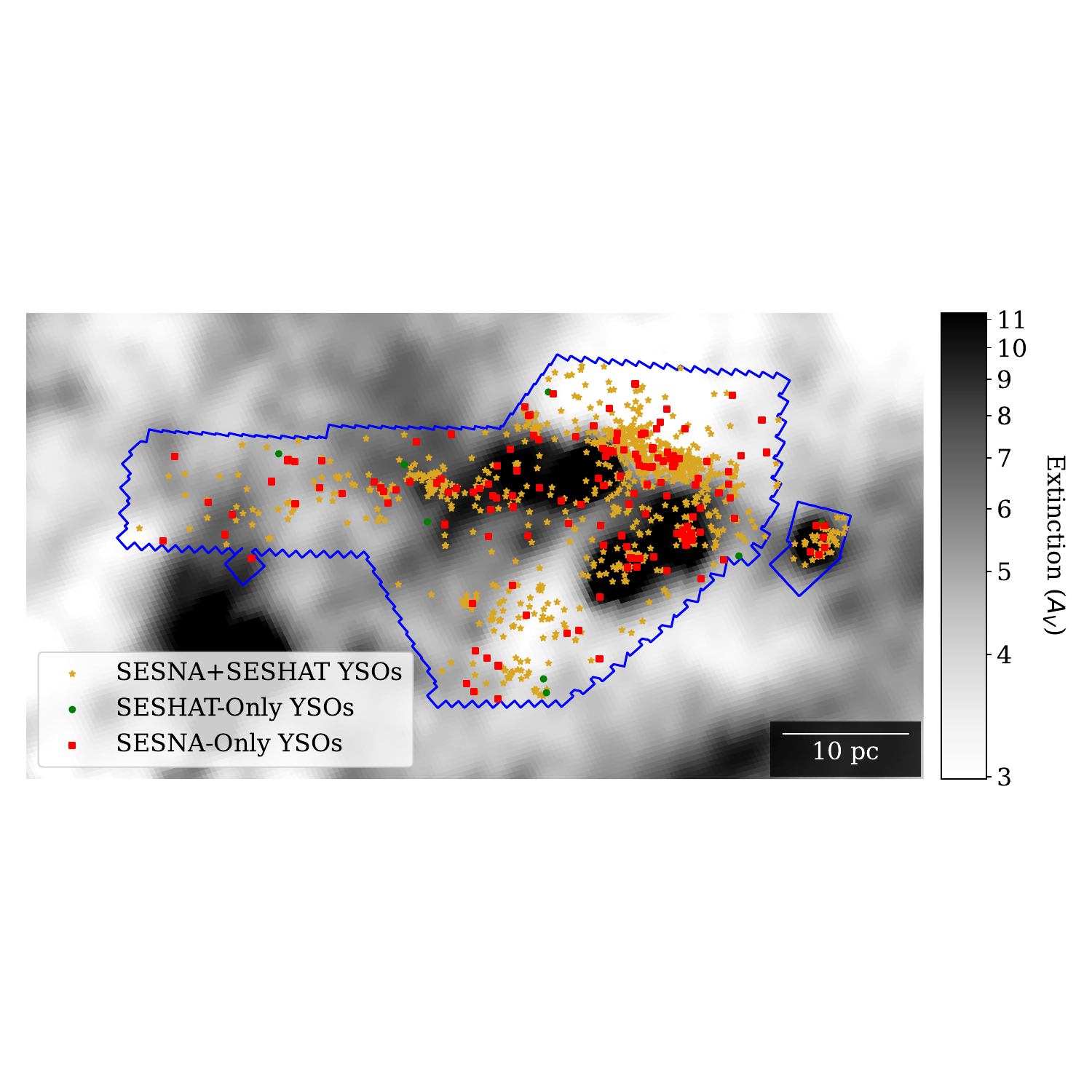}
\figsetgrpnote{The same as Figure~\ref{fig:afgl490}, but now for the Cepheus OB3 region, which is part of the test set.  \label{fig:ceph-ob3}}
\figsetgrpend

\figsetgrpstart
\figsetgrpnum{15.6}
\figsetgrptitle{Chamaeleon}
\figsetplot{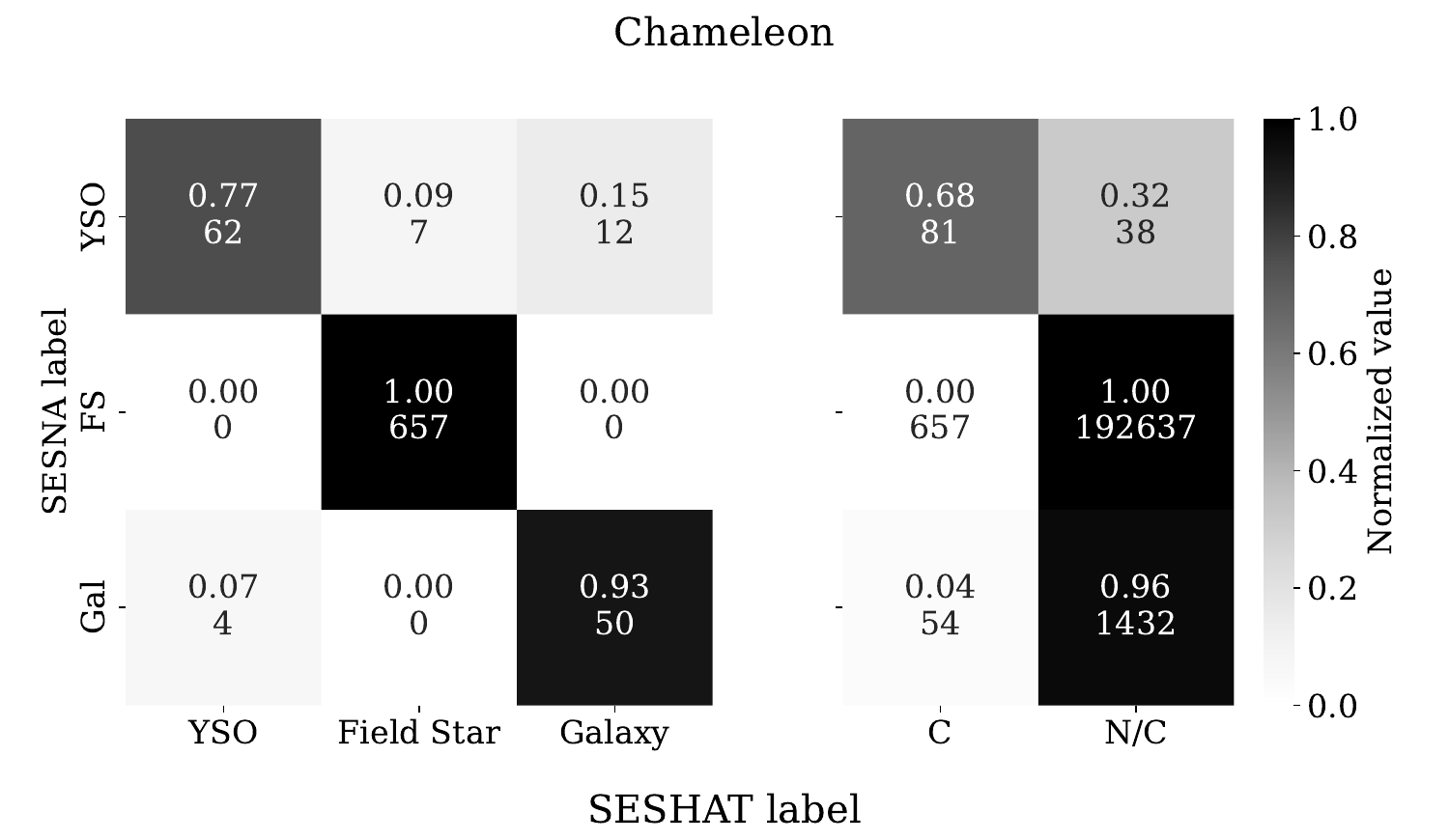}
\figsetplot{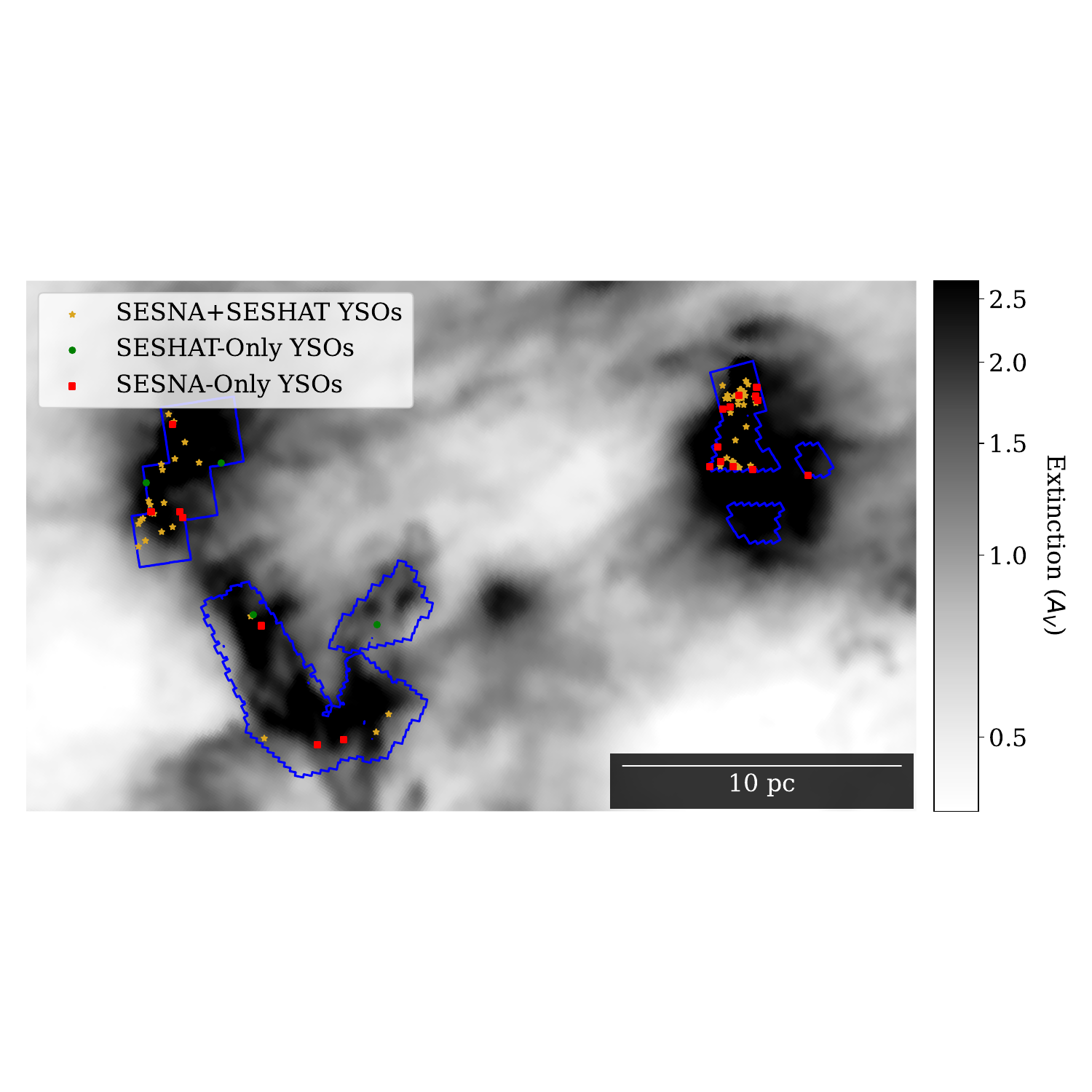}
\figsetgrpnote{The same as Figure~\ref{fig:afgl490}, but now for the Chamaeleon region, which is part of the test set. \label{fig:chamaeleon}}
\figsetgrpend

\figsetgrpstart
\figsetgrpnum{15.7}
\figsetgrptitle{Corona Australis}
\figsetplot{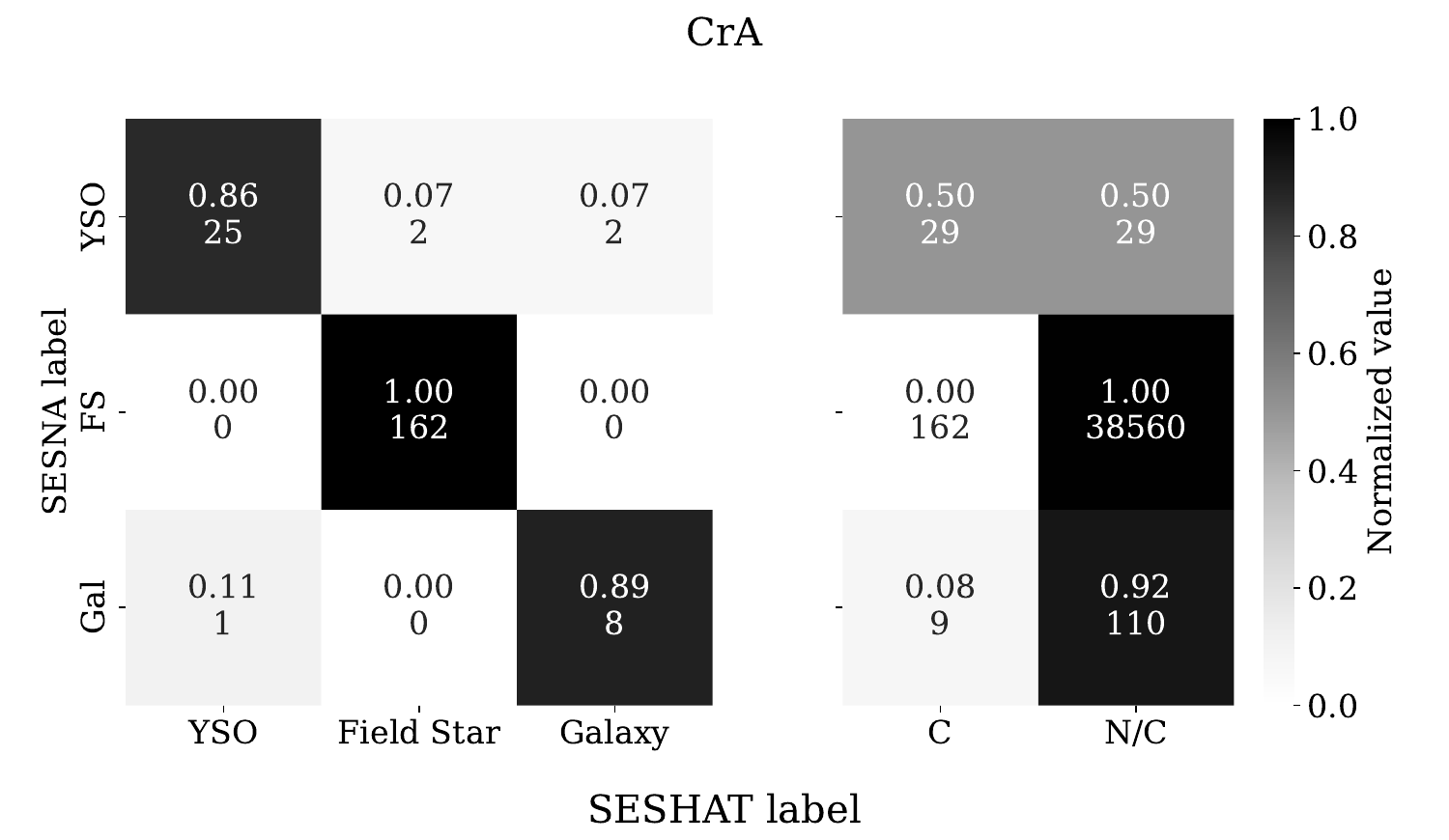}
\figsetplot{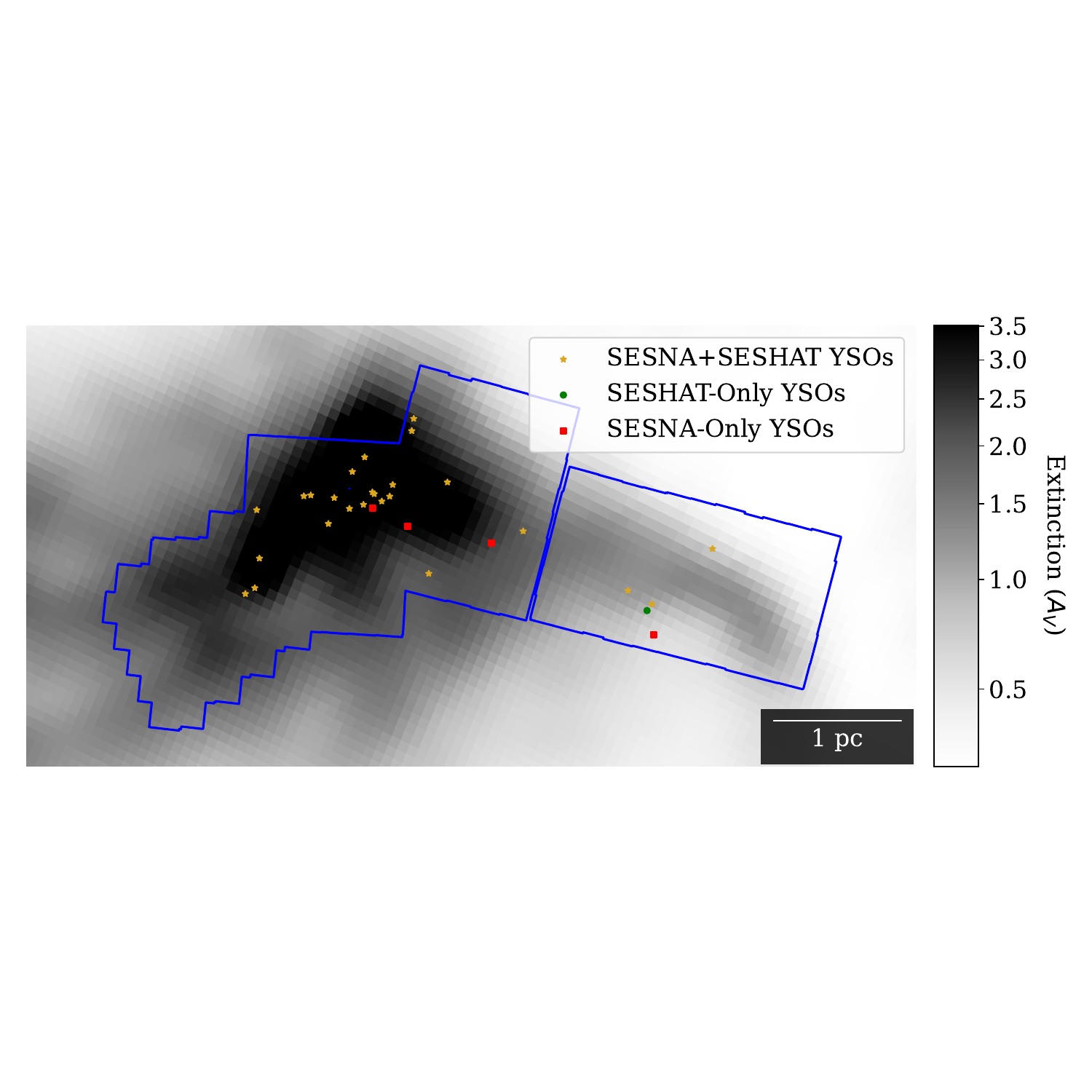}
\figsetgrpnote{The same as Figure~\ref{fig:afgl490}, but now for the Corona Australis region, which is part of set used to define training YSOs. \label{fig:cra}}
\figsetgrpend

\figsetgrpstart
\figsetgrpnum{15.8}
\figsetgrptitle{Cygnus X}
\figsetplot{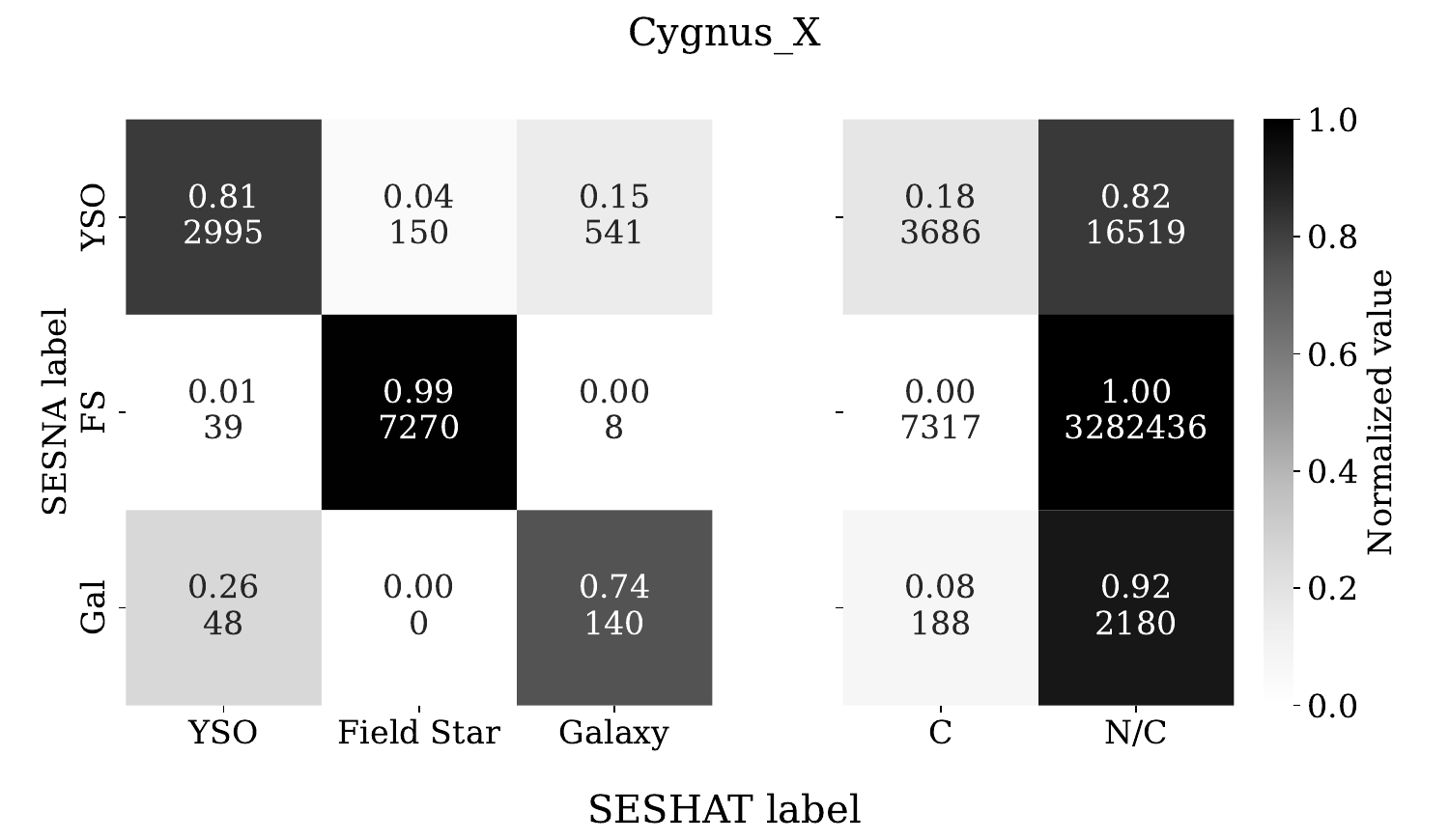}
\figsetplot{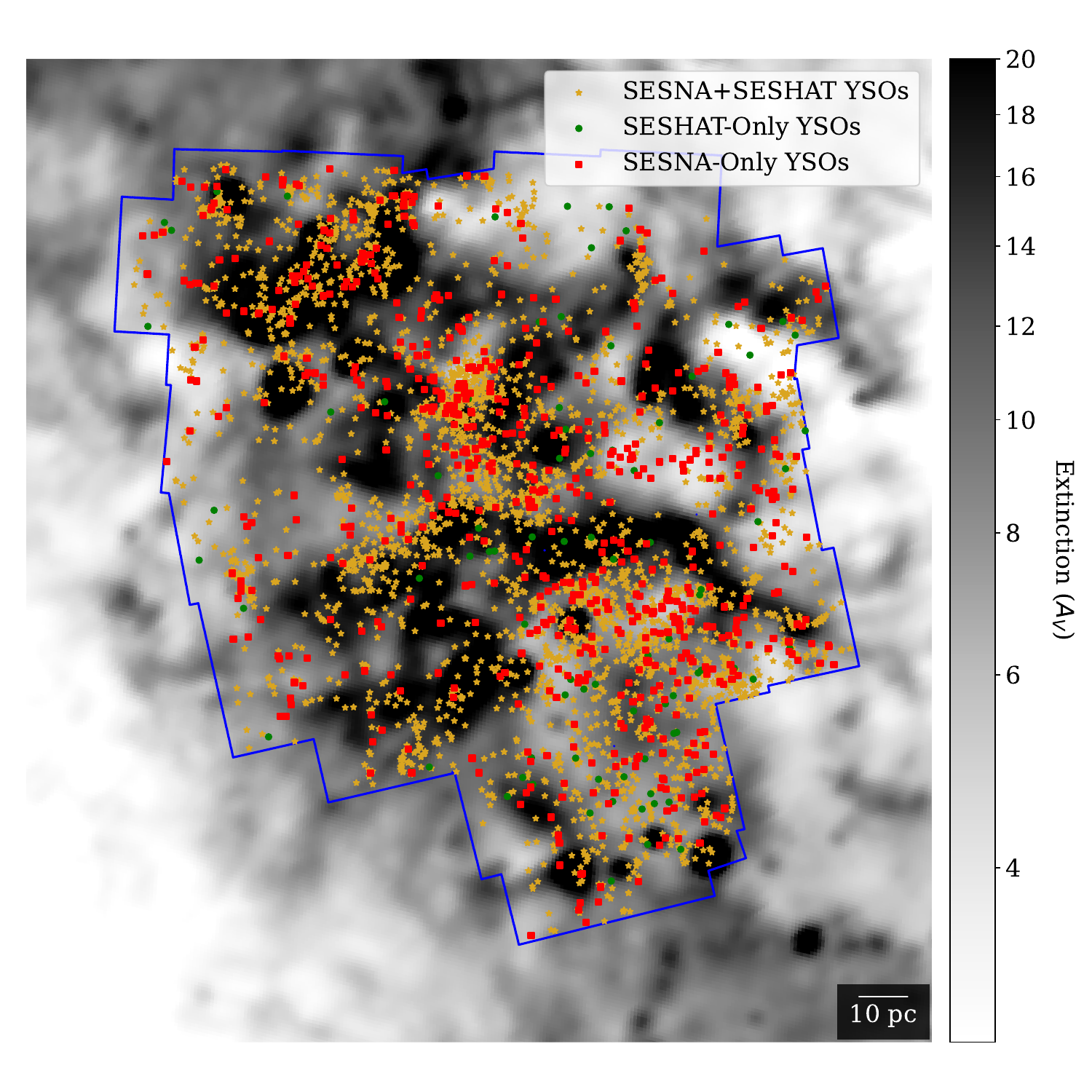}
\figsetgrpnote{The same as Figure~\ref{fig:afgl490}, but now for the Cygnus X region, which is excluded from both test and training sets. \label{fig:cygnusx}}
\figsetgrpend

\figsetgrpstart
\figsetgrpnum{15.9}
\figsetgrptitle{GGD4 CB34}
\figsetplot{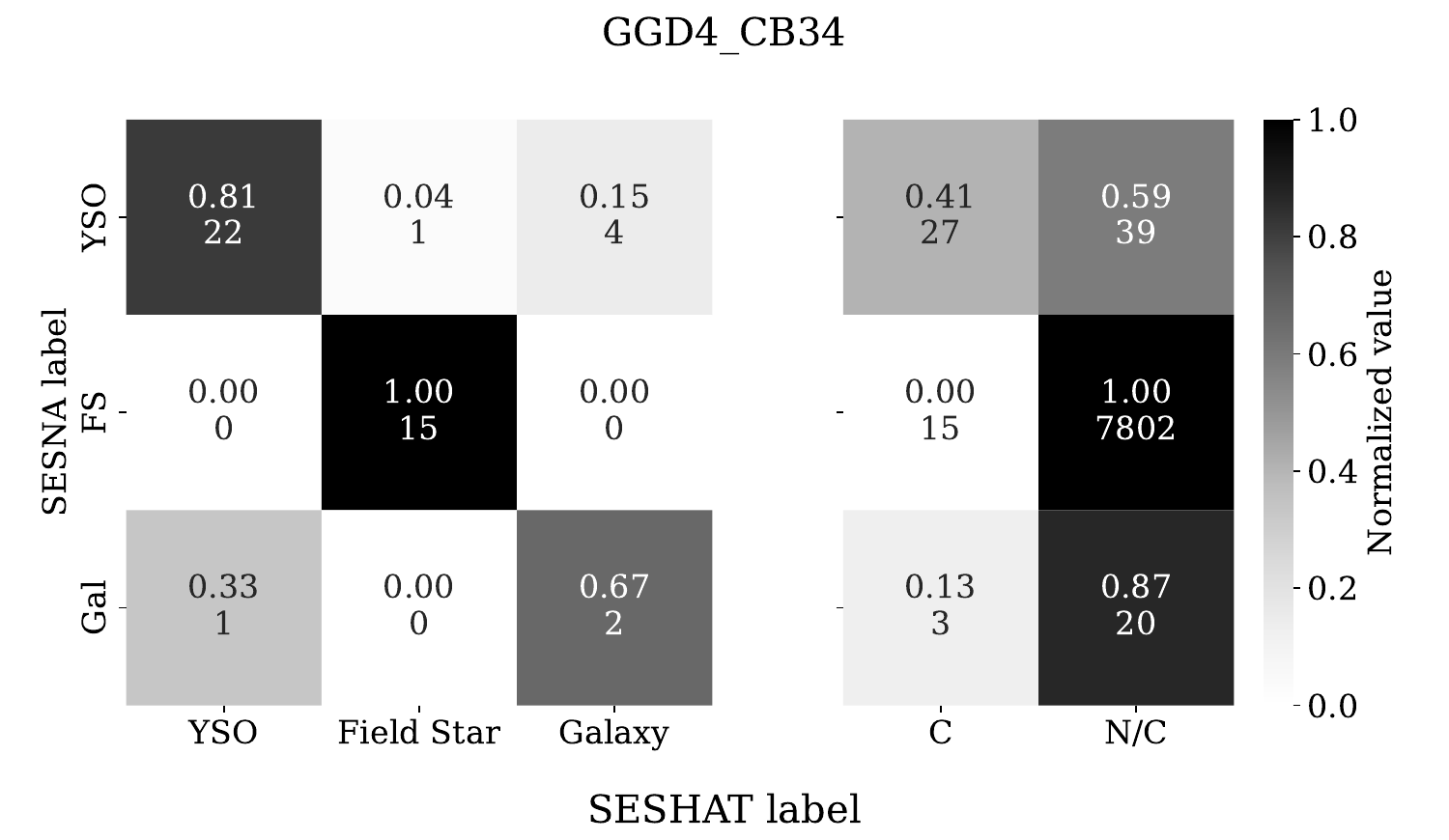}
\figsetplot{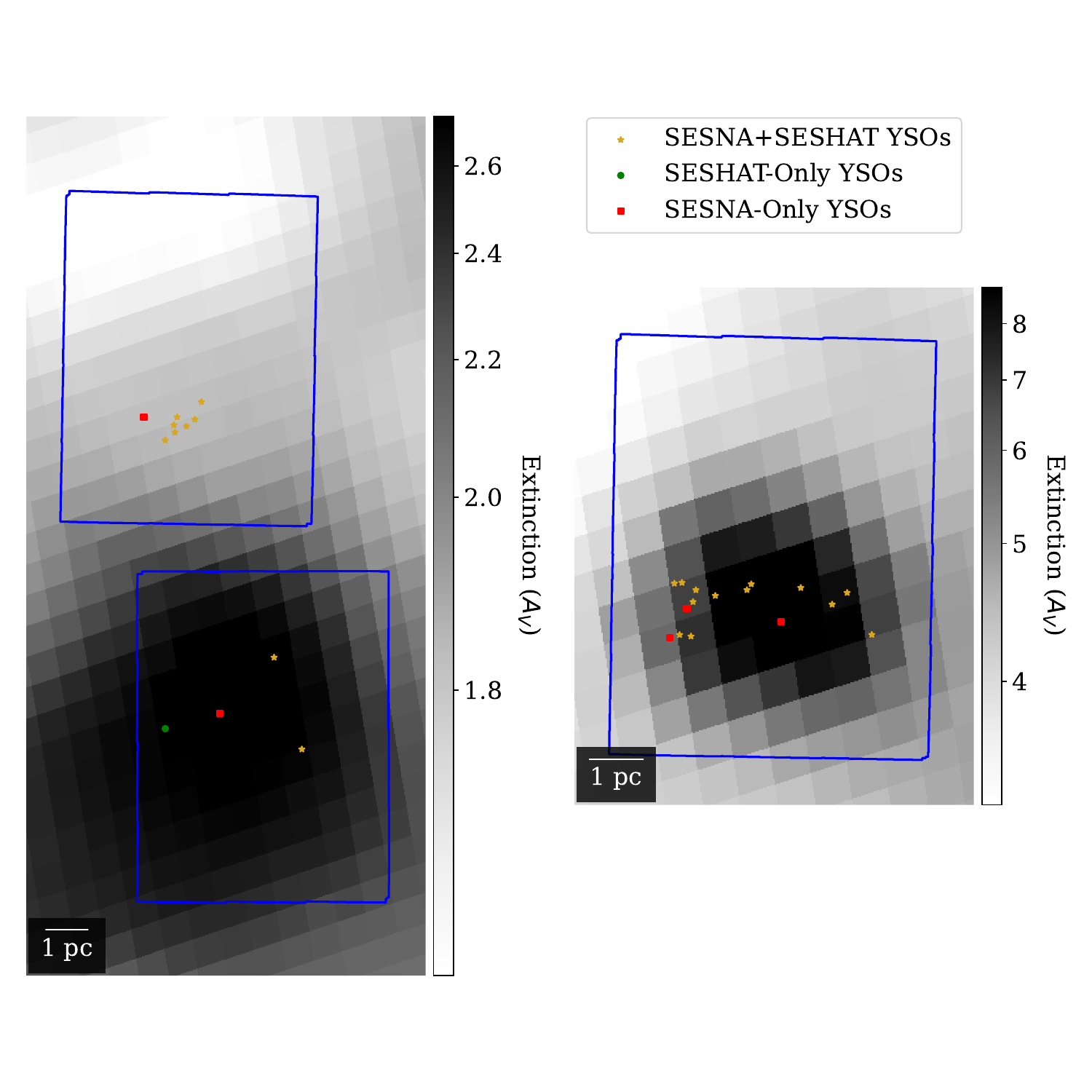}
\figsetgrpnote{The same as Figure~\ref{fig:afgl490}, but now for the GGD4 CB34 region, which is part of the test set. The Spitzer observations are at wide enough separation to require separating the fields into two subplots for better visualization. \label{fig:ggd4_cb34}}
\figsetgrpend

\figsetgrpstart
\figsetgrpnum{15.10}
\figsetgrptitle{IC 5146}
\figsetplot{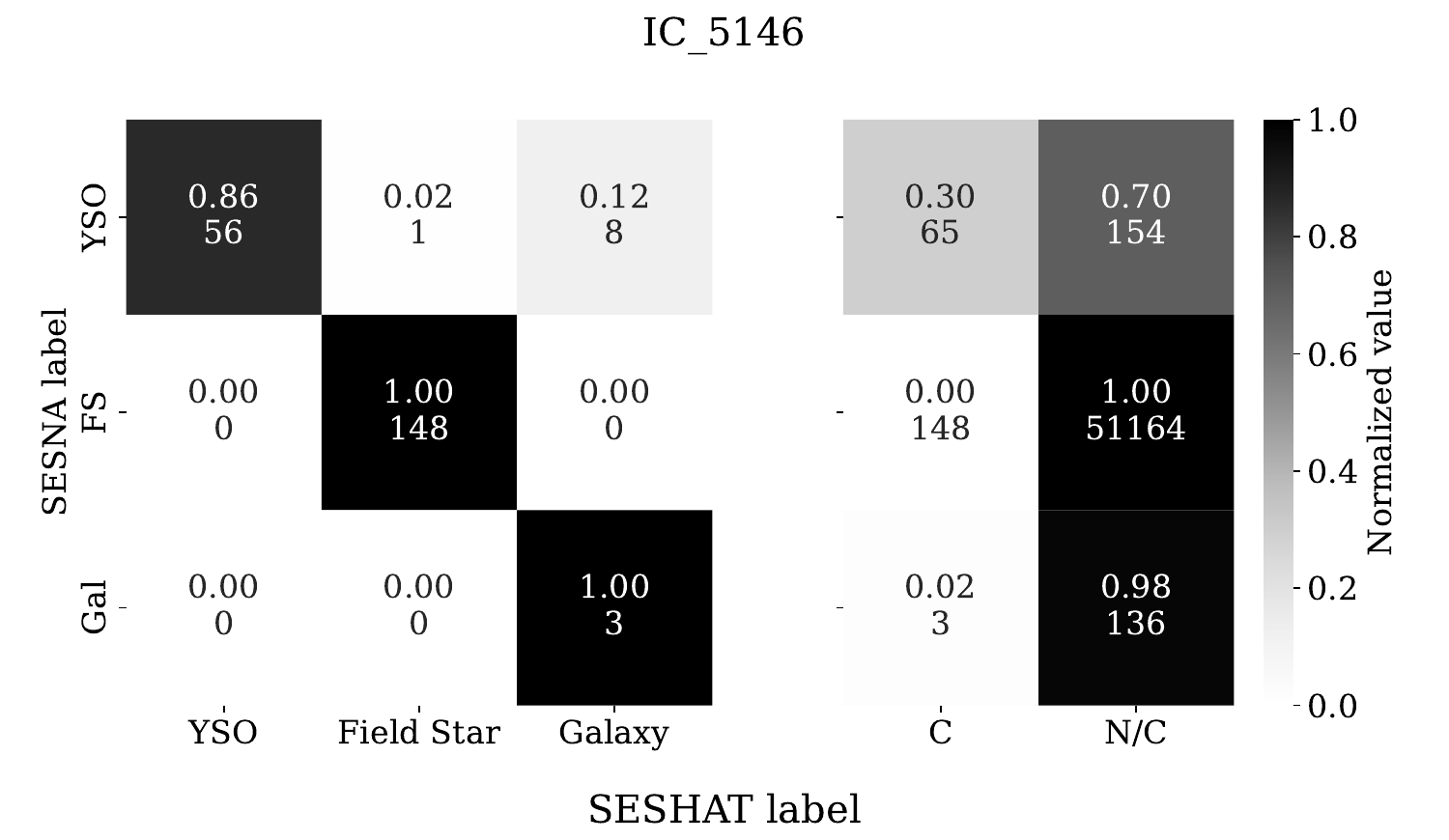}
\figsetplot{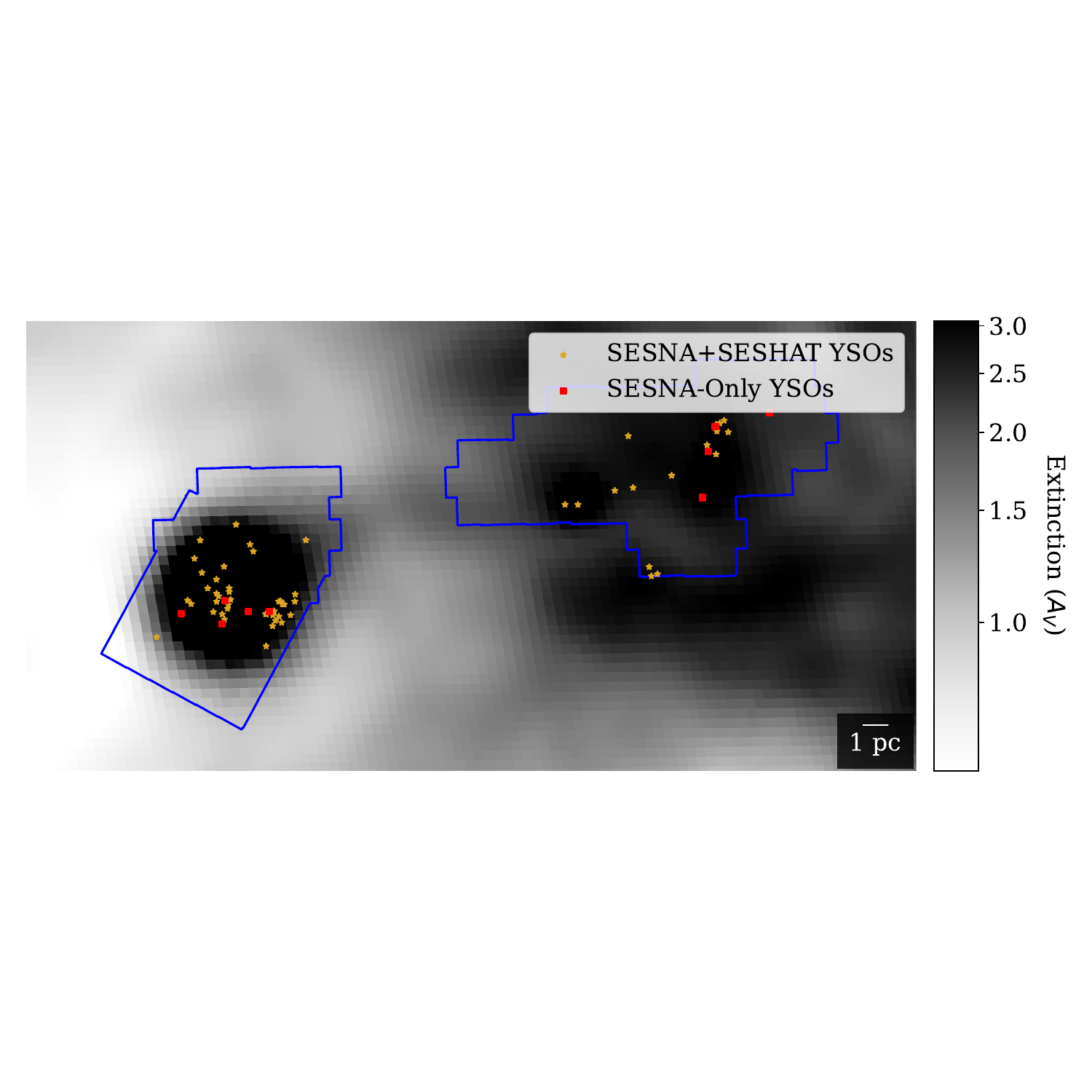}
\figsetgrpnote{The same as Figure~\ref{fig:afgl490}, but now for the IC 5146 region, which is part of the test set.  \label{fig:ic5146}}
\figsetgrpend

\figsetgrpstart
\figsetgrpnum{15.11}
\figsetgrptitle{IRAS 20050+2720}
\figsetplot{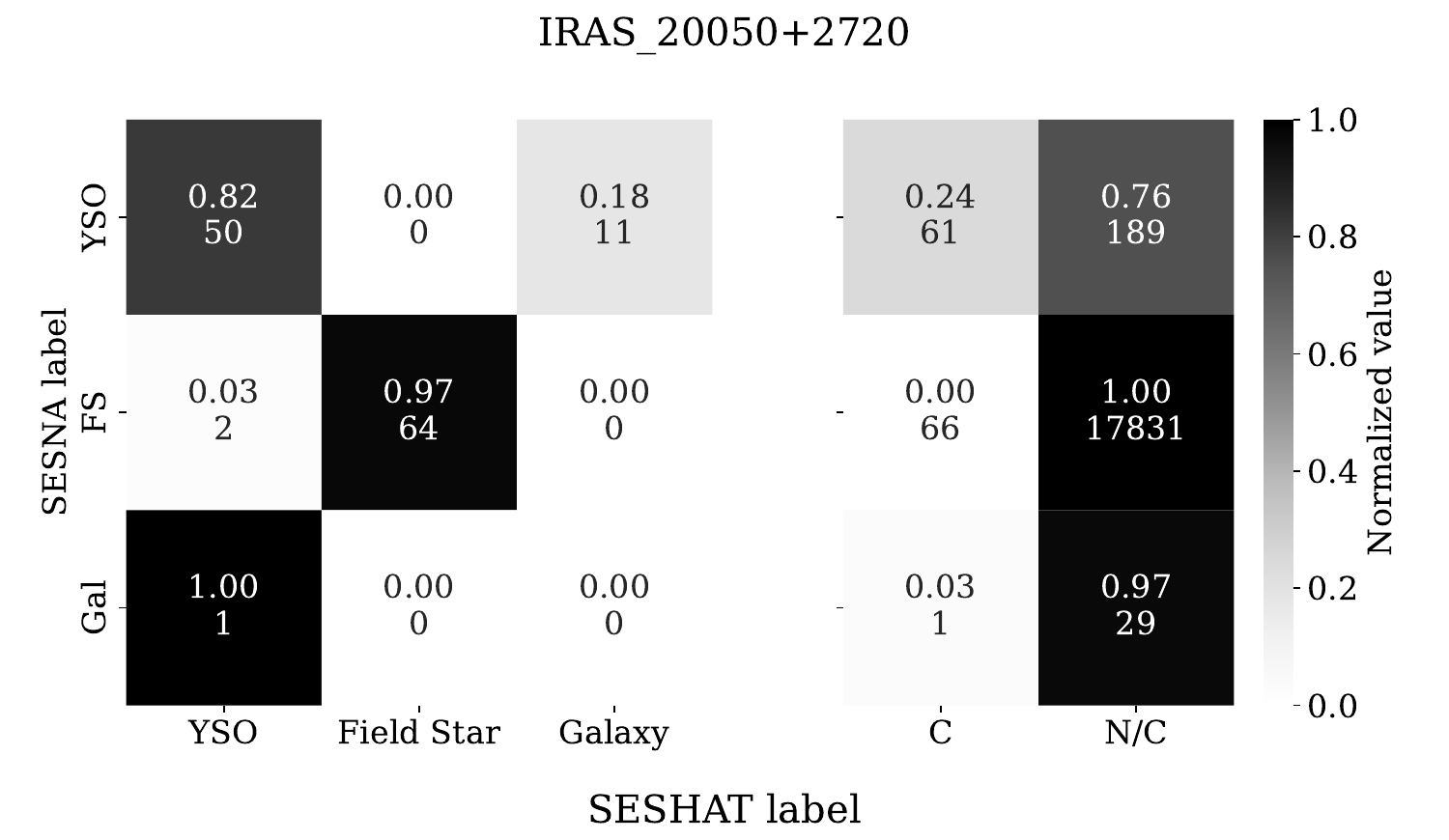}
\figsetplot{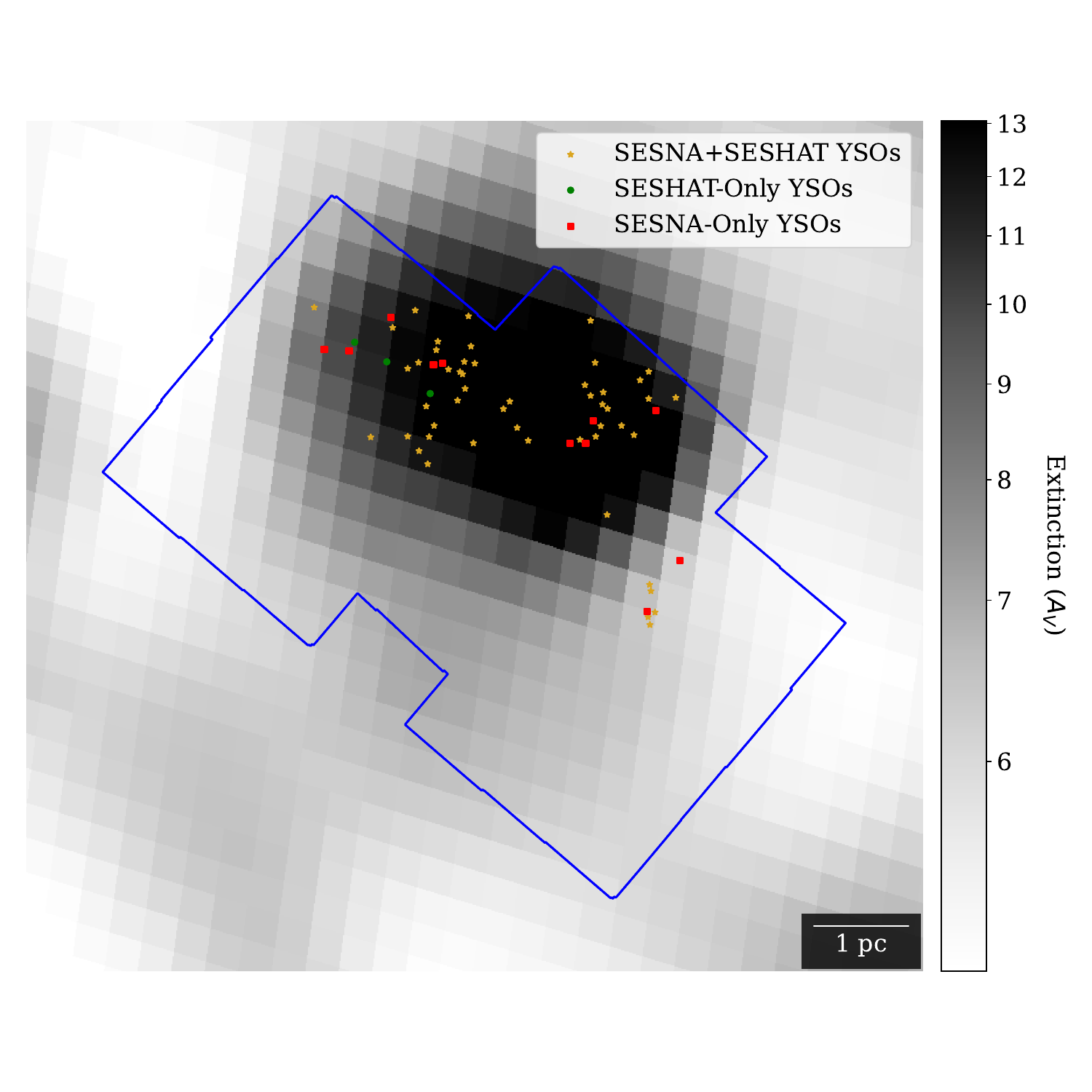}
\figsetgrpnote{The same as Figure~\ref{fig:afgl490}, but now for the IRAS 20050+2720 region, which is part of the test set. \label{fig:iras20050}}
\figsetgrpend

\figsetgrpstart
\figsetgrpnum{15.12}
\figsetgrptitle{L 988}
\figsetplot{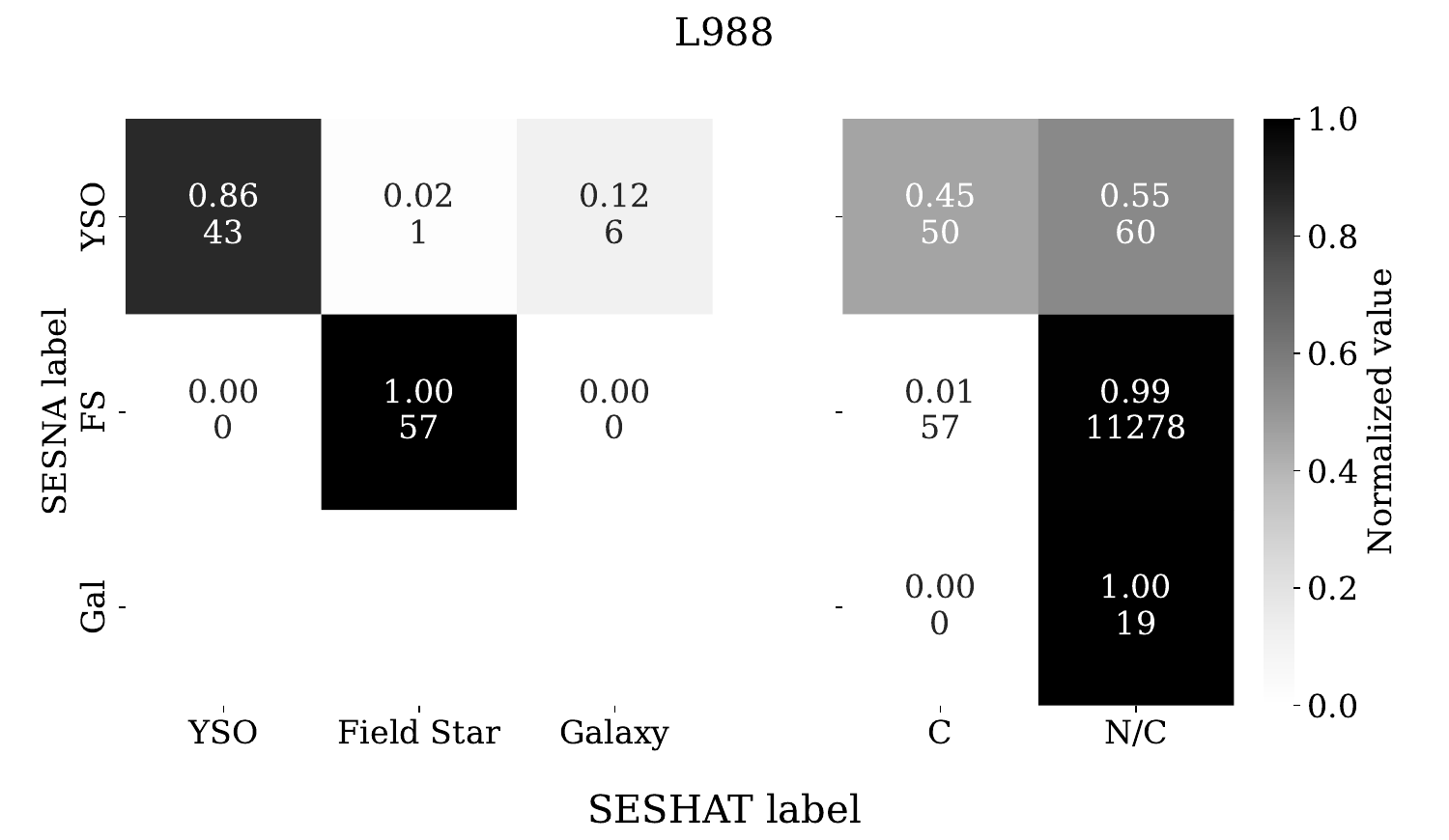}
\figsetplot{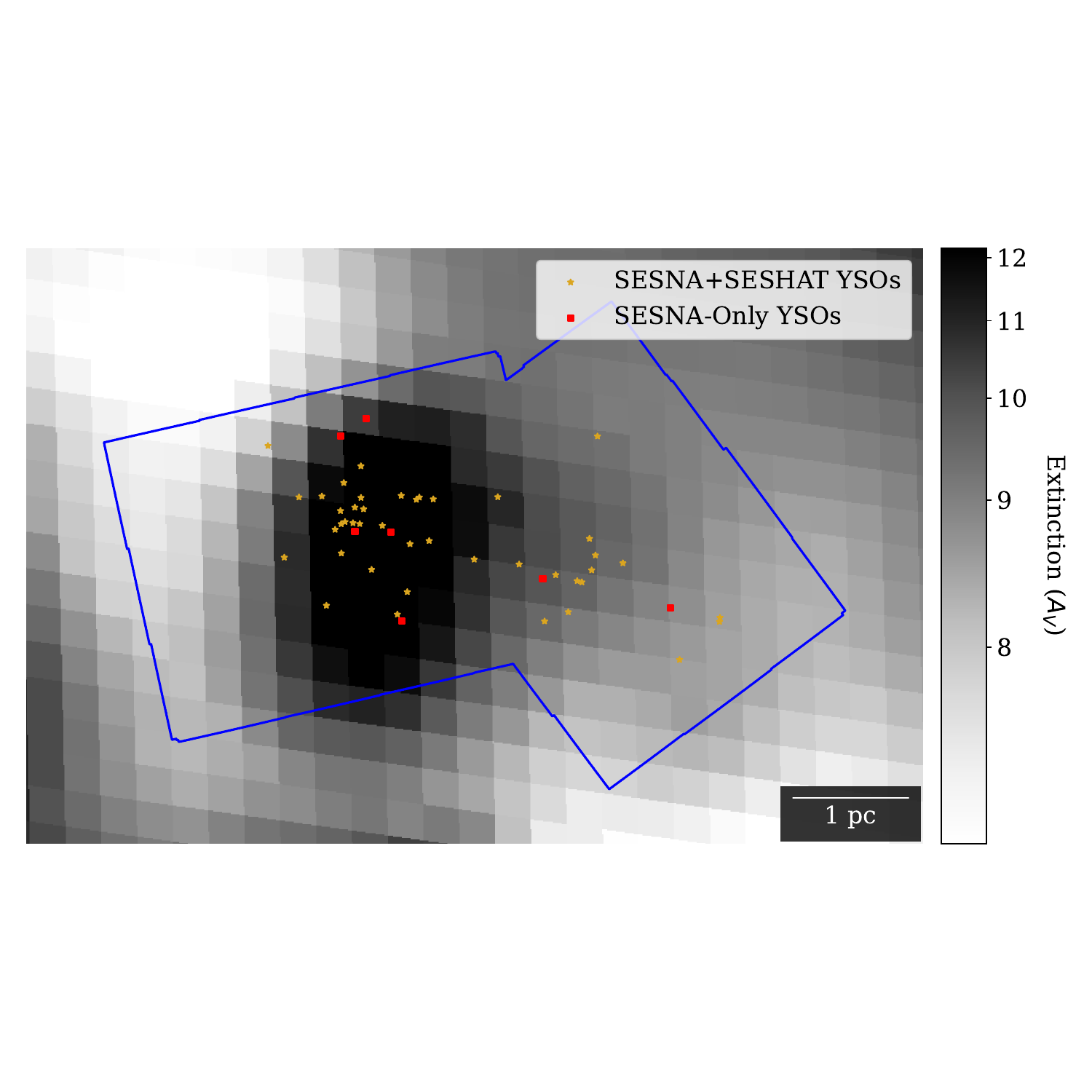}
\figsetgrpnote{The same as Figure~\ref{fig:afgl490}, but now for the L 988 region, which is part of the test set. \label{fig:l988}}
\figsetgrpend

\figsetgrpstart
\figsetgrpnum{15.13}
\figsetgrptitle{Lupus}
\figsetplot{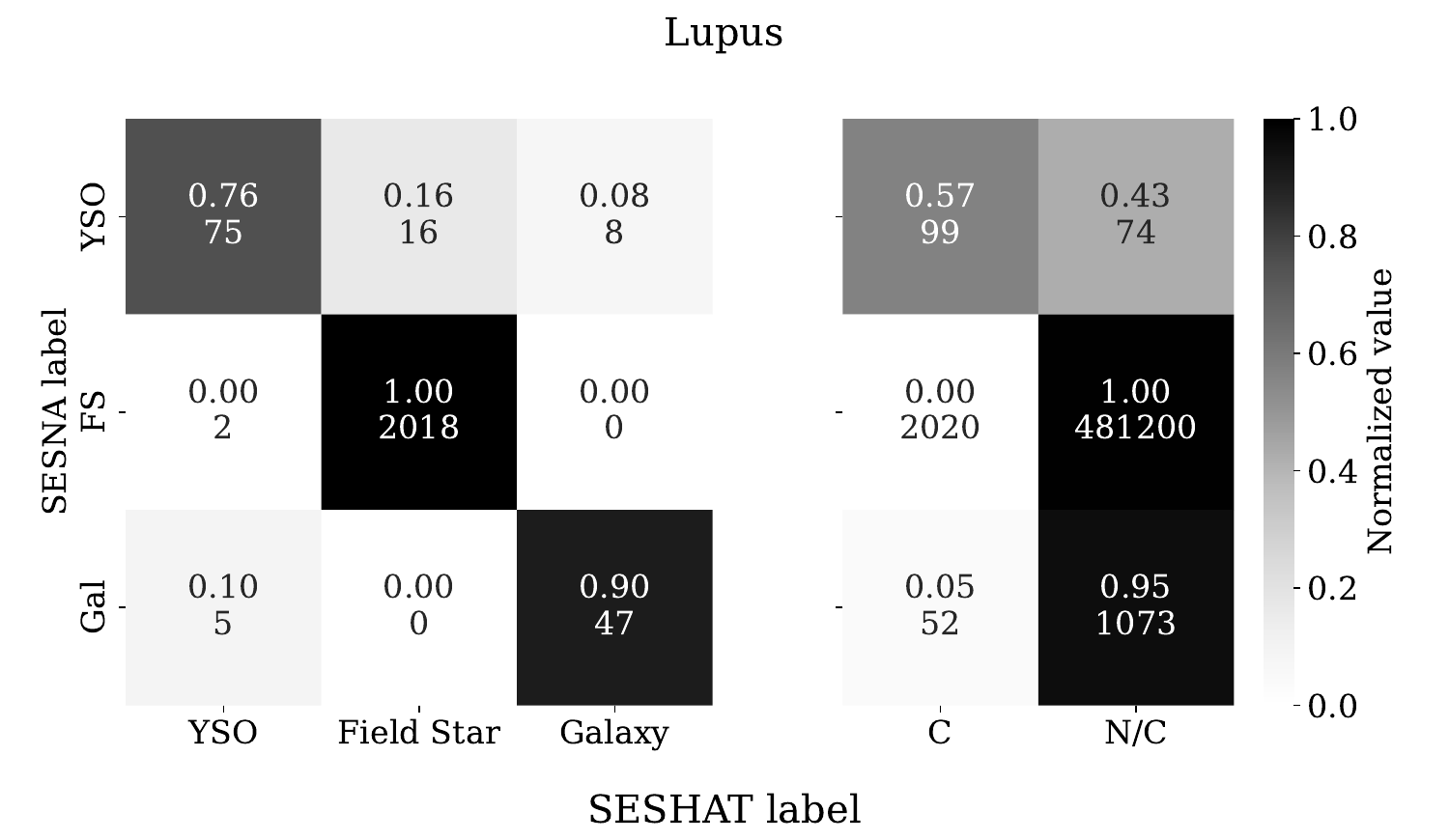}
\figsetplot{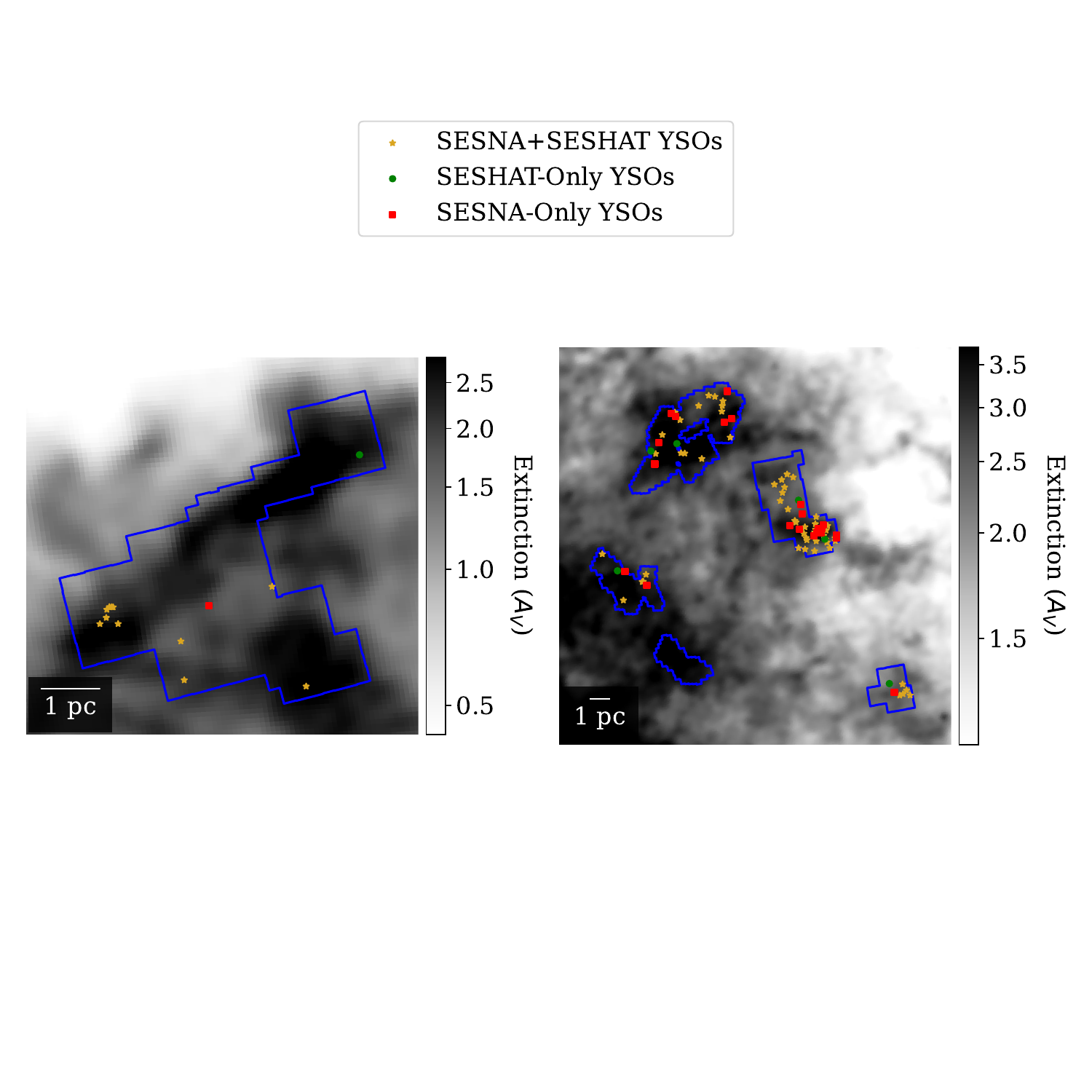}
\figsetgrpnote{The same as Figure~\ref{fig:afgl490}, but now for the Lupus region, which is part of the test set. The Spitzer observations are at wide enough separation to require separating the fields into two subplots for better visualization. \label{fig:lupus}}
\figsetgrpend

\figsetgrpstart
\figsetgrpnum{15.14}
\figsetgrptitle{Mon OB1}
\figsetplot{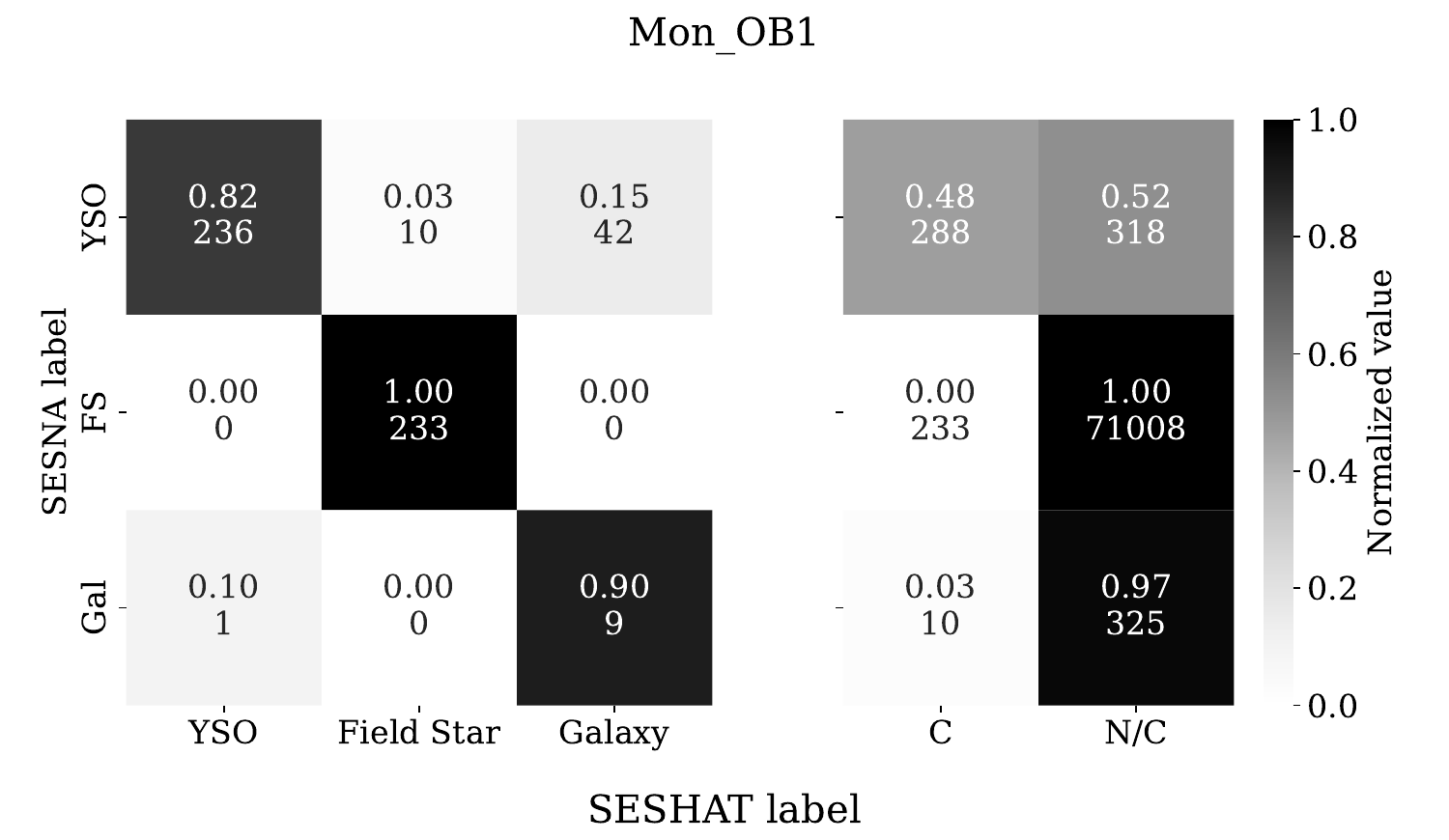}
\figsetplot{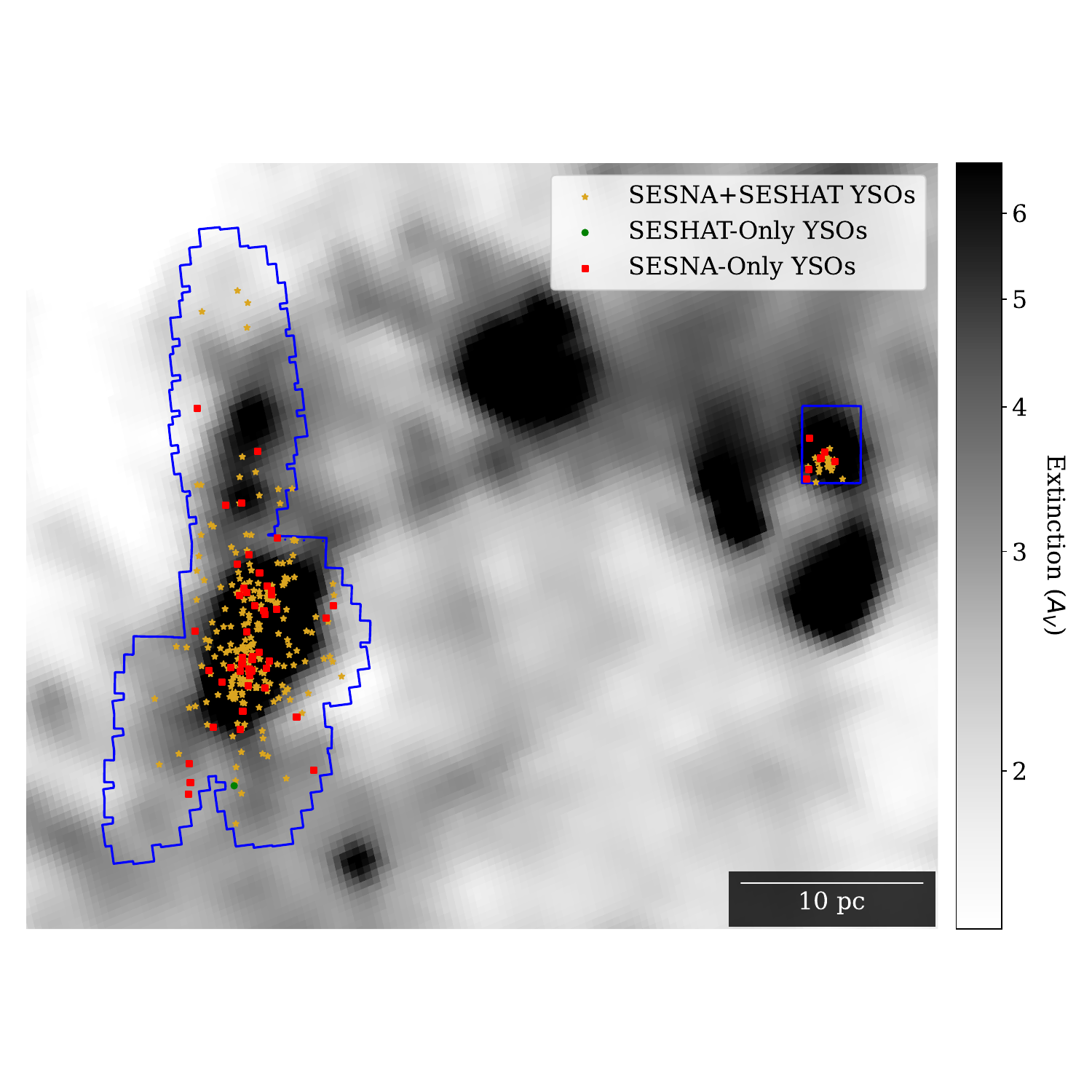}
\figsetgrpnote{The same as Figure~\ref{fig:afgl490}, but now for the Mon OB1 region, which is part of the test set. \label{fig:monob1}}
\figsetgrpend

\figsetgrpstart
\figsetgrpnum{15.15}
\figsetgrptitle{Mon R2}
\figsetplot{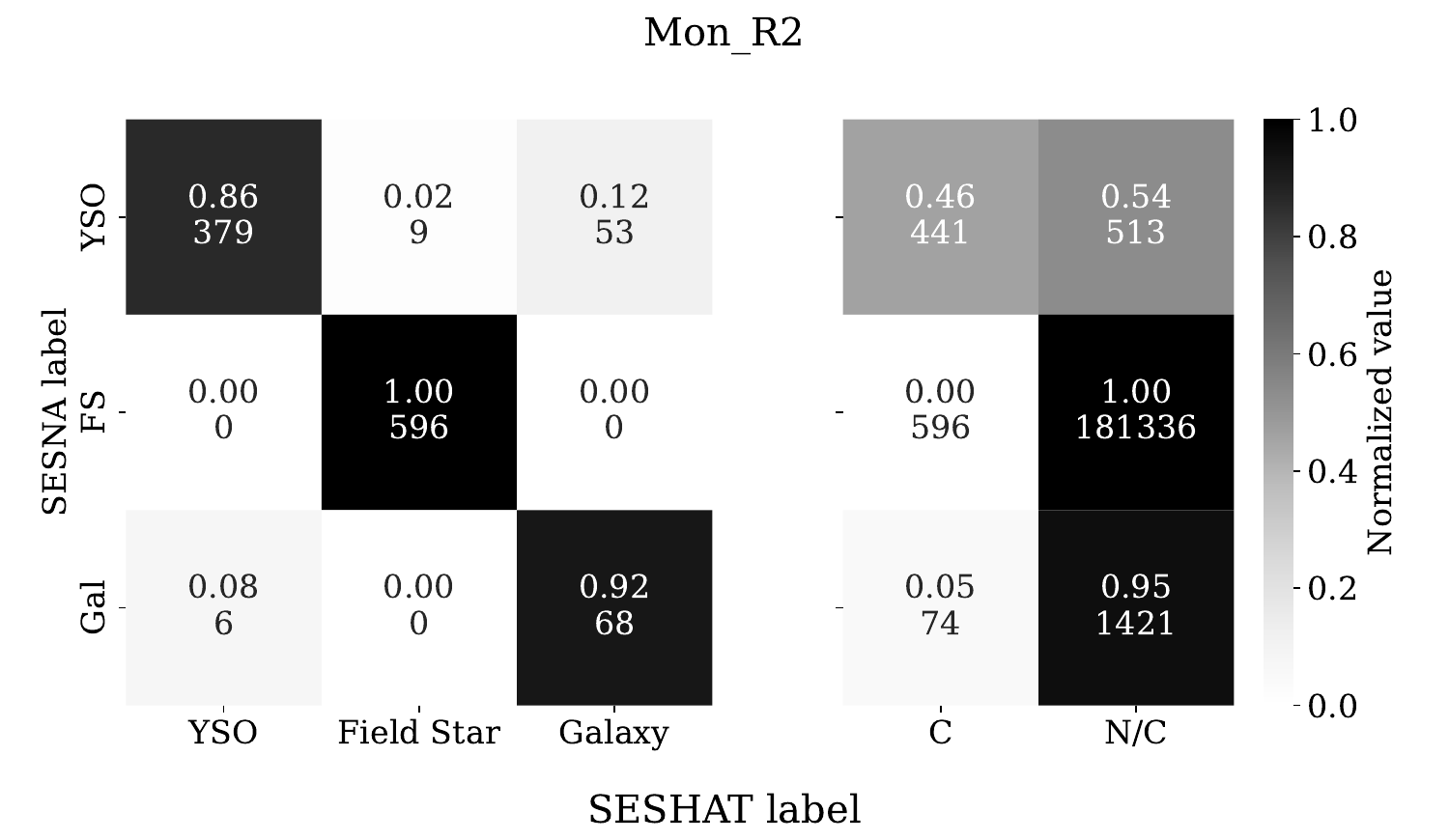}
\figsetplot{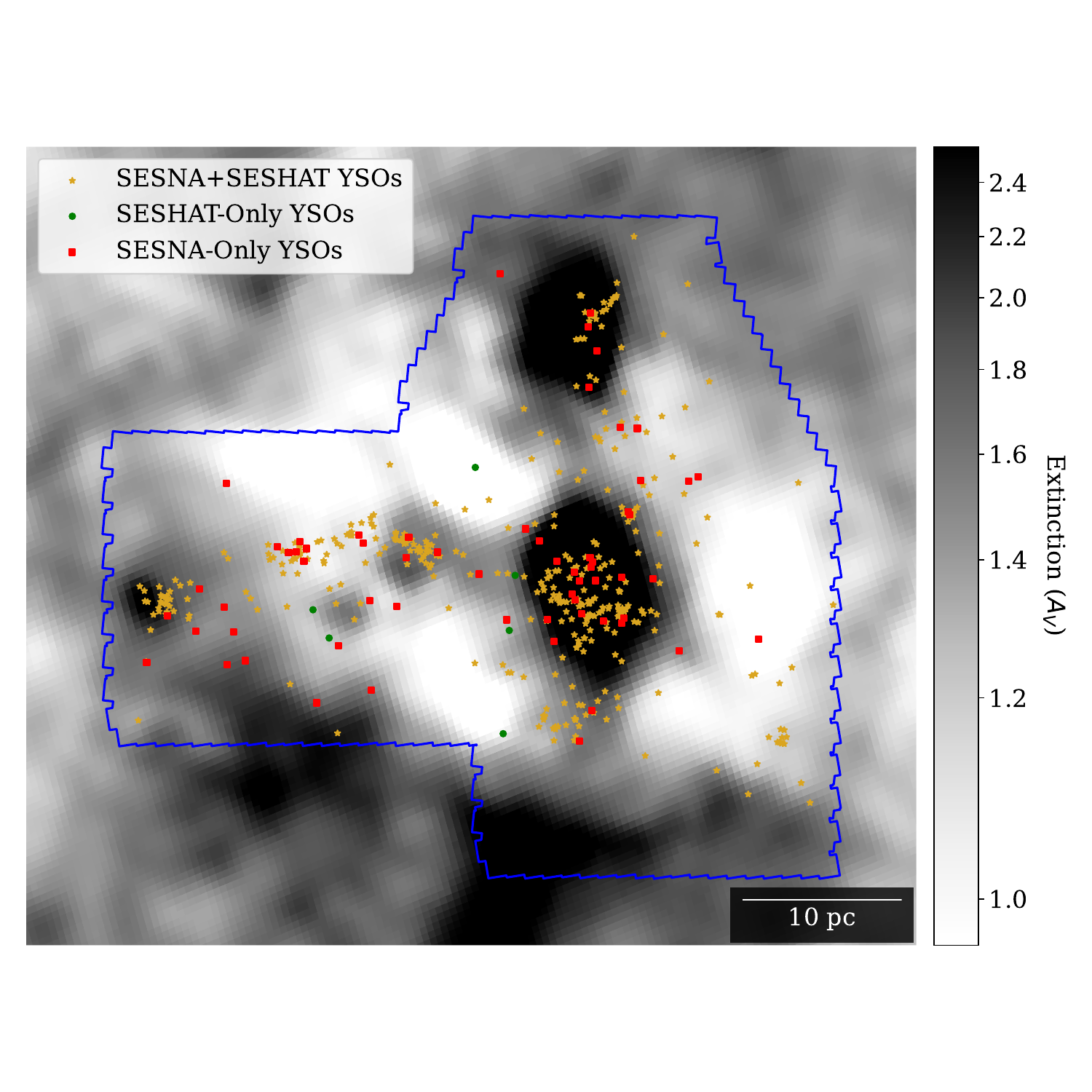}
\figsetgrpnote{The same as Figure~\ref{fig:afgl490}, but now for the Mon R2 region, which is part of the test set.  \label{fig:monr2}}
\figsetgrpend

\figsetgrpstart
\figsetgrpnum{15.16}
\figsetgrptitle{Musca}
\figsetplot{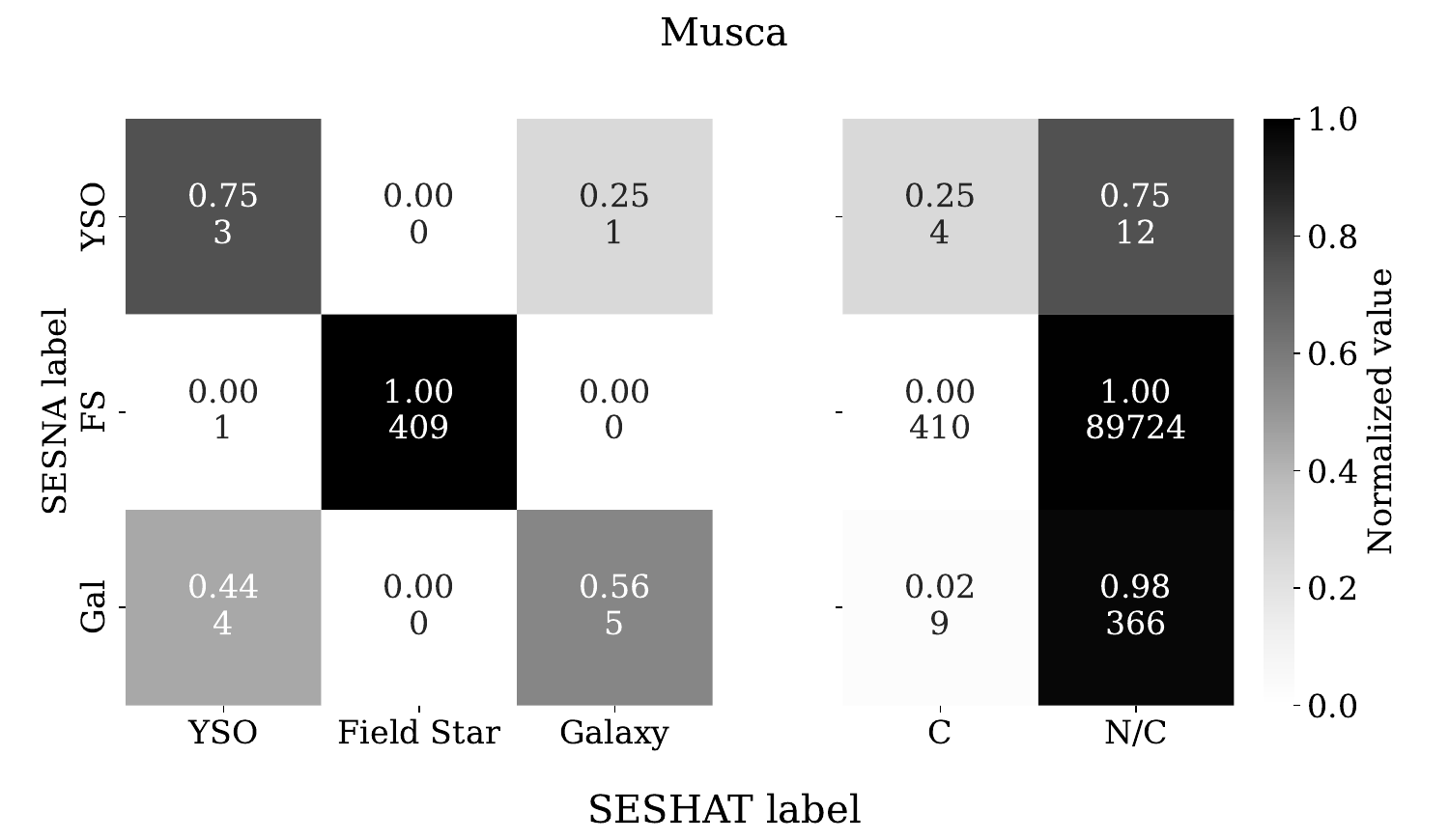}
\figsetplot{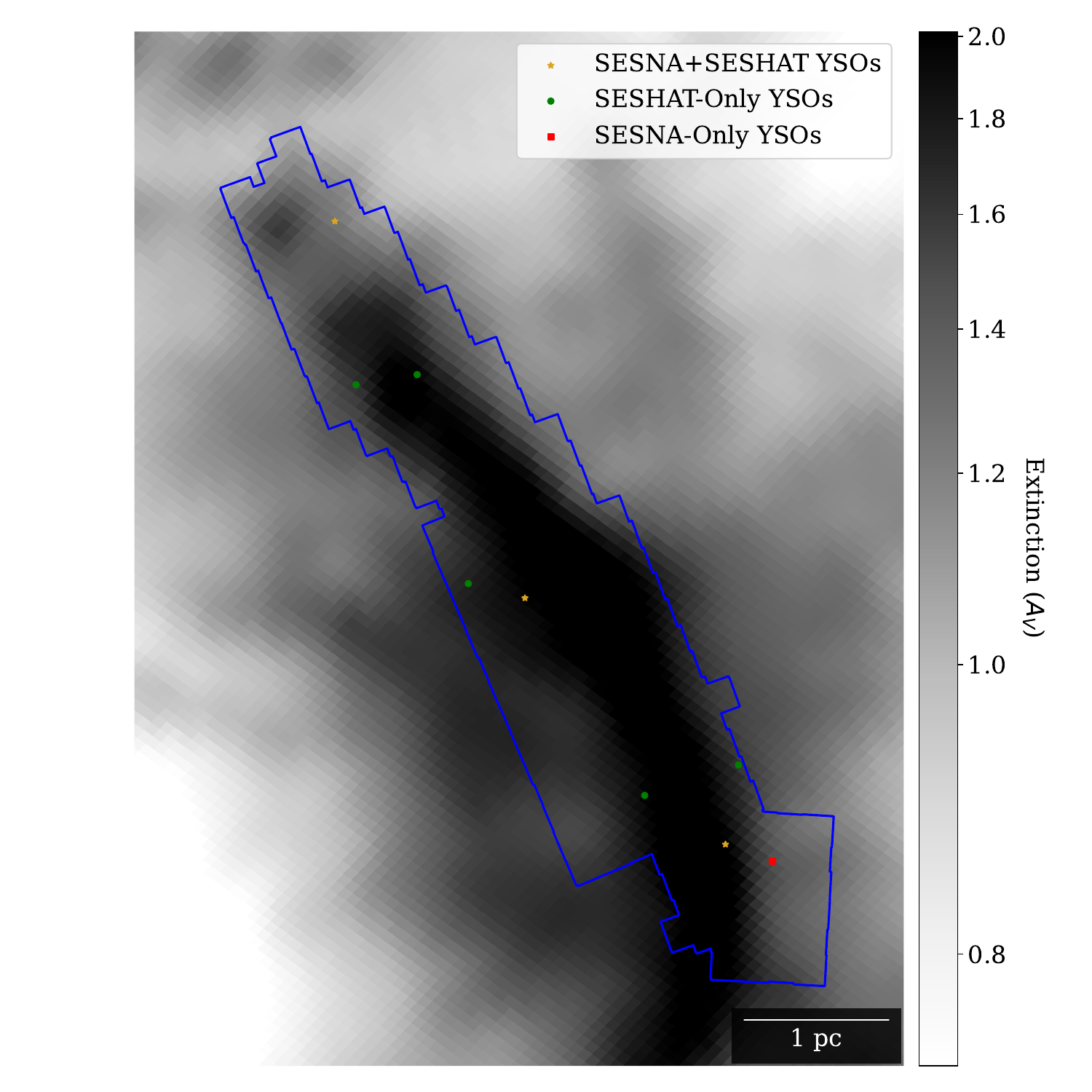}
\figsetgrpnote{The same as Figure~\ref{fig:afgl490}, but now for the Musca region, which is part of the test set. \label{fig:musca}}
\figsetgrpend

\figsetgrpstart
\figsetgrpnum{15.17}
\figsetgrptitle{NGC 7129}
\figsetplot{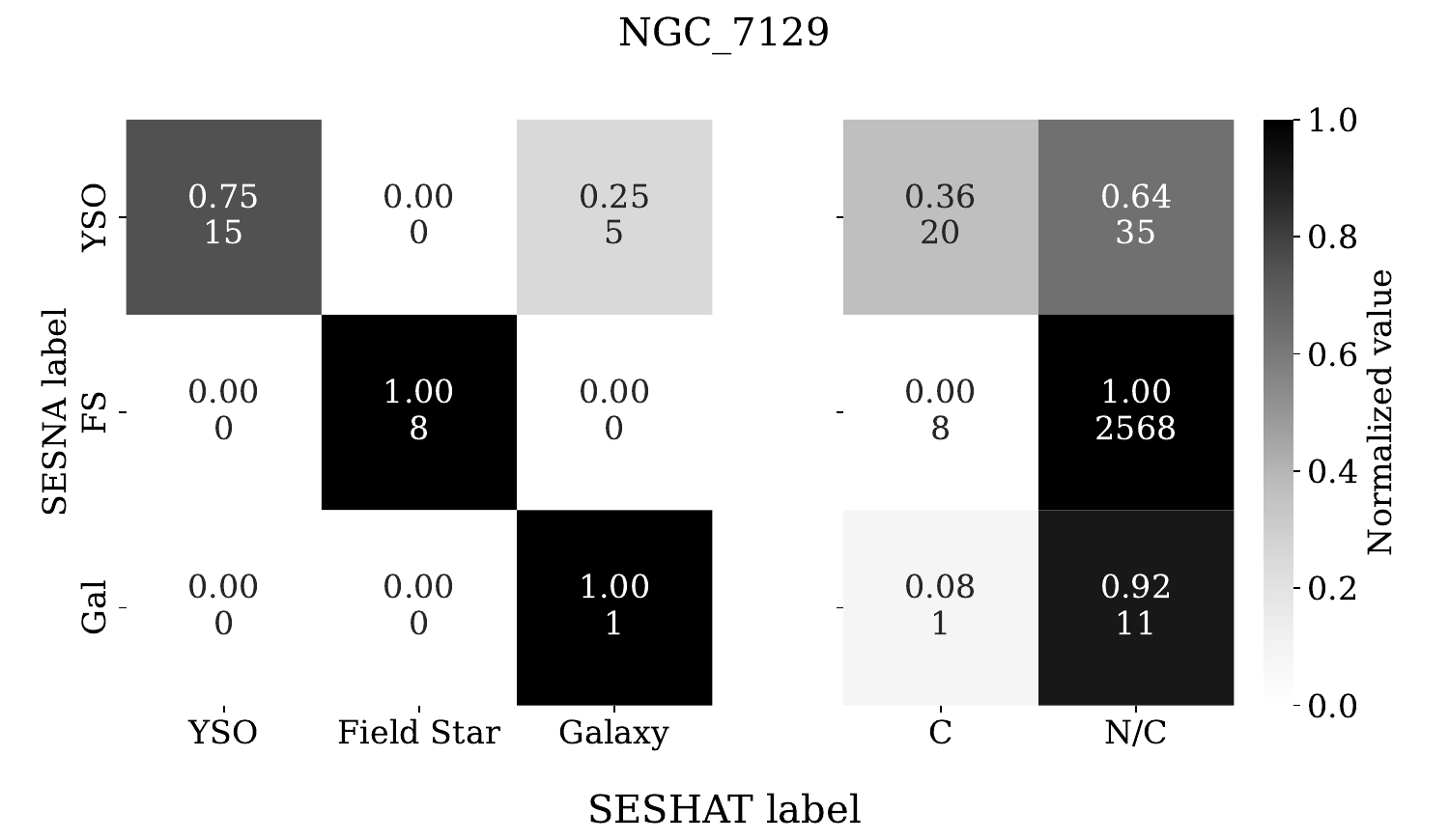}
\figsetplot{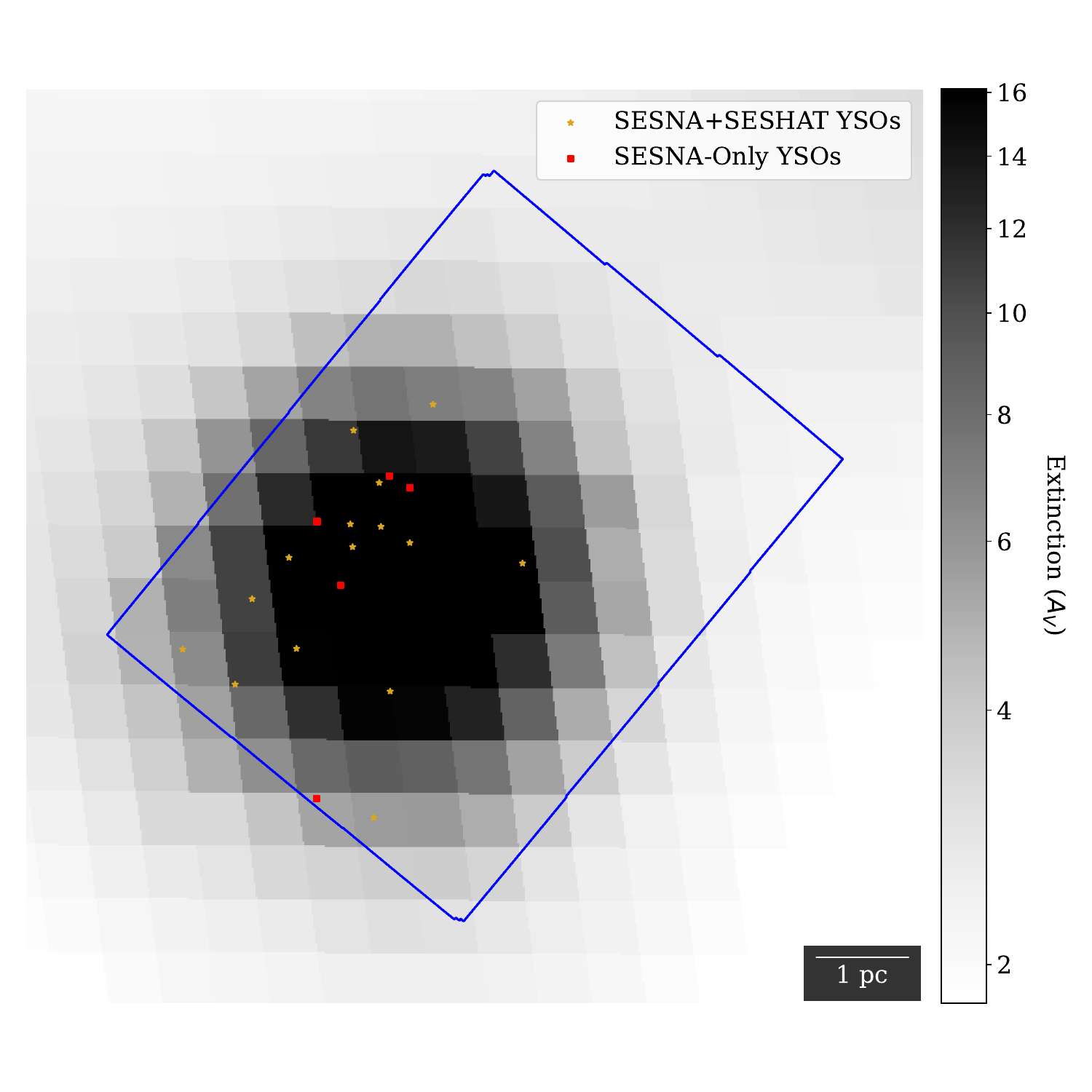}
\figsetgrpnote{The same as Figure~\ref{fig:afgl490}, but now for the NGC 7129 region, which is part of the test set. \label{fig:ngc7129}}
\figsetgrpend

\figsetgrpstart
\figsetgrpnum{15.18}
\figsetgrptitle{North America Nebula}
\figsetplot{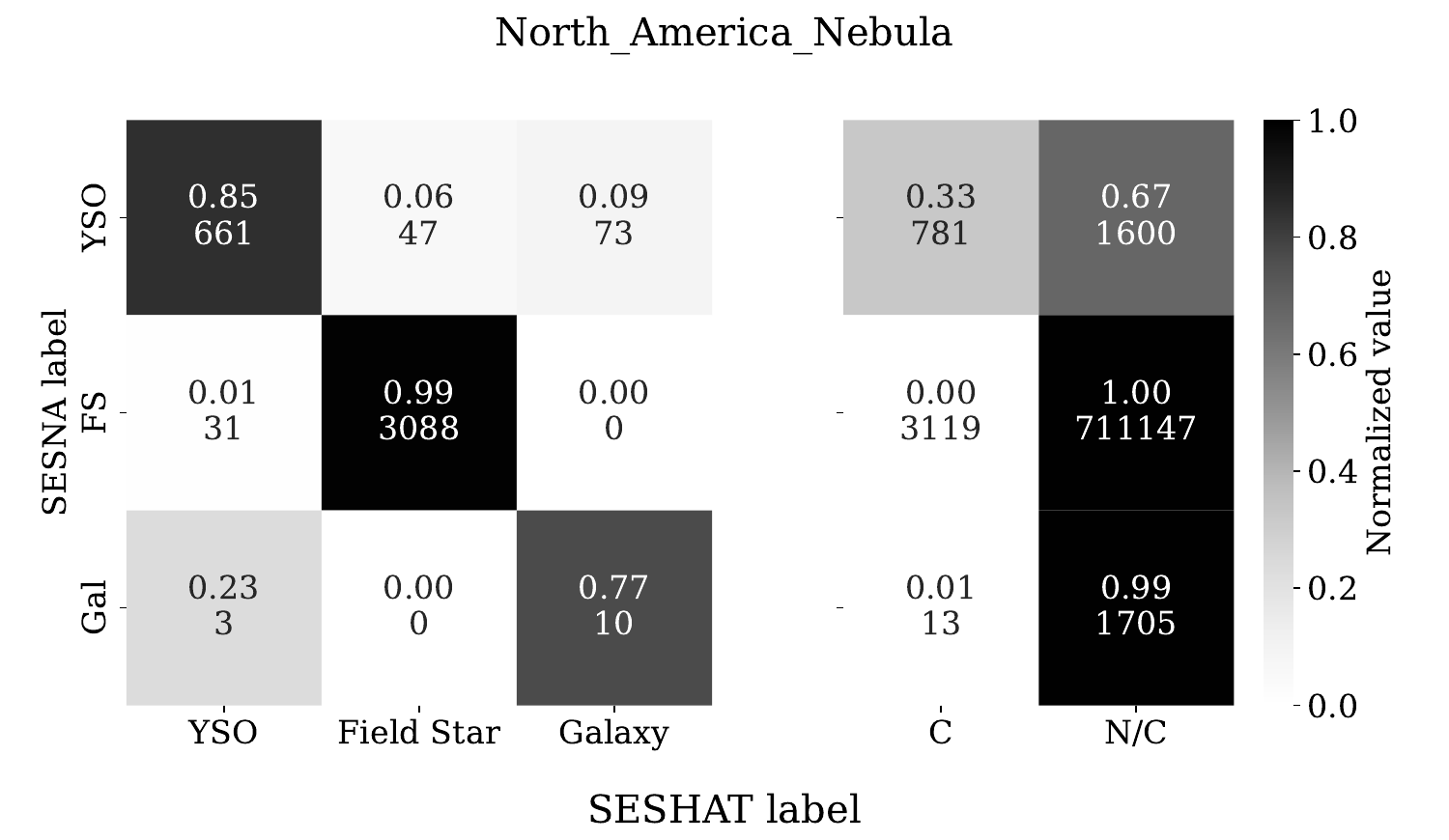}
\figsetplot{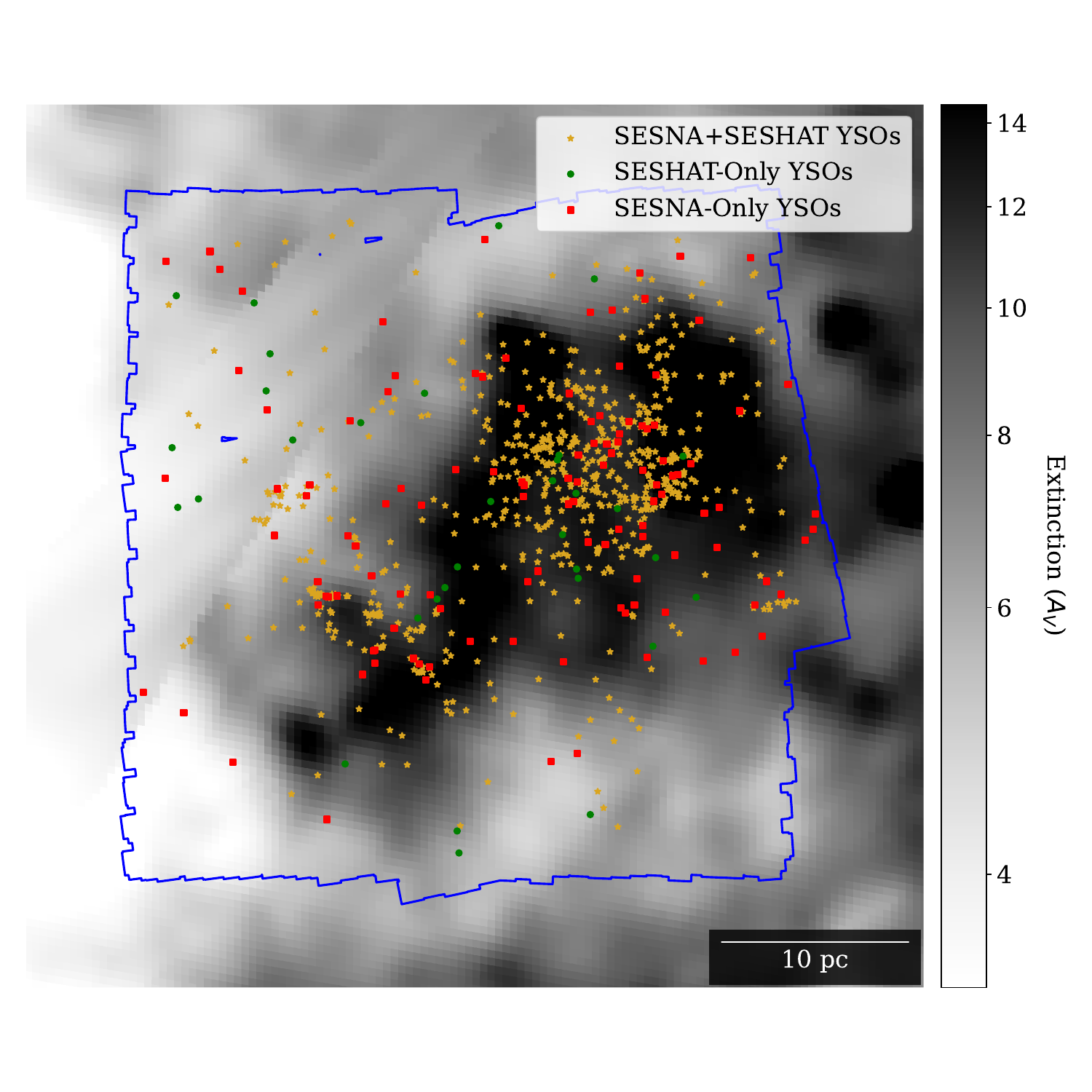}
\figsetgrpnote{The same as Figure~\ref{fig:afgl490}, but now for the North America Nebula, which is part of the test set. \label{fig:northamerica}}
\figsetgrpend

\figsetgrpstart
\figsetgrpnum{15.19}
\figsetgrptitle{Ophiuchus}
\figsetplot{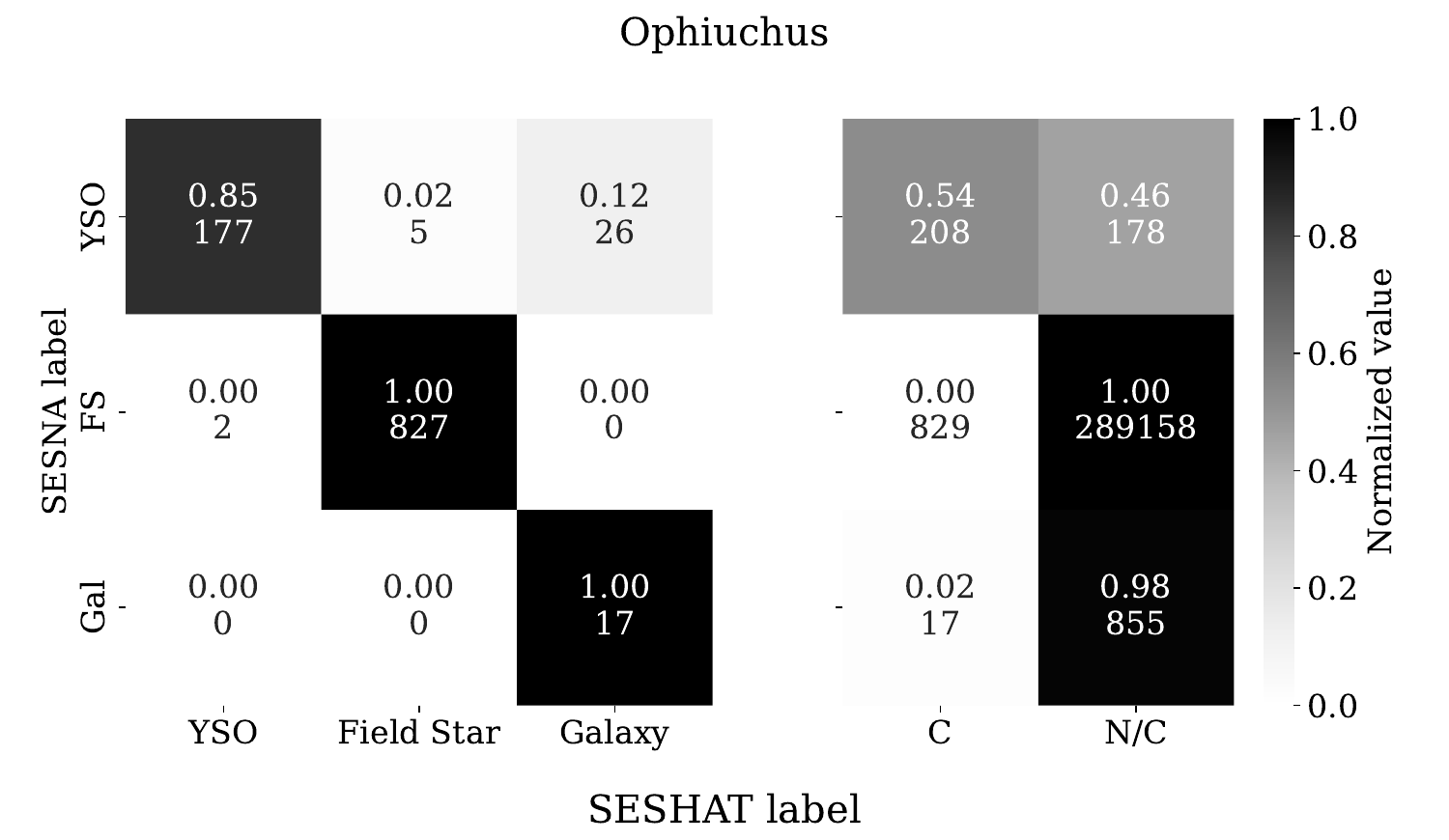}
\figsetplot{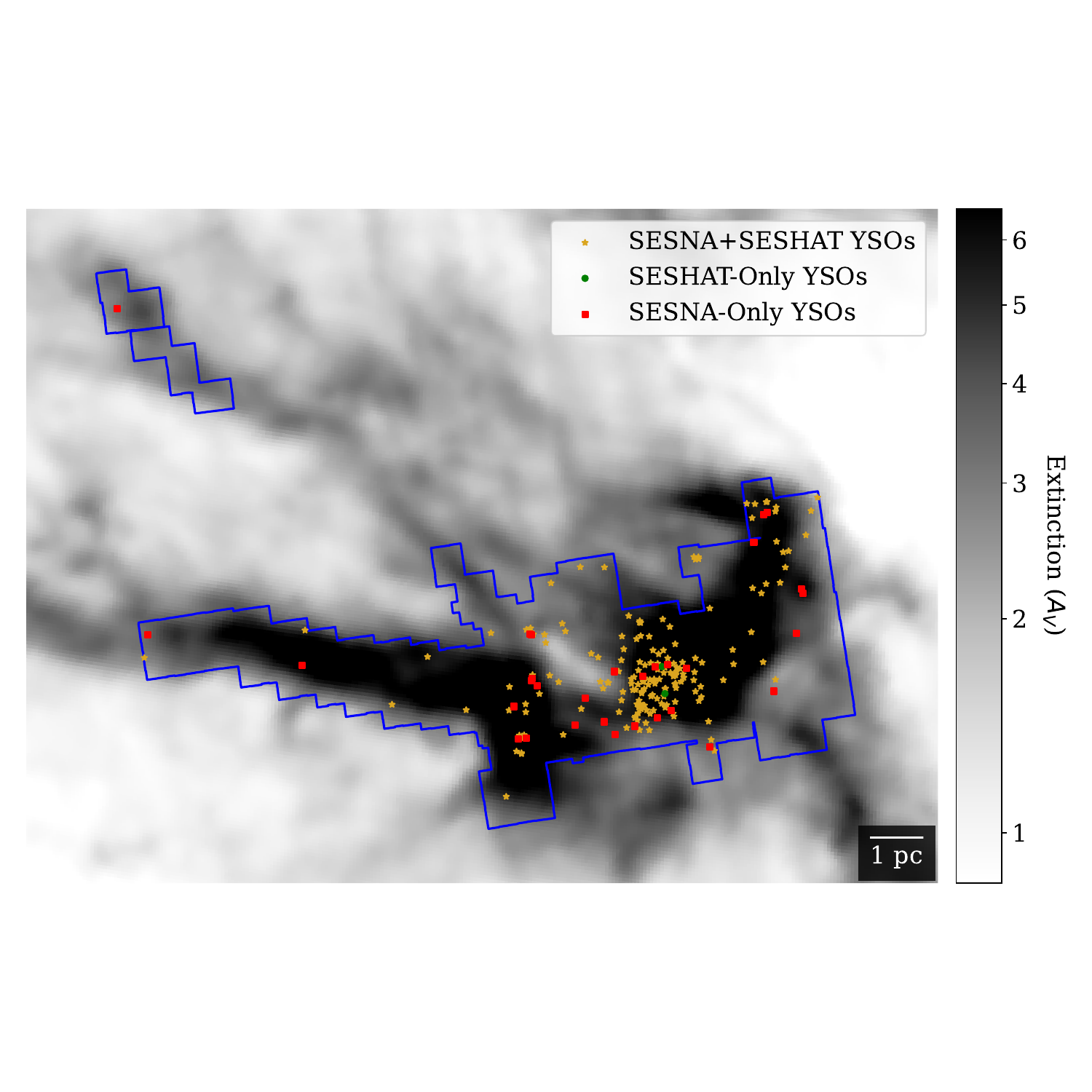}
\figsetgrpnote{The same as Figure~\ref{fig:afgl490}, but now for Ophiuchus, which is part of set used to define training YSOs. \label{fig:ophiuchus}}
\figsetgrpend

\figsetgrpstart
\figsetgrpnum{15.20}
\figsetgrptitle{Orion A}
\figsetplot{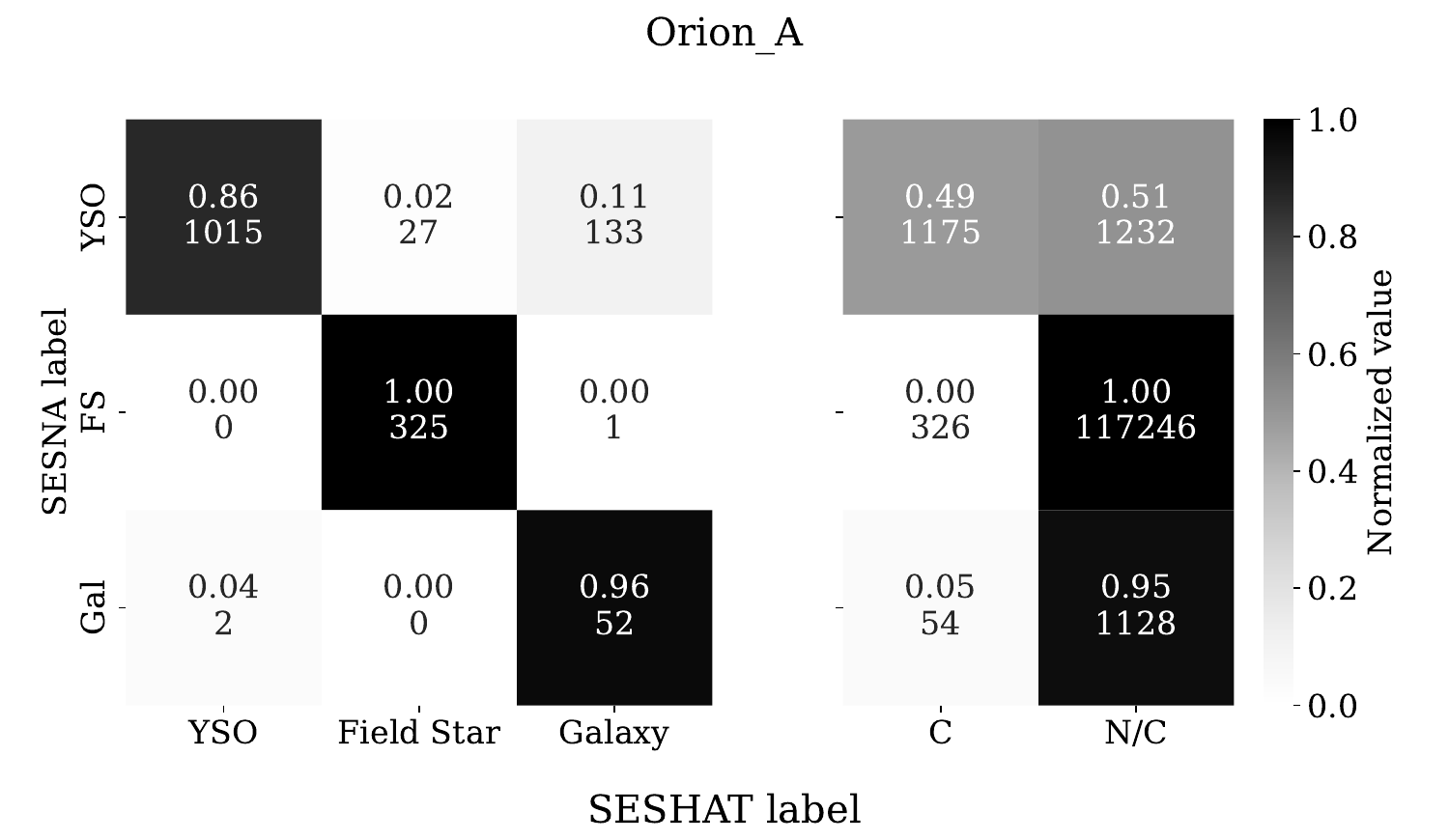}
\figsetplot{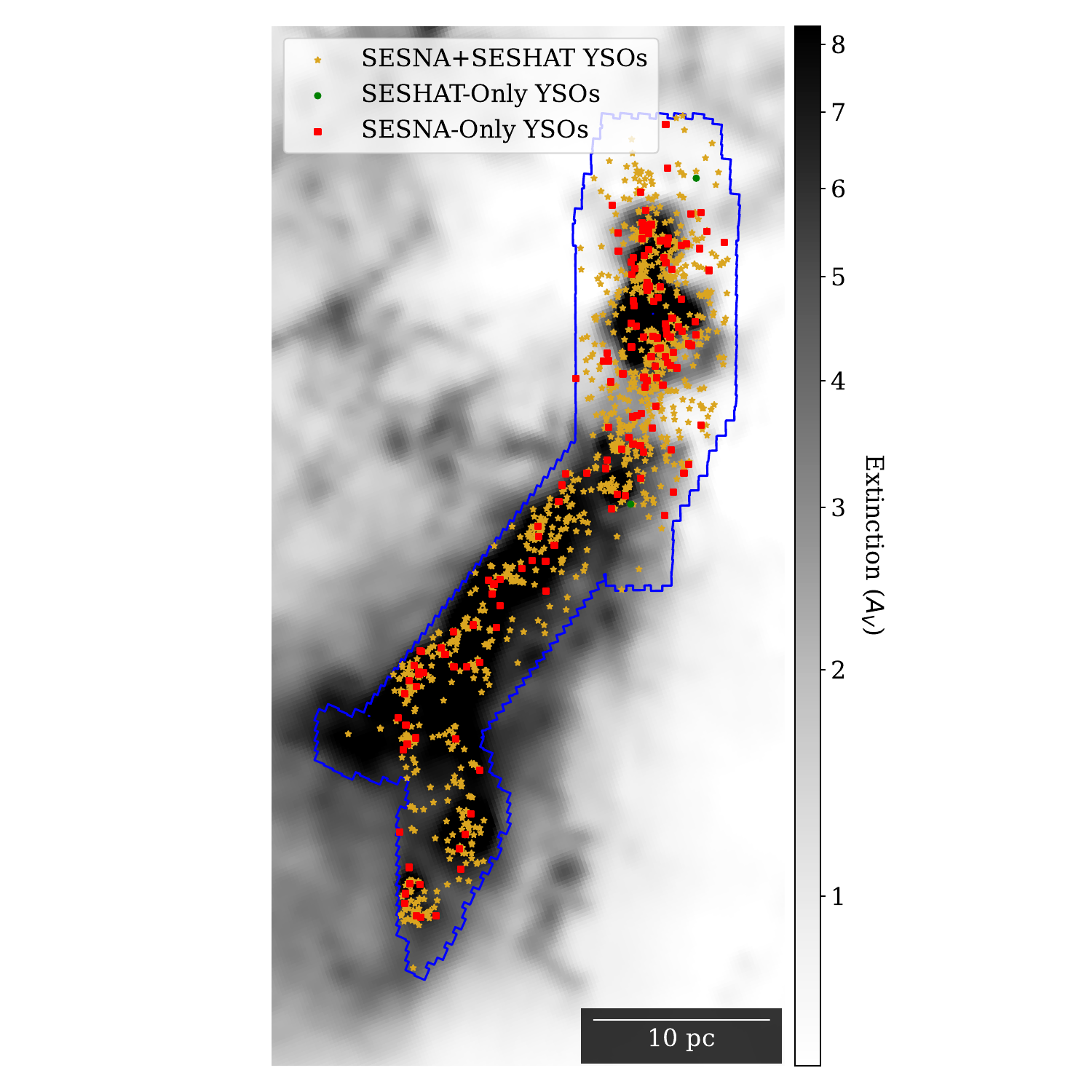}
\figsetgrpnote{The same as Figure~\ref{fig:afgl490}, but now for Orion A, which is part of set used to define training YSOs. \label{fig:oriona}}
\figsetgrpend

\figsetgrpstart
\figsetgrpnum{15.21}
\figsetgrptitle{Orion B}
\figsetplot{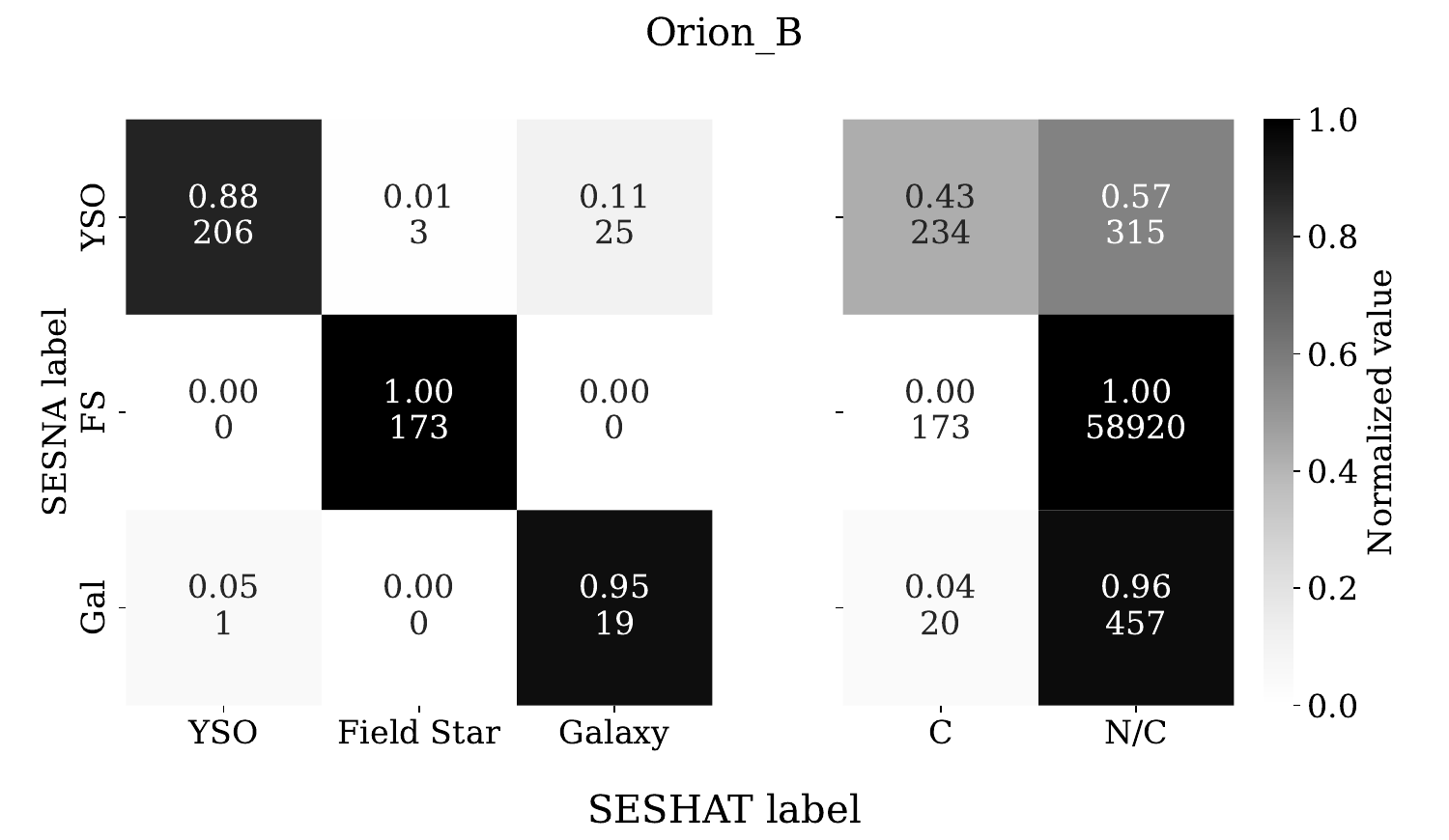}
\figsetplot{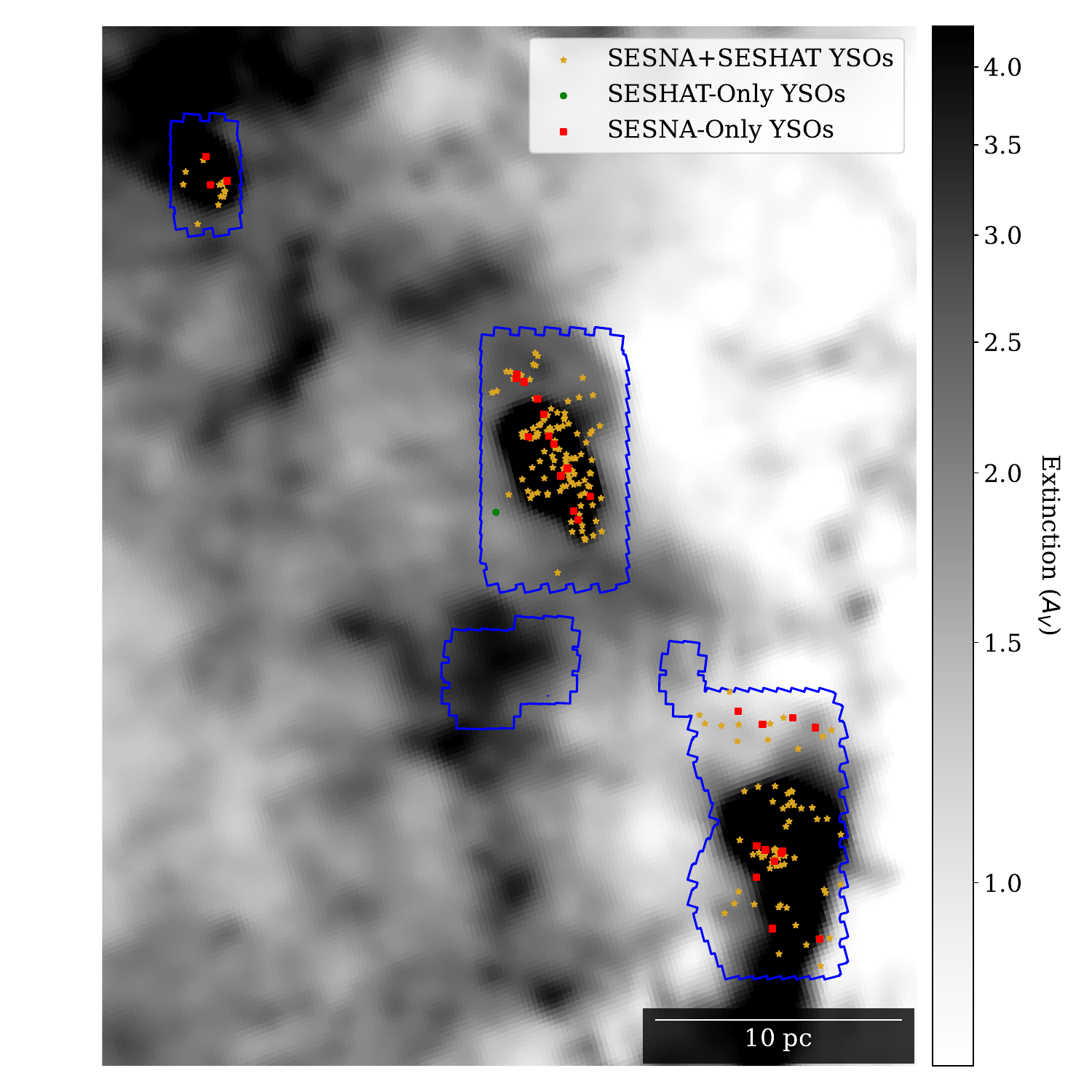}
\figsetgrpnote{The same as Figure~\ref{fig:afgl490}, but now for Orion B, which is part of set used to define training YSOs. \label{fig:orionb}}
\figsetgrpend

\figsetgrpstart
\figsetgrpnum{15.22}
\figsetgrptitle{Perseus}
\figsetplot{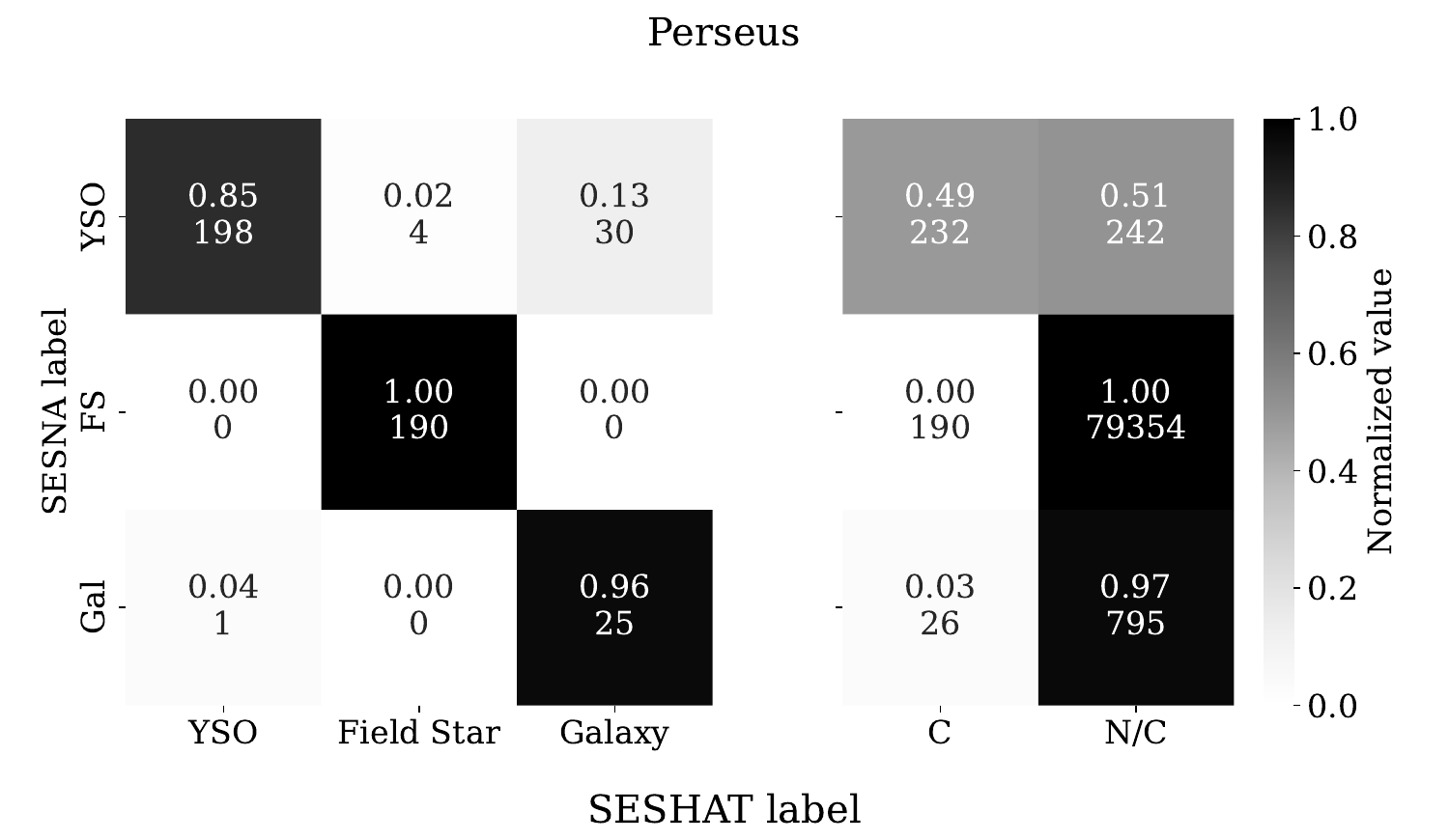}
\figsetplot{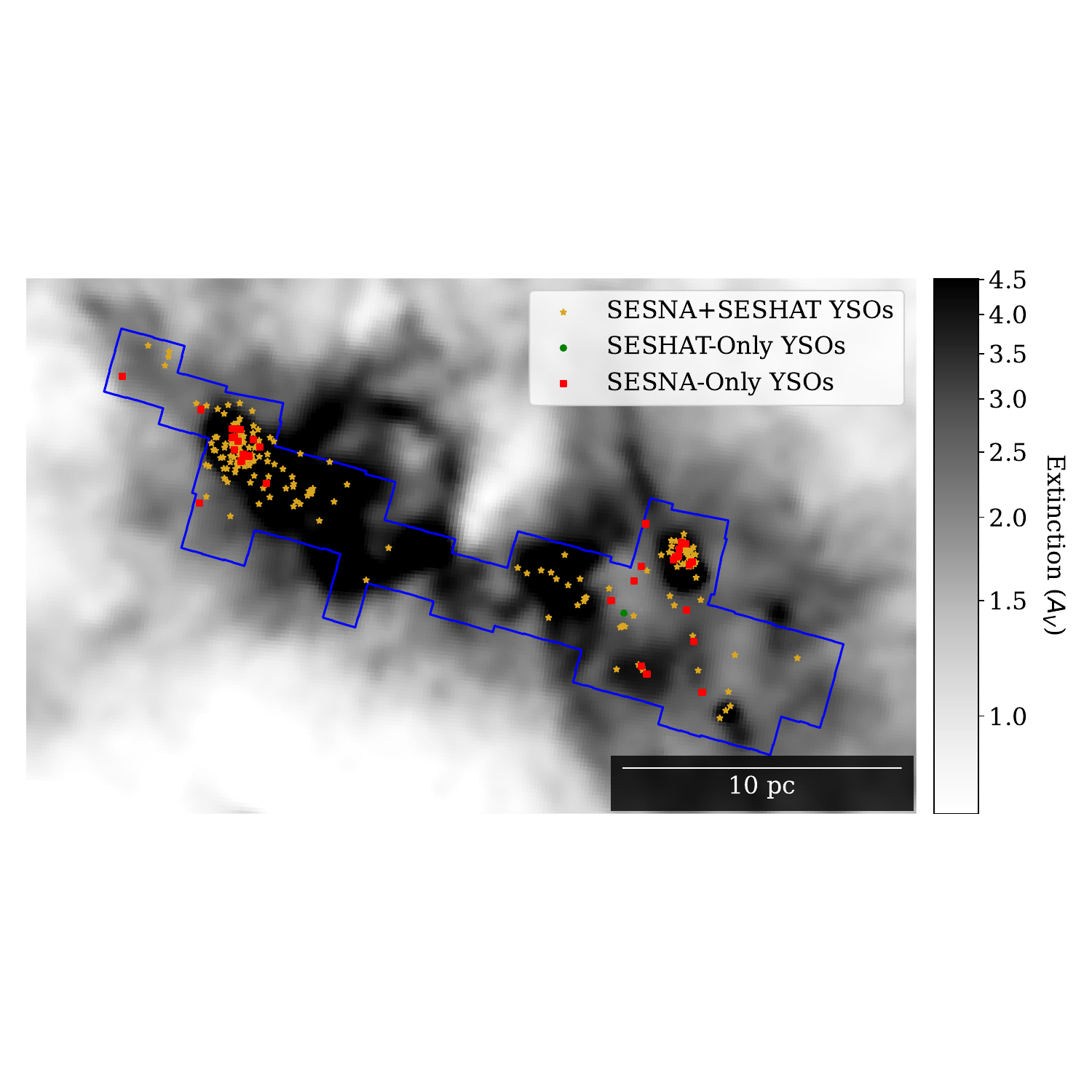}
\figsetgrpnote{The same as Figure~\ref{fig:afgl490}, but now for the Perseus region, which is part of the test set. \label{fig:perseus}}
\figsetgrpend

\figsetgrpstart
\figsetgrpnum{15.23}
\figsetgrptitle{Pipe}
\figsetplot{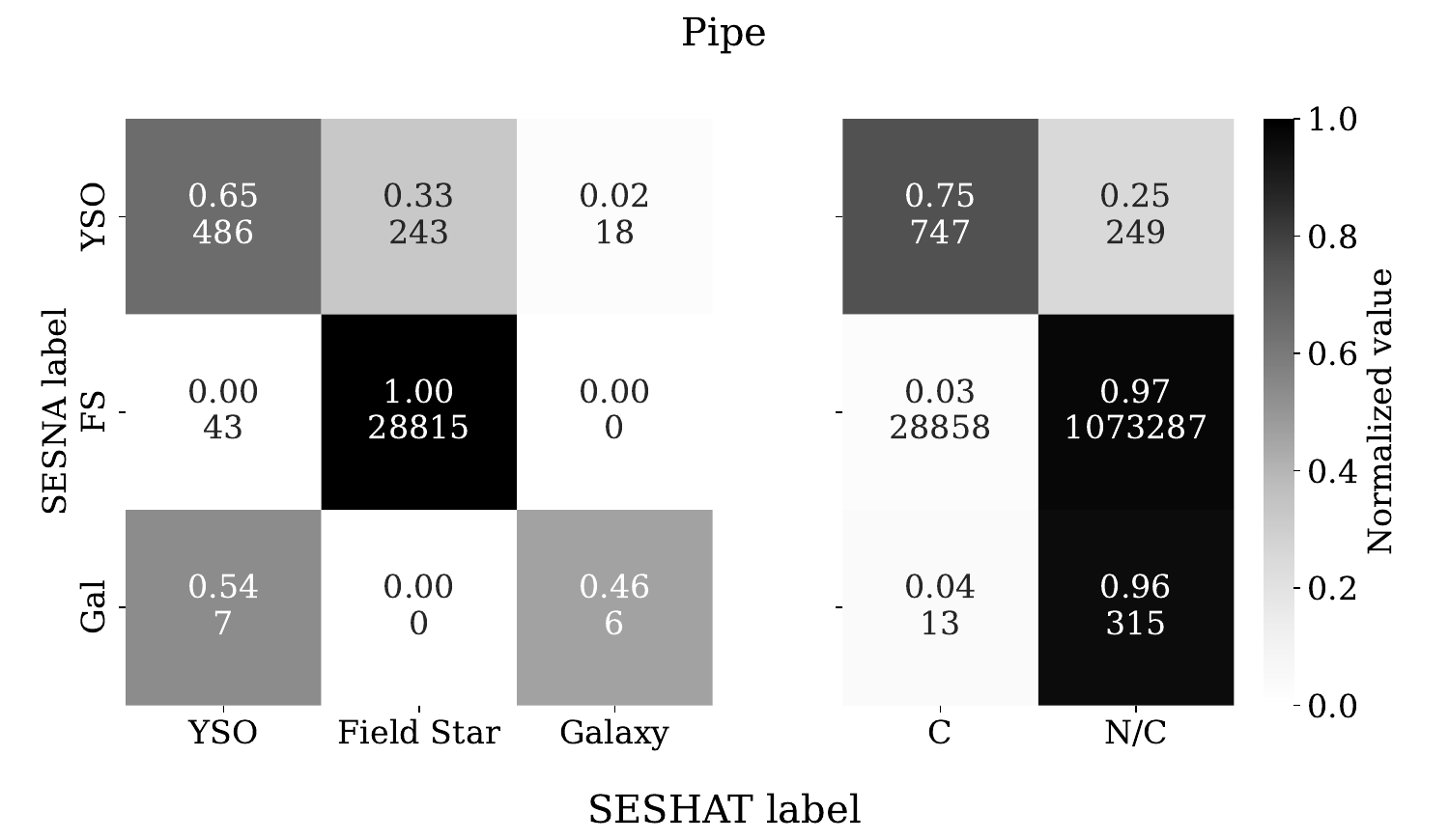}
\figsetplot{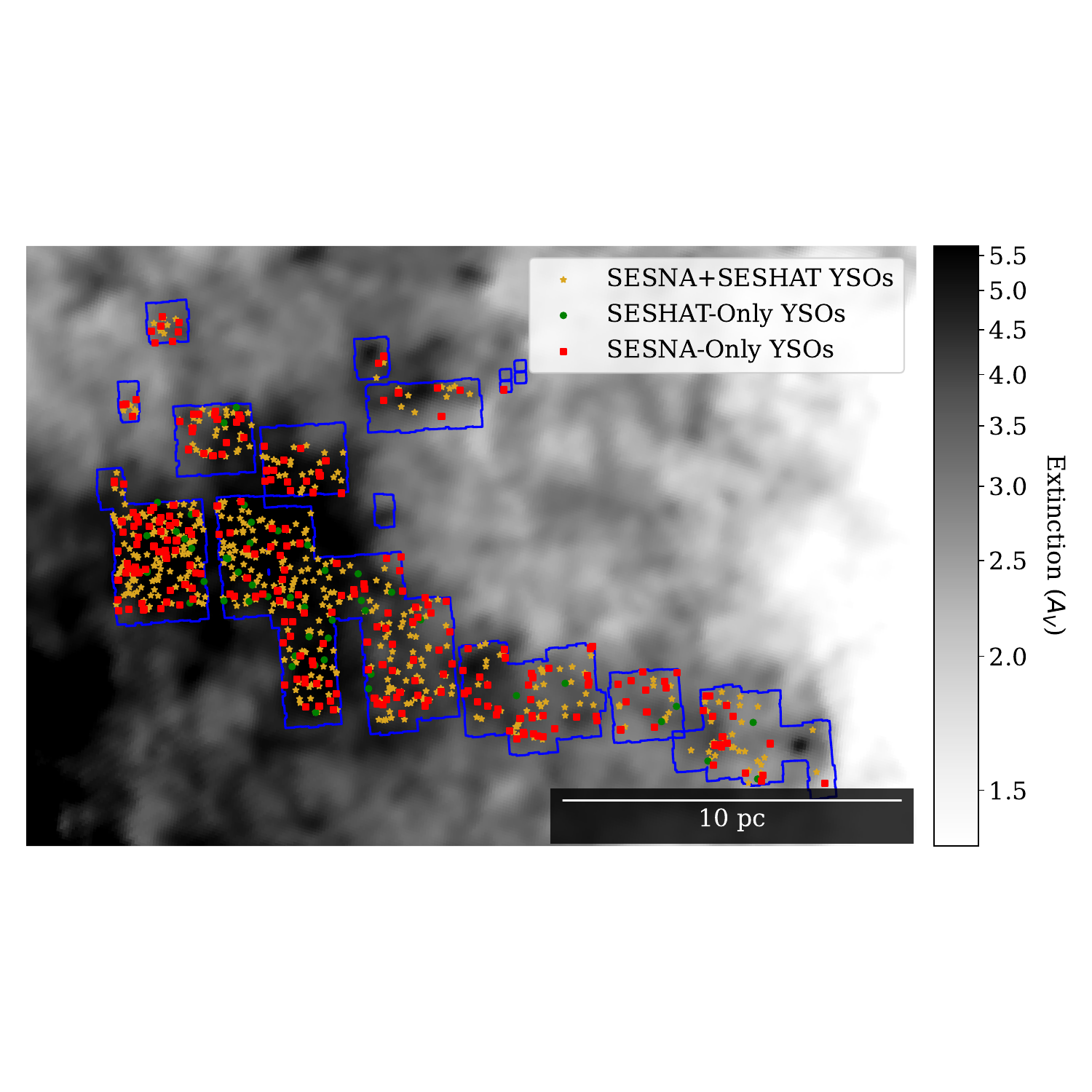}
\figsetgrpnote{The same as Figure~\ref{fig:afgl490}, but now for the Pipe region, which is excluded from both test and training sets. NOTE: there is no estimate for the contamination rate by field stars, which, though generally easy to distinguish from YSOs, becomes significant when looking directly into the Galactic plane, as is done here. \label{fig:pipe}}
\figsetgrpend

\figsetgrpstart
\figsetgrpnum{15.24}
\figsetgrptitle{S 131}
\figsetplot{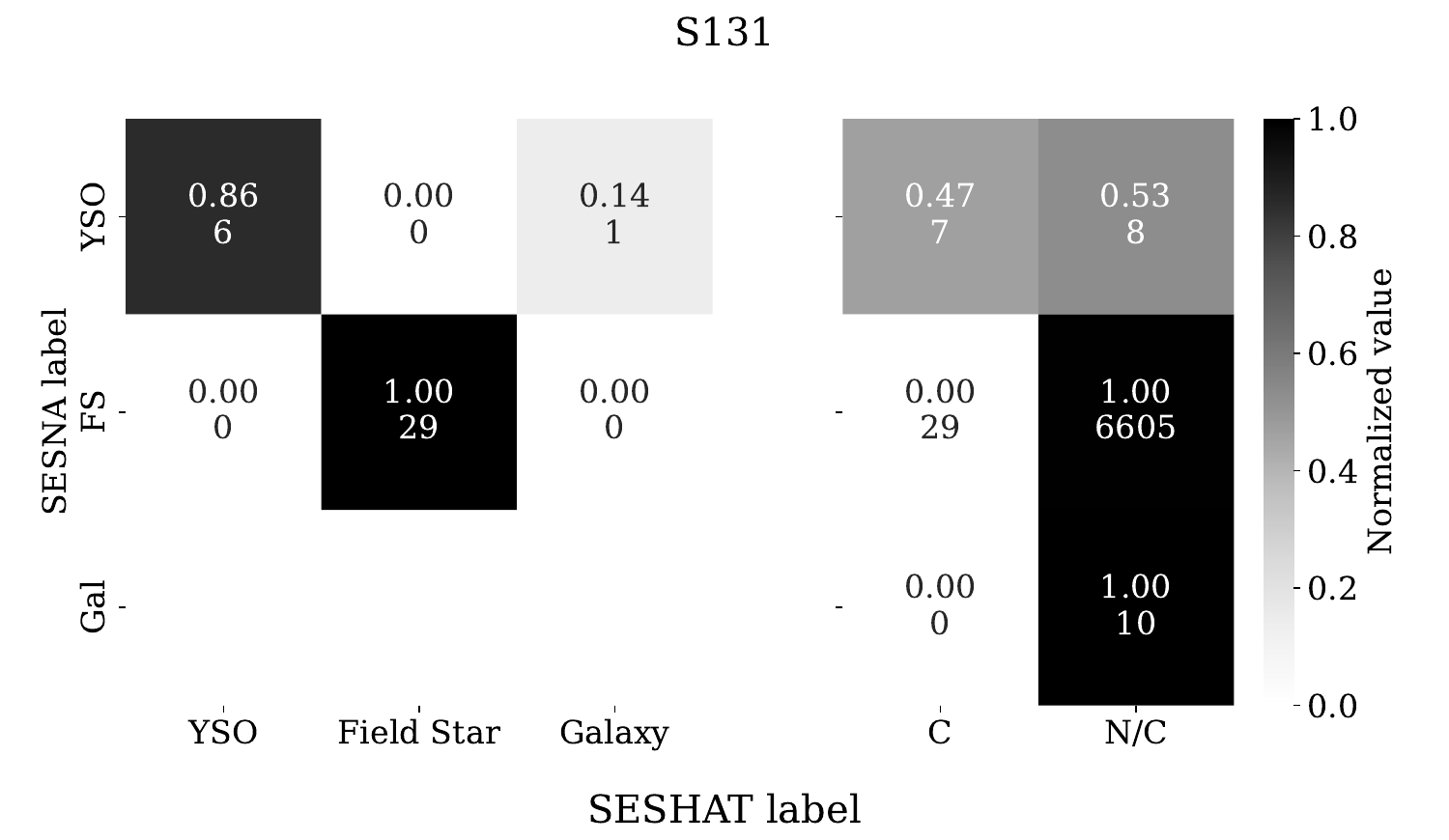}
\figsetplot{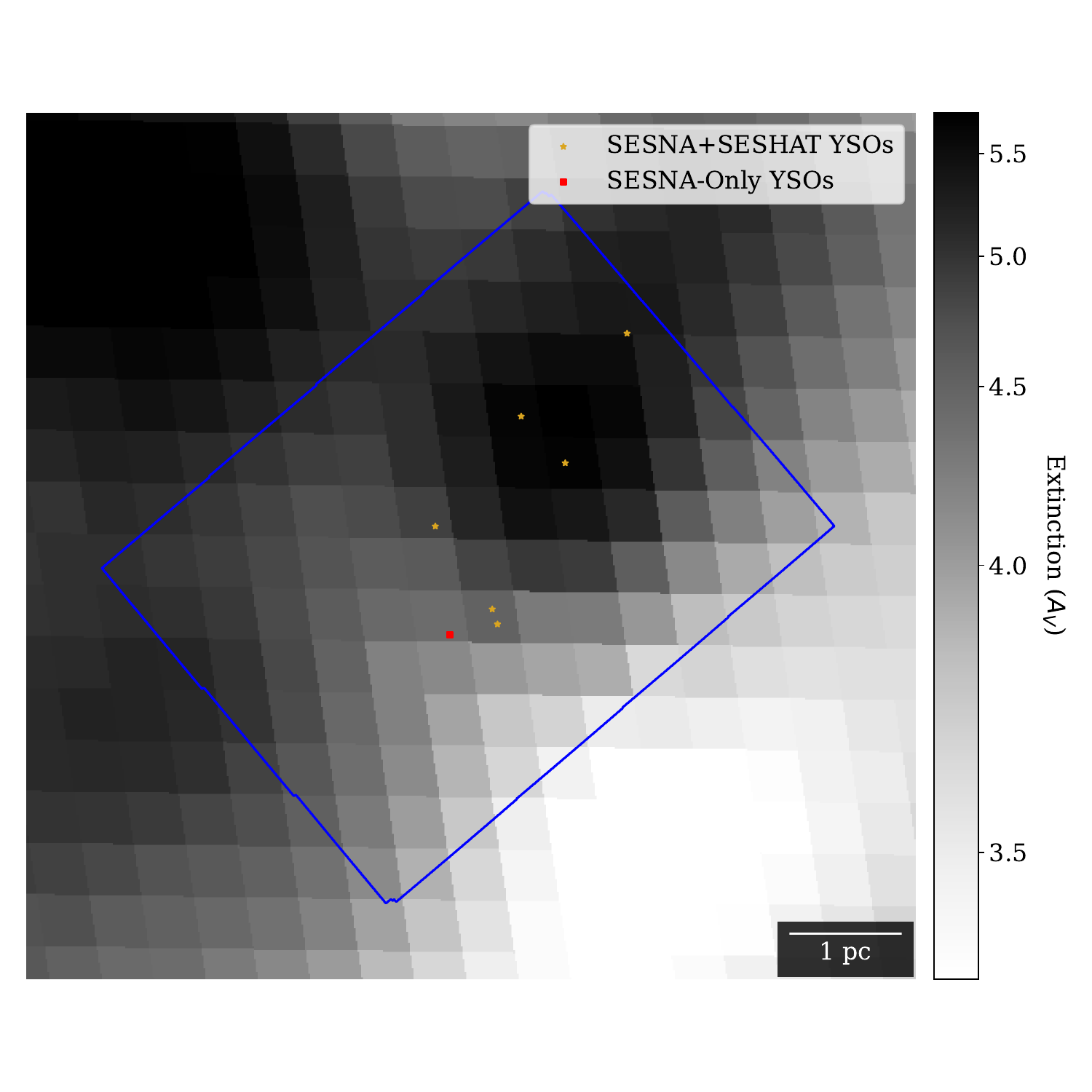}
\figsetgrpnote{The same as Figure~\ref{fig:afgl490}, but now for the S 131 region, which is part of the test set. \label{fig:s131}}
\figsetgrpend

\figsetgrpstart
\figsetgrpnum{15.25}
\figsetgrptitle{S 140}
\figsetplot{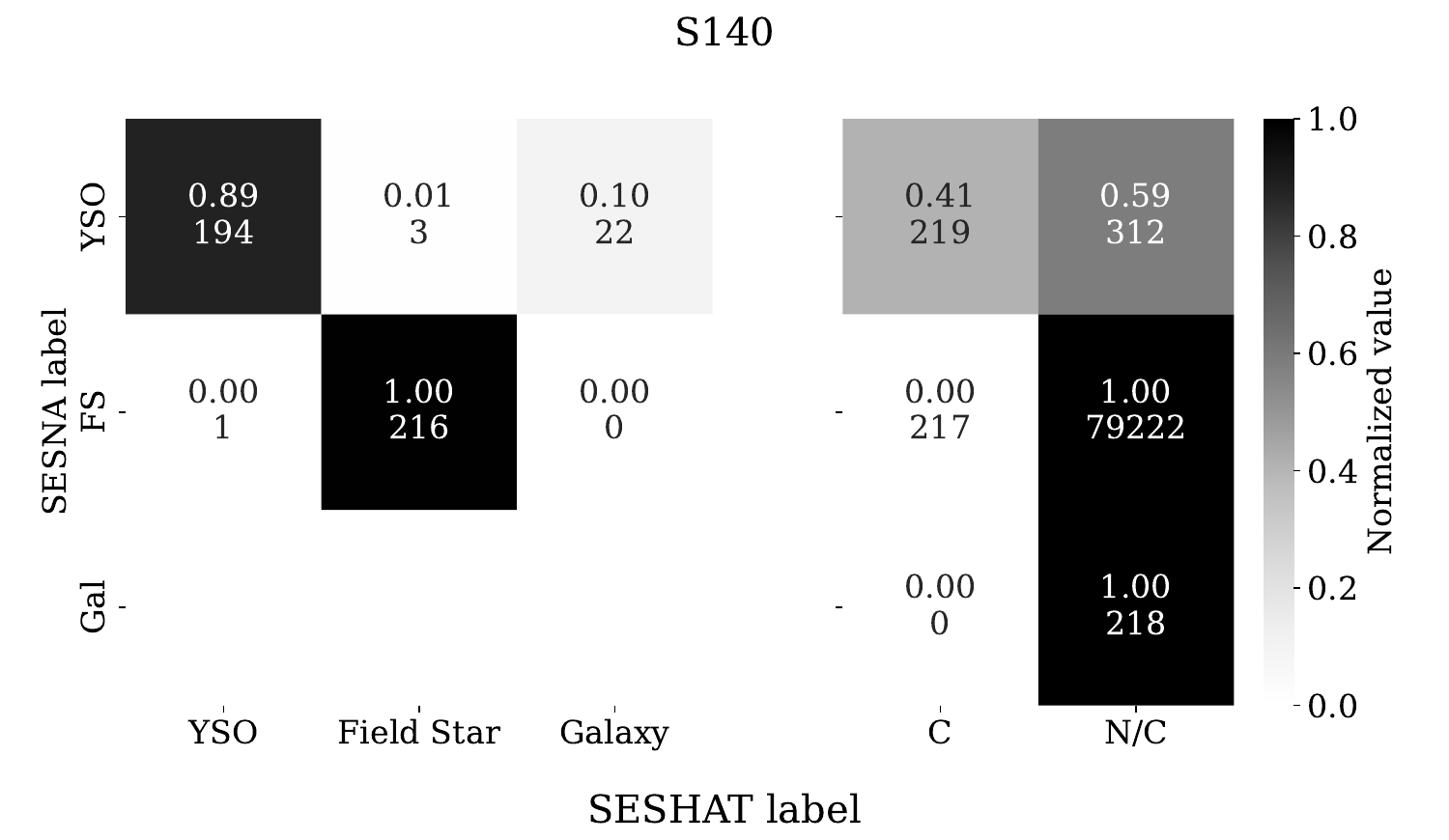}
\figsetplot{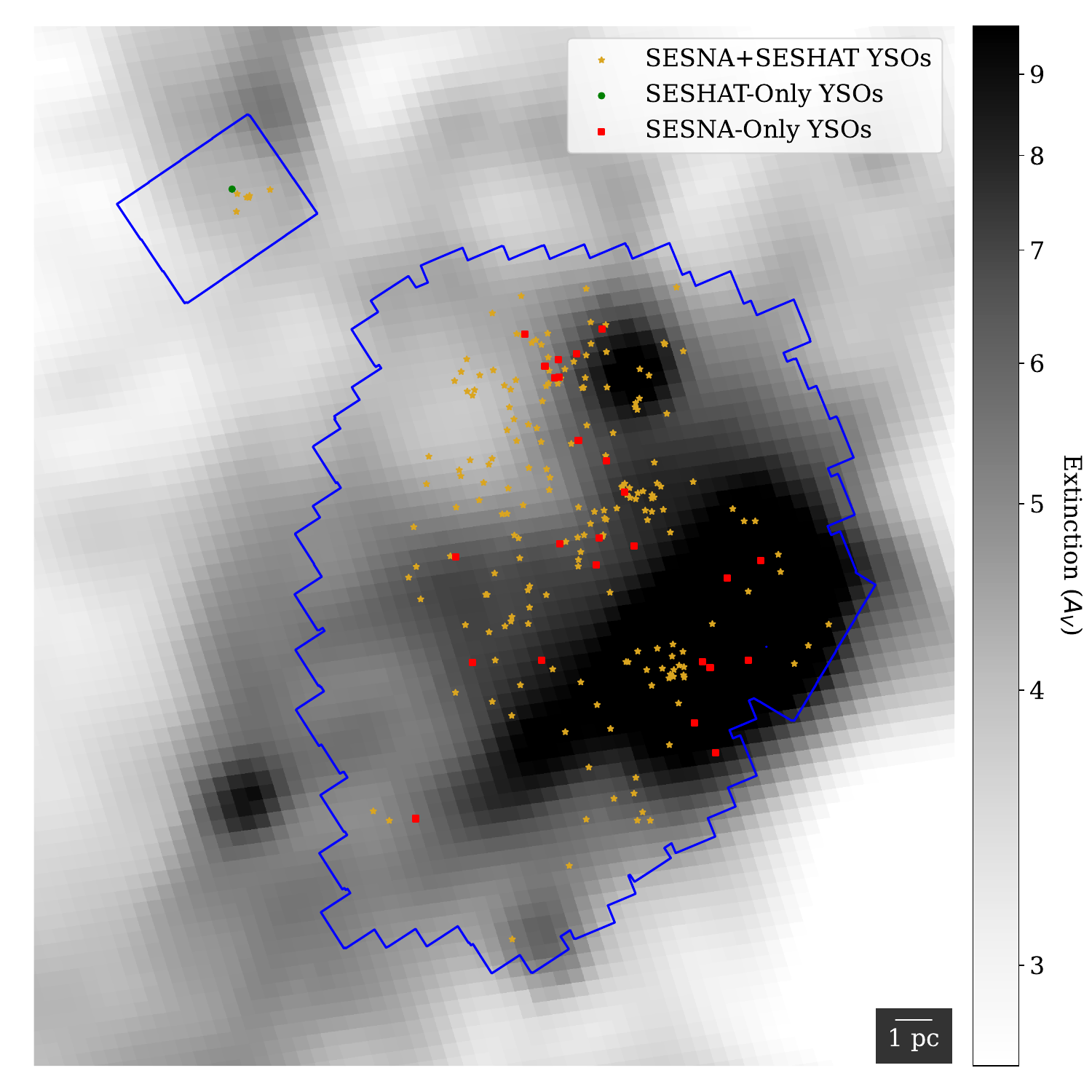}
\figsetgrpnote{The same as Figure~\ref{fig:afgl490}, but now for the S 140 region, which is part of the test set. \label{fig:s140}}
\figsetgrpend

\figsetgrpstart
\figsetgrpnum{15.26}
\figsetgrptitle{S 171}
\figsetplot{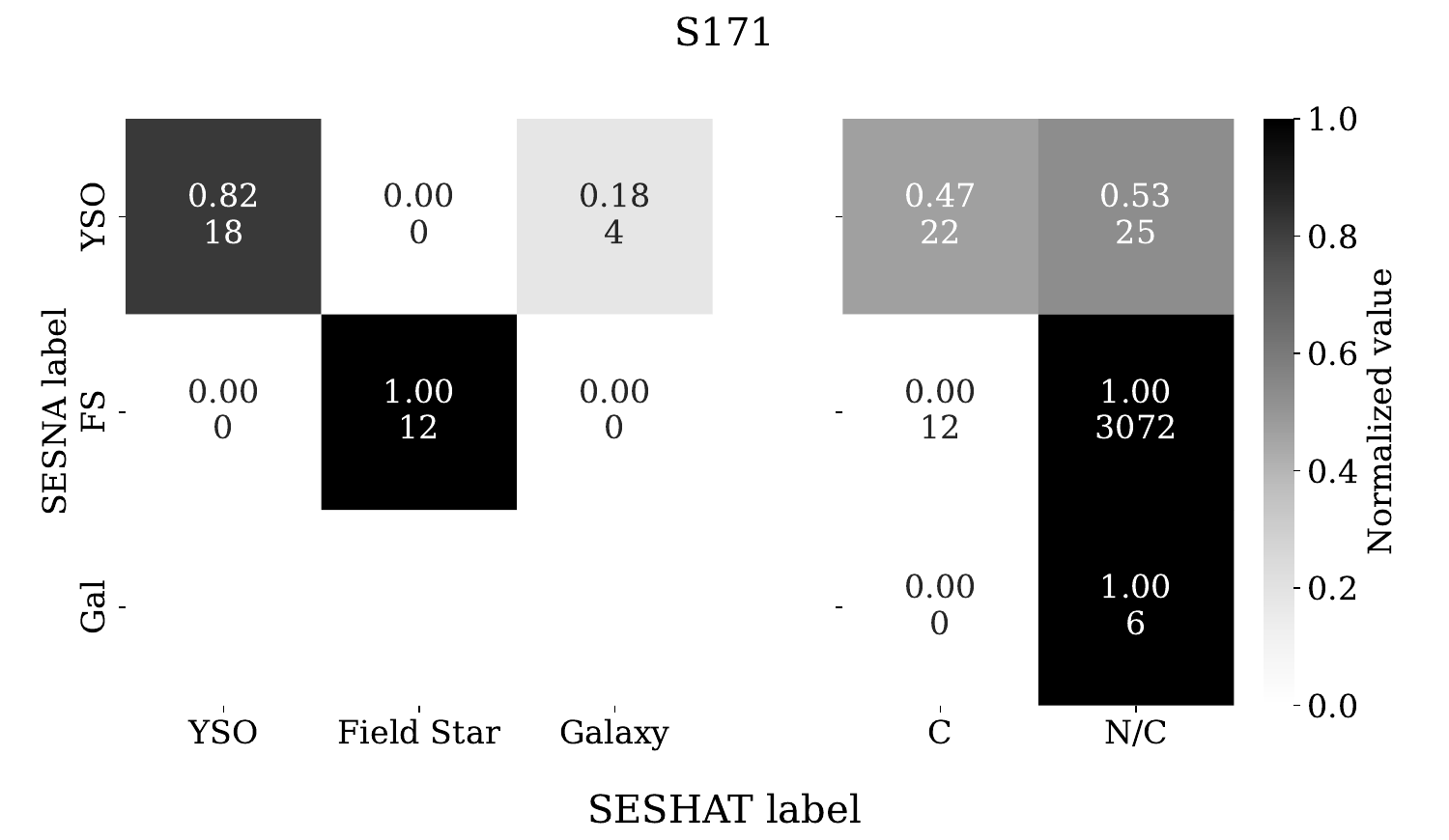}
\figsetplot{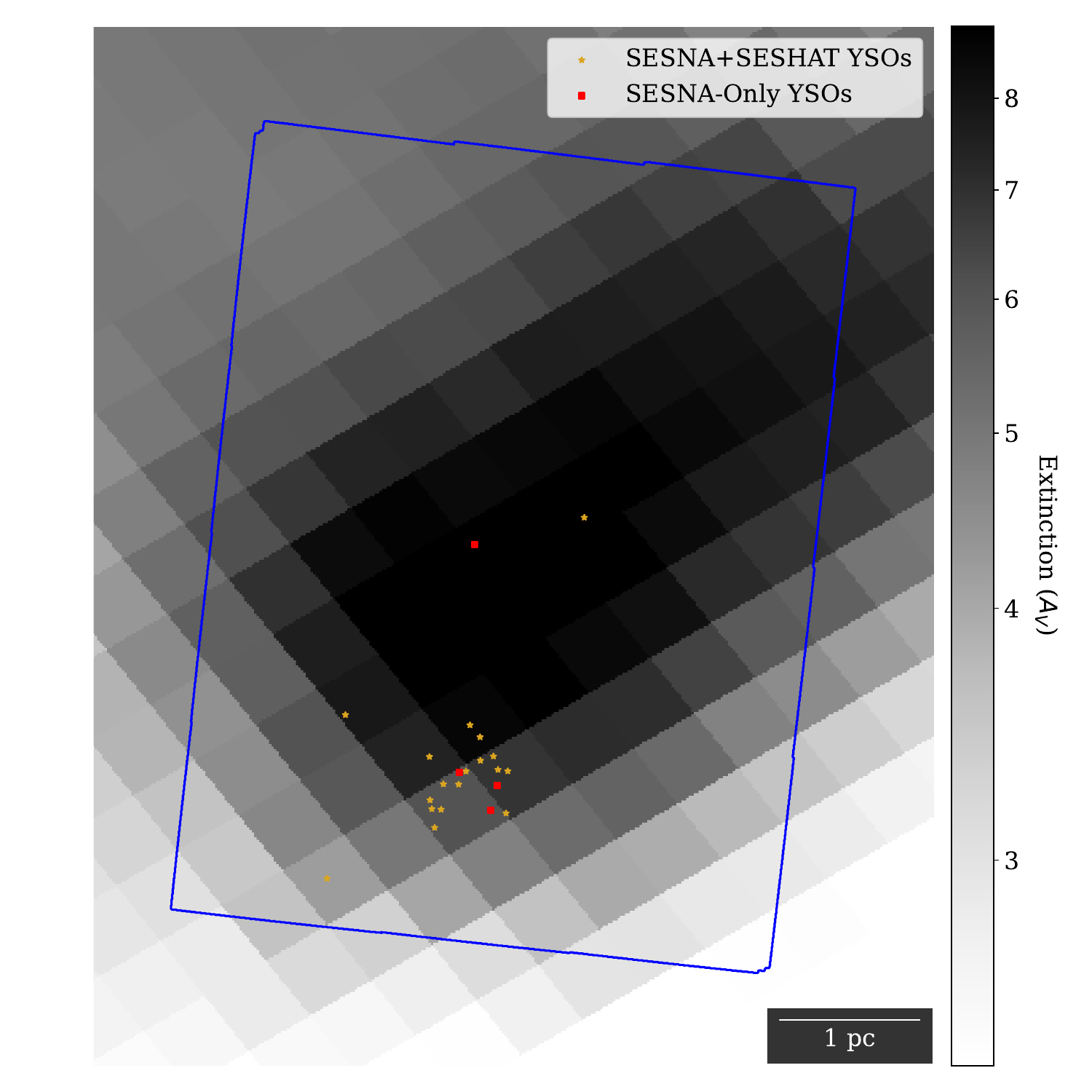}
\figsetgrpnote{The same as Figure~\ref{fig:afgl490}, but now for the S 171 region, which is part of the test set. \label{fig:s171}}
\figsetgrpend

\figsetgrpstart
\figsetgrpnum{15.27}
\figsetgrptitle{Scorpius}
\figsetplot{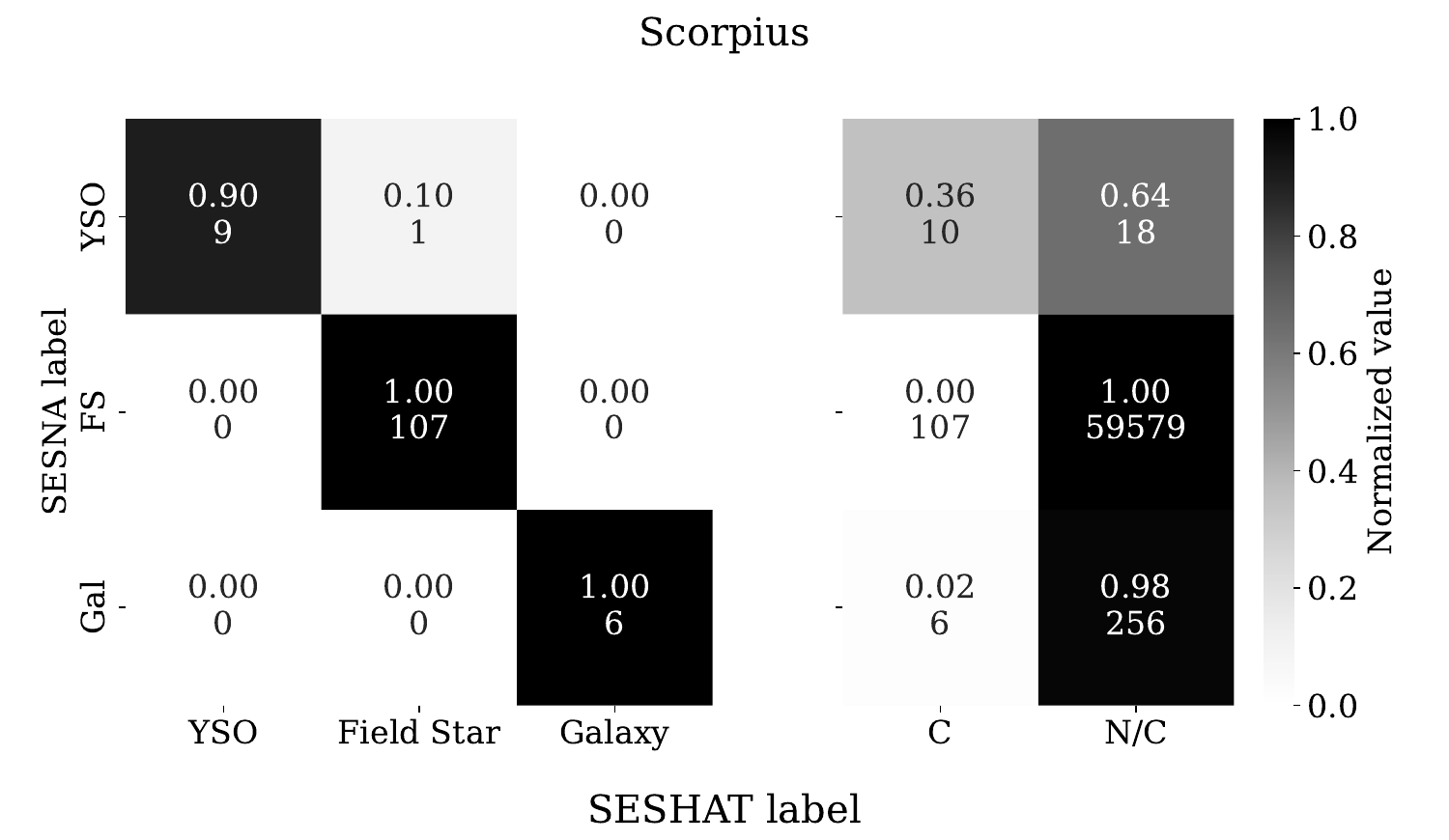}
\figsetplot{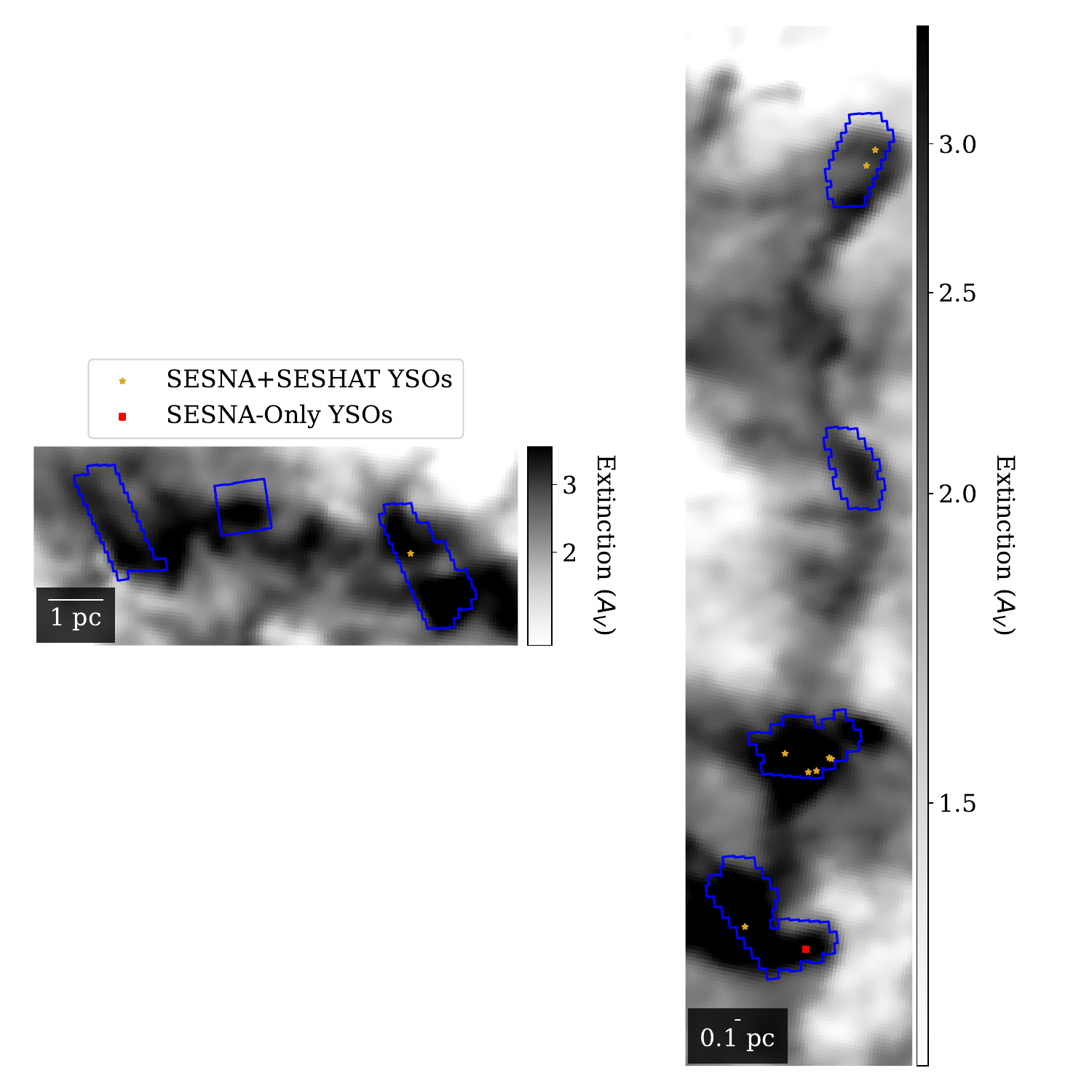}
\figsetgrpnote{The same as Figure~\ref{fig:afgl490}, but now for the Scorpius region, which is part of the test set. The Spitzer observations are at wide enough separation to require separating the fields into two subplots for better visualization. \label{fig:scorpius}}
\figsetgrpend

\figsetgrpstart
\figsetgrpnum{15.28}
\figsetgrptitle{Taurus}
\figsetplot{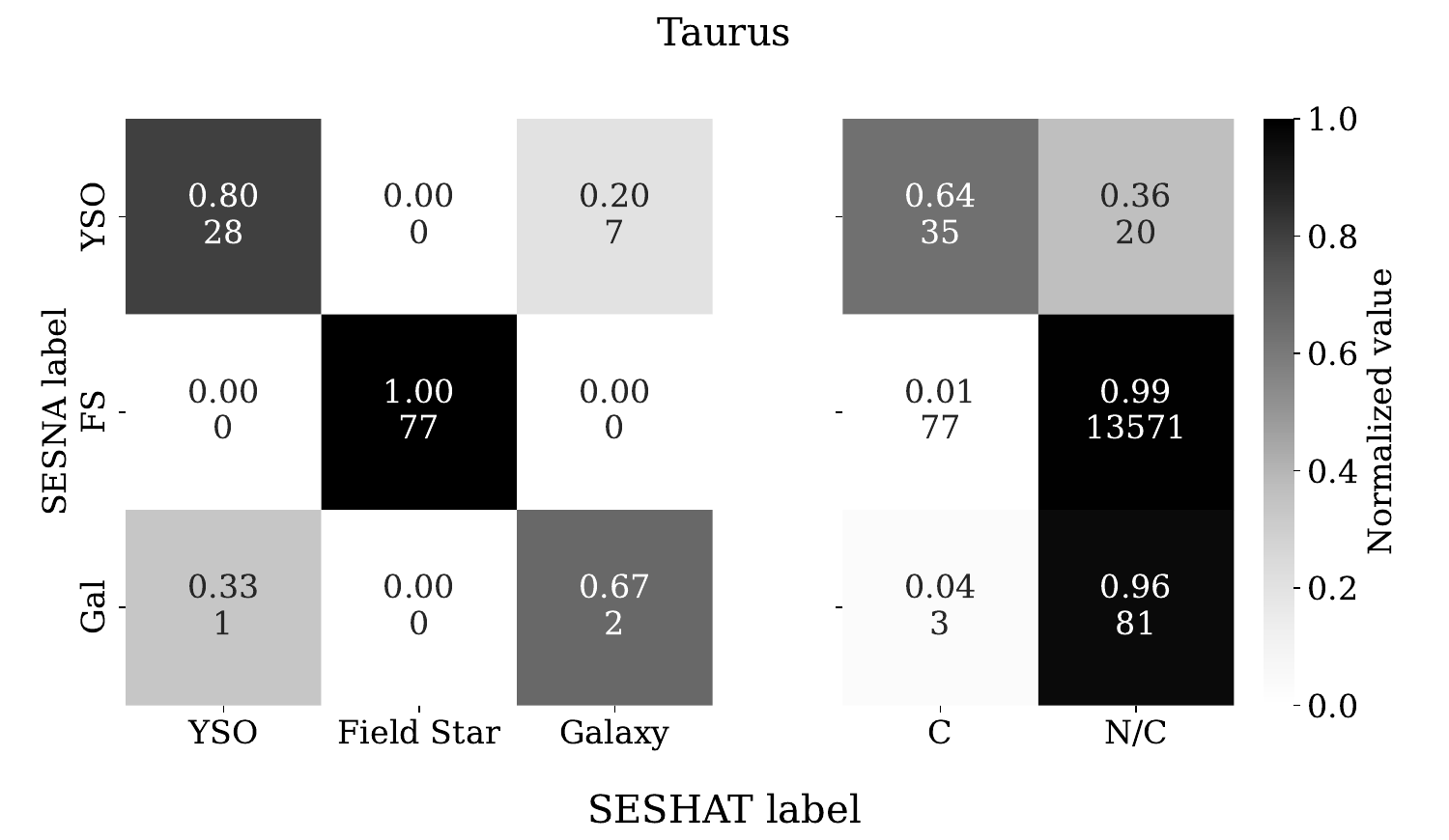}
\figsetplot{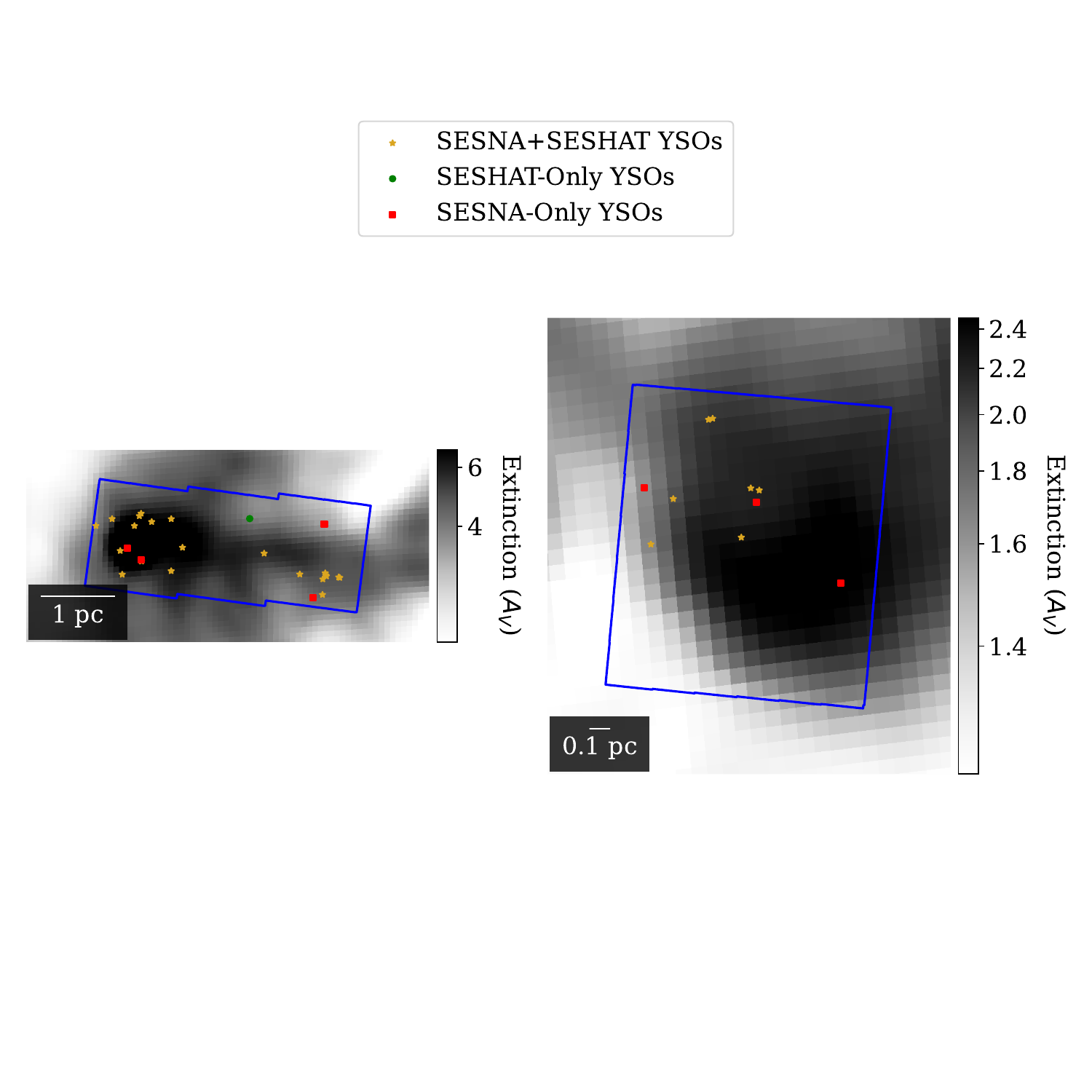}
\figsetgrpnote{The same as Figure~\ref{fig:afgl490}, but now for Taurus, which is part of the set used to define training YSOs. The Spitzer observations are at wide enough separation to require separating the fields into multiple subplots for better visualization. \label{fig:taurus}}
\figsetgrpend

\figsetgrpstart
\figsetgrpnum{15.29}
\figsetgrptitle{Vela D}
\figsetplot{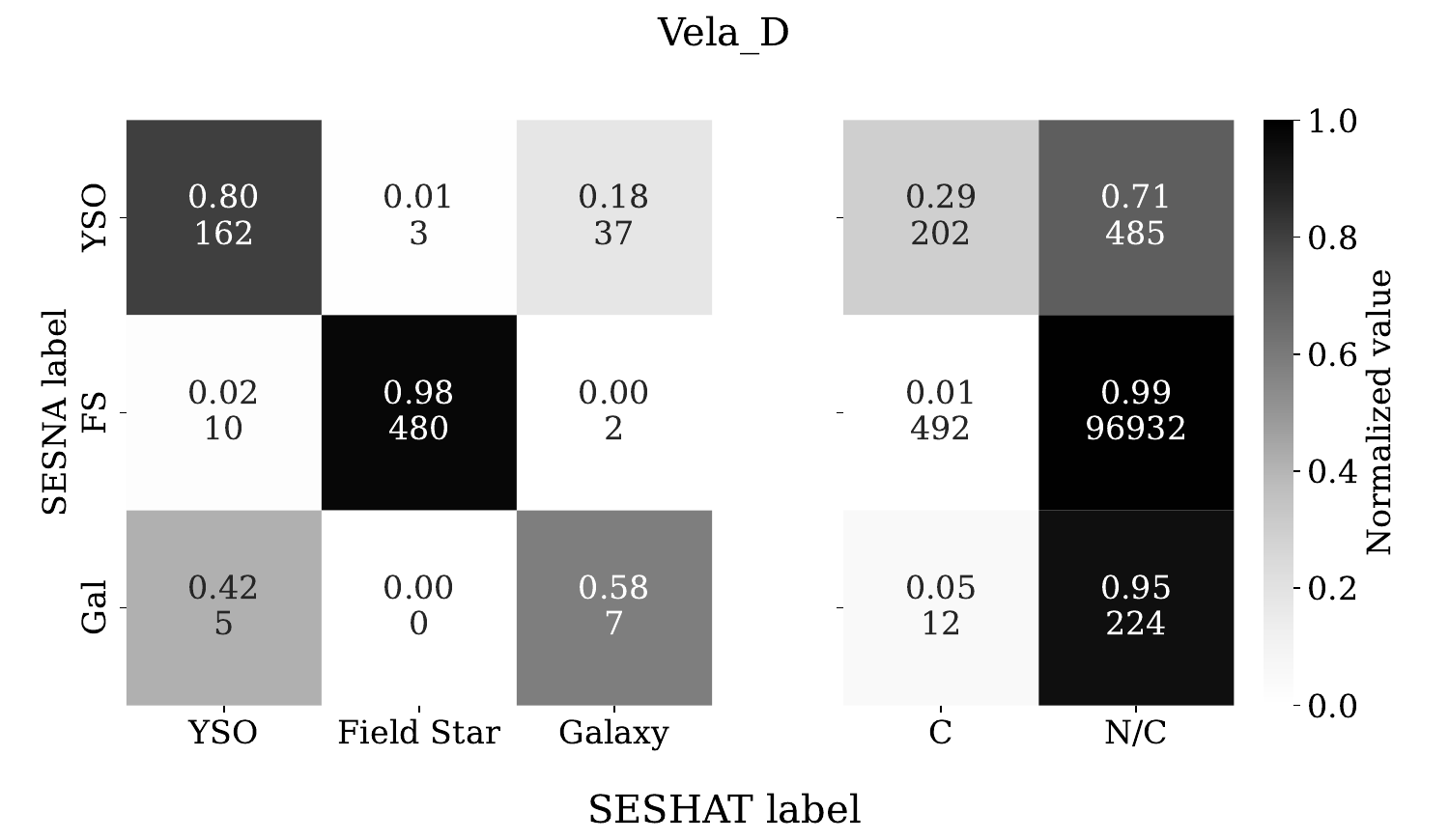}
\figsetplot{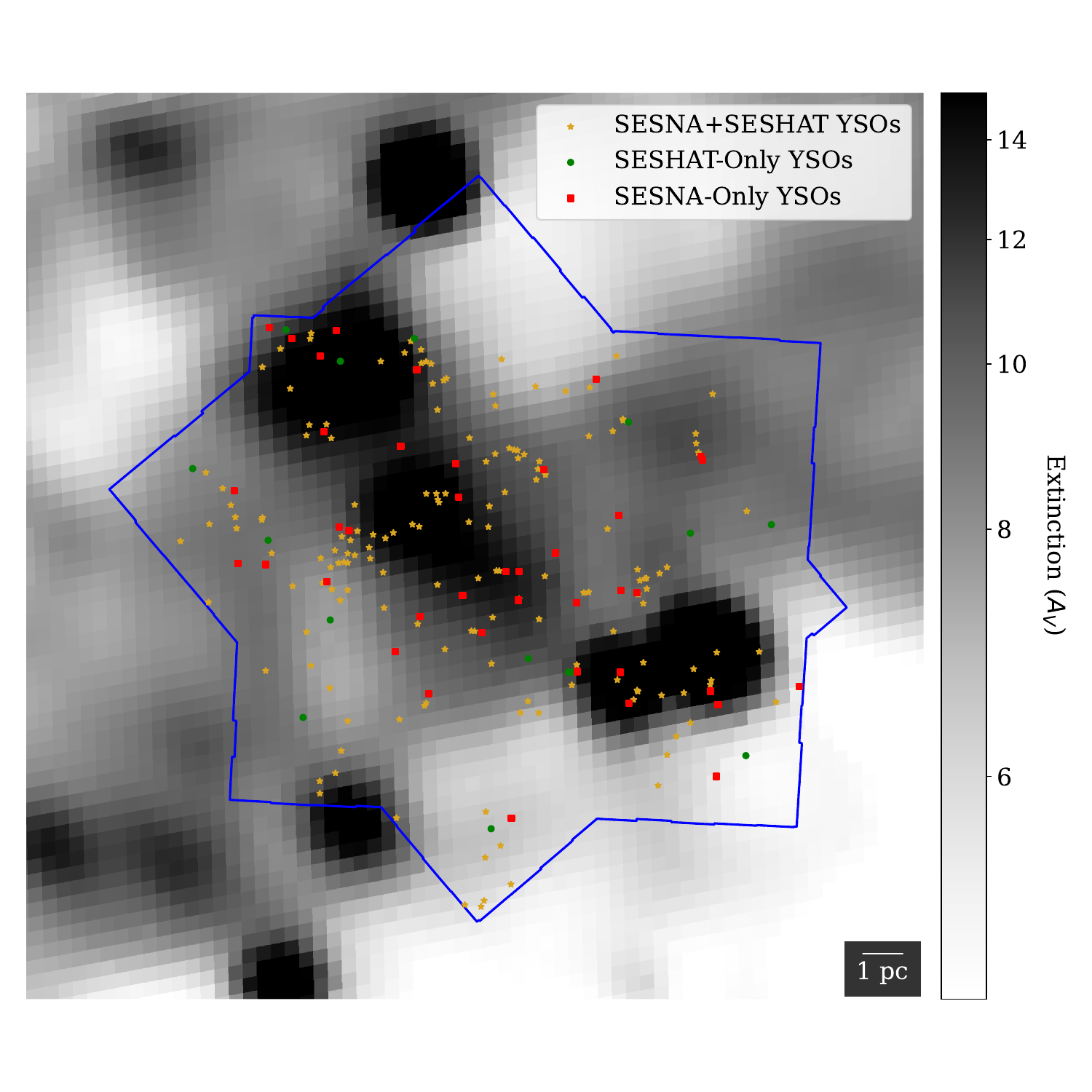}
\figsetgrpnote{The same as Figure~\ref{fig:afgl490}, but now for the Vela D region, which is part of the test set. \label{fig:velad}}
\figsetgrpend

\figsetend

%% file: main.bbl
\begin{thebibliography}{}
\expandafter\ifx\csname natexlab\endcsname\relax\def\natexlab#1{#1}\fi
\providecommand{\url}[1]{\href{#1}{#1}}
\providecommand{\dodoi}[1]{doi:~\href{http://doi.org/#1}{\nolinkurl{#1}}}
\providecommand{\doeprint}[1]{\href{http://ascl.net/#1}{\nolinkurl{http://ascl.net/#1}}}
\providecommand{\doarXiv}[1]{\href{https://arxiv.org/abs/#1}{\nolinkurl{https://arxiv.org/abs/#1}}}

\bibitem[{S. {Alberts} {et~al.}(2024){Alberts}, {Lyu}, {Shivaei}, {Rieke}, {Perez-Gonzalez}, {Bonventura}, {Zhu}, {Helton}, {Ji}, {Morrison}, {Robertson}, {Stone}, {Sun}, {Williams}, \& {Willmer}}]{SMILESAlberts2024}
{Alberts}, S., {Lyu}, J., {Shivaei}, I., {et~al.} 2024, \bibinfo{title}{{SMILES Initial Data Release: Unveiling the Obscured Universe with MIRI Multi-band Imaging},} arXiv e-prints, arXiv:2405.15972, \dodoi{10.48550/arXiv.2405.15972}

\bibitem[{S. {Blouin}(2024){Blouin}}]{Blouin2024}
{Blouin}, S. 2024, \bibinfo{title}{{White dwarf fundamentals},} arXiv e-prints, arXiv:2409.03941, \dodoi{10.48550/arXiv.2409.03941}

\bibitem[{S. {Blouin} {et~al.}(2018){Blouin}, {Dufour}, \& {Allard}}]{Blouin2018}
{Blouin}, S., {Dufour}, P., \& {Allard}, N.~F. 2018, \bibinfo{title}{{A New Generation of Cool White Dwarf Atmosphere Models. I. Theoretical Framework and Applications to DZ Stars},} \apj, 863, 184, \dodoi{10.3847/1538-4357/aad4a9}

\bibitem[{L. {Bonne} {et~al.}(2020){Bonne}, {Bontemps}, {Schneider}, {Clarke}, {Arzoumanian}, {Fukui}, {Tachihara}, {Csengeri}, {Guesten}, {Ohama}, {Okamoto}, {Simon}, {Yahia}, \& {Yamamoto}}]{Bonne2020}
{Bonne}, L., {Bontemps}, S., {Schneider}, N., {et~al.} 2020, \bibinfo{title}{{Formation of the Musca filament: evidence for asymmetries in the accretion flow due to a cloud-cloud collision},} \aap, 644, A27, \dodoi{10.1051/0004-6361/202038281}

\bibitem[{M. {Boquien} {et~al.}(2019){Boquien}, {Burgarella}, {Roehlly}, {Buat}, {Ciesla}, {Corre}, {Inoue}, \& {Salas}}]{Boquien2019}
{Boquien}, M., {Burgarella}, D., {Roehlly}, Y., {et~al.} 2019, \bibinfo{title}{{CIGALE: a python Code Investigating GALaxy Emission},} \aap, 622, A103, \dodoi{10.1051/0004-6361/201834156}

\bibitem[{A. {Bressan} {et~al.}(2012){Bressan}, {Marigo}, {Girardi}, {Salasnich}, {Dal Cero}, {Rubele}, \& {Nanni}}]{Bressan2012}
{Bressan}, A., {Marigo}, P., {Girardi}, L., {et~al.} 2012, \bibinfo{title}{{PARSEC: stellar tracks and isochrones with the PAdova and TRieste Stellar Evolution Code},} \mnras, 427, 127, \dodoi{10.1111/j.1365-2966.2012.21948.x}

\bibitem[{D. {Burgarella} {et~al.}(2005){Burgarella}, {Buat}, \& {Iglesias-P{\'a}ramo}}]{Burgarella2005}
{Burgarella}, D., {Buat}, V., \& {Iglesias-P{\'a}ramo}, J. 2005, \bibinfo{title}{{Star formation and dust attenuation properties in galaxies from a statistical ultraviolet-to-far-infrared analysis},} \mnras, 360, 1413, \dodoi{10.1111/j.1365-2966.2005.09131.x}

\bibitem[{D. {Burgarella} {et~al.}(2020){Burgarella}, {Nanni}, {Hirashita}, {Theul{\'e}}, {Inoue}, \& {Takeuchi}}]{Burgarella2020}
{Burgarella}, D., {Nanni}, A., {Hirashita}, H., {et~al.} 2020, \bibinfo{title}{{Observational and theoretical constraints on the formation and early evolution of the first dust grains in galaxies at 5 < z < 10},} \aap, 637, A32, \dodoi{10.1051/0004-6361/201937143}

\bibitem[{B. {Burkhart}(2018){Burkhart}}]{Burkhart2018}
{Burkhart}, B. 2018, \bibinfo{title}{{The Star Formation Rate in the Gravoturbulent Interstellar Medium},} \apj, 863, 118, \dodoi{10.3847/1538-4357/aad002}

\bibitem[{C.~M. {Casey} {et~al.}(2023){Casey}, {Kartaltepe}, {Drakos}, {Franco}, {Harish}, {Paquereau}, {Ilbert}, {Rose}, {Cox}, {Nightingale}, {Robertson}, {Silverman}, {Koekemoer}, {Massey}, {McCracken}, {Rhodes}, {Akins}, {Allen}, {Amvrosiadis}, {Arango-Toro}, {Bagley}, {Bongiorno}, {Capak}, {Champagne}, {Chartab}, {Ch{\'a}vez Ortiz}, {Chworowsky}, {Cooke}, {Cooper}, {Darvish}, {Ding}, {Faisst}, {Finkelstein}, {Fujimoto}, {Gentile}, {Gillman}, {Gould}, {Gozaliasl}, {Hayward}, {He}, {Hemmati}, {Hirschmann}, {Jahnke}, {Jin}, {Khostovan}, {Kokorev}, {Lambrides}, {Laigle}, {Larson}, {Leung}, {Liu}, {Liaudat}, {Long}, {Magdis}, {Mahler}, {Mainieri}, {Manning}, {Maraston}, {Martin}, {McCleary}, {McKinney}, {McPartland}, {Mobasher}, {Pattnaik}, {Renzini}, {Rich}, {Sanders}, {Sattari}, {Scognamiglio}, {Scoville}, {Sheth}, {Shuntov}, {Sparre}, {Suzuki}, {Talia}, {Toft}, {Trakhtenbrot}, {Urry}, {Valentino}, {Vanderhoof}, {Vardoulaki}, {Weaver}, {Whitaker}, {Wilkins}, {Yang}, \& {Zavala}}]{Cosmos2023}
{Casey}, C.~M., {Kartaltepe}, J.~S., {Drakos}, N.~E., {et~al.} 2023, \bibinfo{title}{{COSMOS-Web: An Overview of the JWST Cosmic Origins Survey},} \apj, 954, 31, \dodoi{10.3847/1538-4357/acc2bc}

\bibitem[{F. {Castelli} \& R.~L. {Kurucz}(2003){Castelli} \& {Kurucz}}]{Castelli2003}
{Castelli}, F., \& {Kurucz}, R.~L. 2003, in Modelling of Stellar Atmospheres, ed. N.~{Piskunov}, W.~W. {Weiss}, \& D.~F. {Gray}, Vol. 210, A20, \dodoi{10.48550/arXiv.astro-ph/0405087}

\bibitem[{N.~V. Chawla {et~al.}(2002)Chawla, Bowyer, Hall, \& Kegelmeyer}]{SMOTE2002}
Chawla, N.~V., Bowyer, K.~W., Hall, L.~O., \& Kegelmeyer, W.~P. 2002, \bibinfo{title}{SMOTE: synthetic minority over-sampling technique,} Journal of artificial intelligence research, 16, 321

\bibitem[{A.~Y.~A. {Chen} {et~al.}(2025){Chen}, {Goto}, {Wu}, {Ling}, {Kim}, {Ho}, {Kilerci}, {Uno}, {Phan}, {Lin}, {Yang}, \& {Hashimoto}}]{Chenbd2025}
{Chen}, A. Y.~A., {Goto}, T., {Wu}, C. K.~W., {et~al.} 2025, \bibinfo{title}{{Brown dwarf number density in the JWST COSMOS-Web field},} \pasa, 42, e042, \dodoi{10.1017/pasa.2025.25}

\bibitem[{T. {Chen} \& C. {Guestrin}(2016){Chen} \& {Guestrin}}]{xgboost}
{Chen}, T., \& {Guestrin}, C. 2016, \bibinfo{title}{{XGBoost: A Scalable Tree Boosting System},} arXiv e-prints, arXiv:1603.02754, \dodoi{10.48550/arXiv.1603.02754}

\bibitem[{Y.-K. {Chiang}(2023){Chiang}}]{Chiang2023}
{Chiang}, Y.-K. 2023, \bibinfo{title}{{Corrected SFD: A More Accurate Galactic Dust Map with Minimal Extragalactic Contamination},} \apj, 958, 118, \dodoi{10.3847/1538-4357/acf4a1}

\bibitem[{D. {Cornu} \& J. {Montillaud}(2021){Cornu} \& {Montillaud}}]{Cornu2021}
{Cornu}, D., \& {Montillaud}, J. 2021, \bibinfo{title}{{A neural network-based methodology to select young stellar object candidates from IR surveys},} \aap, 647, A116, \dodoi{10.1051/0004-6361/202038516}

\bibitem[{B.~L. {Crompvoets} {et~al.}(2024){Crompvoets}, {Di Francesco}, {Teimoorinia}, \& {Preibisch}}]{Crompvoets2024}
{Crompvoets}, B.~L., {Di Francesco}, J., {Teimoorinia}, H., \& {Preibisch}, T. 2024, \bibinfo{title}{{Climbing the Cliffs: Classifying Young Stellar Objects in the Cosmic Cliffs JWST Data Using a Probabilistic Random Forest},} \aj, 168, 63, \dodoi{10.3847/1538-3881/ad51fc}

\bibitem[{M.~M. {Dunham} {et~al.}(2015){Dunham}, {Allen}, {Evans}, {Broekhoven-Fiene}, {Cieza}, {Di Francesco}, {Gutermuth}, {Harvey}, {Hatchell}, {Heiderman}, {Huard}, {Johnstone}, {Kirk}, {Matthews}, {Miller}, {Peterson}, \& {Young}}]{Dunham2015}
{Dunham}, M.~M., {Allen}, L.~E., {Evans}, Neal~J., I., {et~al.} 2015, \bibinfo{title}{{Young Stellar Objects in the Gould Belt},} \apjs, 220, 11, \dodoi{10.1088/0067-0049/220/1/11}

\bibitem[{S. {Foschino} {et~al.}(2019){Foschino}, {Bern{\'e}}, \& {Joblin}}]{Foschino2019}
{Foschino}, S., {Bern{\'e}}, O., \& {Joblin}, C. 2019, \bibinfo{title}{{Learning mid-IR emission spectra of polycyclic aromatic hydrocarbon populations from observations},} \aap, 632, A84, \dodoi{10.1051/0004-6361/201935085}

\bibitem[{G. {Green}(2018){Green}}]{Green2018DustMaps}
{Green}, G. 2018, \bibinfo{title}{{dustmaps: A Python interface for maps of interstellar dust},} The Journal of Open Source Software, 3, 695, \dodoi{10.21105/joss.00695}

\bibitem[{J.~D. {Green} {et~al.}(2024){Green}, {Pontoppidan}, {Reiter}, {Watson}, {Shenoy}, {Manoj}, \& {Narang}}]{Green2024}
{Green}, J.~D., {Pontoppidan}, K.~M., {Reiter}, M., {et~al.} 2024, \bibinfo{title}{{Why Are (Almost) All the Protostellar Outflows Aligned in Serpens Main?},} \apj, 972, 5, \dodoi{10.3847/1538-4357/ad5a02}

\bibitem[{A. {Gupta} \& W.-P. {Chen}(2022){Gupta} \& {Chen}}]{Gupta2022}
{Gupta}, A., \& {Chen}, W.-P. 2022, \bibinfo{title}{{Interplay between Young Stars and Molecular Clouds in the Ophiuchus Star-forming Complex},} \aj, 163, 233, \dodoi{10.3847/1538-3881/ac5cc8}

\bibitem[{R.~A. {Gutermuth} {et~al.}(2009){Gutermuth}, {Megeath}, {Myers}, {Allen}, {Pipher}, \& {Fazio}}]{Gutermuth2009}
{Gutermuth}, R.~A., {Megeath}, S.~T., {Myers}, P.~C., {et~al.} 2009, \bibinfo{title}{{A Spitzer Survey of Young Stellar Clusters Within One Kiloparsec of the Sun: Cluster Core Extraction and Basic Structural Analysis},} \apjs, 184, 18, \dodoi{10.1088/0067-0049/184/1/18}

\bibitem[{R.~A. {Gutermuth} {et~al.}(2008){Gutermuth}, {Myers}, {Megeath}, {Allen}, {Pipher}, {Muzerolle}, {Porras}, {Winston}, \& {Fazio}}]{Gutermuth2008}
{Gutermuth}, R.~A., {Myers}, P.~C., {Megeath}, S.~T., {et~al.} 2008, \bibinfo{title}{{Spitzer Observations of NGC 1333: A Study of Structure and Evolution in a Nearby Embedded Cluster},} \apj, 674, 336, \dodoi{10.1086/524722}

\bibitem[{T. {Hirota} {et~al.}(2008){Hirota}, {Ando}, {Bushimata}, {Choi}, {Honma}, {Imai}, {Iwadate}, {Jike}, {Kameno}, {Kameya}, {Kamohara}, {Kan-Ya}, {Kawaguchi}, {Kijima}, {Kim}, {Kobayashi}, {Kuji}, {Kurayama}, {Manabe}, {Matsui}, {Matsumoto}, {Miyaji}, {Miyazaki}, {Nagayama}, {Nakagawa}, {Namikawa}, {Nyu}, {Oh}, {Omodaka}, {Oyama}, {Sakai}, {Sasao}, {Sato}, {Sato}, {Shibata}, {Tamura}, {Ueda}, \& {Yamashita}}]{Hirota2008}
{Hirota}, T., {Ando}, K., {Bushimata}, T., {et~al.} 2008, \bibinfo{title}{{Astrometry of H$_{2}$O Masers in Nearby Star-Forming Regions with VERA III. IRAS 22198+6336 in Lynds1204G},} \pasj, 60, 961, \dodoi{10.1093/pasj/60.5.961}

\bibitem[{J. {Kainulainen} {et~al.}(2009){Kainulainen}, {Beuther}, {Henning}, \& {Plume}}]{Kainulainen2009}
{Kainulainen}, J., {Beuther}, H., {Henning}, T., \& {Plume}, R. 2009, \bibinfo{title}{{Probing the evolution of molecular cloud structure. From quiescence to birth},} \aap, 508, L35, \dodoi{10.1051/0004-6361/200913605}

\bibitem[{M. {Kilic} {et~al.}(2006){Kilic}, {von Hippel}, {Mullally}, {Reach}, {Kuchner}, {Winget}, \& {Burrows}}]{Kilic2006}
{Kilic}, M., {von Hippel}, T., {Mullally}, F., {et~al.} 2006, \bibinfo{title}{{The Mystery Deepens: Spitzer Observations of Cool White Dwarfs},} \apj, 642, 1051, \dodoi{10.1086/501042}

\bibitem[{P. {Kroupa}(2001){Kroupa}}]{Kroupa2001}
{Kroupa}, P. 2001, \bibinfo{title}{{On the variation of the initial mass function},} \mnras, 322, 231, \dodoi{10.1046/j.1365-8711.2001.04022.x}

\bibitem[{M.~A. {Kuhn} {et~al.}(2021){Kuhn}, {de Souza}, {Krone-Martins}, {Castro-Ginard}, {Ishida}, {Povich}, {Hillenbrand}, \& {COIN Collaboration}}]{Kuhn2021}
{Kuhn}, M.~A., {de Souza}, R.~S., {Krone-Martins}, A., {et~al.} 2021, \bibinfo{title}{{SPICY: The Spitzer/IRAC Candidate YSO Catalog for the Inner Galactic Midplane},} \apjs, 254, 33, \dodoi{10.3847/1538-4365/abe465}

\bibitem[{C.~J. {Lada} {et~al.}(2009){Lada}, {Lombardi}, \& {Alves}}]{Lada2009}
{Lada}, C.~J., {Lombardi}, M., \& {Alves}, J.~F. 2009, \bibinfo{title}{{The California Molecular Cloud},} \apj, 703, 52, \dodoi{10.1088/0004-637X/703/1/52}

\bibitem[{C.~J. {Lada} {et~al.}(2010){Lada}, {Lombardi}, \& {Alves}}]{Lada2010}
{Lada}, C.~J., {Lombardi}, M., \& {Alves}, J.~F. 2010, \bibinfo{title}{{On the Star Formation Rates in Molecular Clouds},} \apj, 724, 687, \dodoi{10.1088/0004-637X/724/1/687}

\bibitem[{M.~A. {Limbach} {et~al.}(2022){Limbach}, {Vanderburg}, {Stevenson}, {Blouin}, {Morley}, {Lustig-Yaeger}, {Soares-Furtado}, \& {Janson}}]{Limbach2022}
{Limbach}, M.~A., {Vanderburg}, A., {Stevenson}, K.~B., {et~al.} 2022, \bibinfo{title}{{A new method for finding nearby white dwarfs exoplanets and detecting biosignatures},} \mnras, 517, 2622, \dodoi{10.1093/mnras/stac2823}

\bibitem[{R. {Liseau} {et~al.}(1992){Liseau}, {Lorenzetti}, {Nisini}, {Spinoglio}, \& {Moneti}}]{Liseau1992}
{Liseau}, R., {Lorenzetti}, D., {Nisini}, B., {Spinoglio}, L., \& {Moneti}, A. 1992, \bibinfo{title}{{Star formation in the VELA molecular clouds. I. The IRAS-bright class I sources.},} \aap, 265, 577

\bibitem[{P.~W. {Lucas} {et~al.}(2008){Lucas}, {Hoare}, {Longmore}, {Schr{\"o}der}, {Davis}, {Adamson}, {Bandyopadhyay}, {de Grijs}, {Smith}, {Gosling}, {Mitchison}, {G{\'a}sp{\'a}r}, {Coe}, {Tamura}, {Parker}, {Irwin}, {Hambly}, {Bryant}, {Collins}, {Cross}, {Evans}, {Gonzalez-Solares}, {Hodgkin}, {Lewis}, {Read}, {Riello}, {Sutorius}, {Lawrence}, {Drew}, {Dye}, \& {Thompson}}]{Lucas2008}
{Lucas}, P.~W., {Hoare}, M.~G., {Longmore}, A., {et~al.} 2008, \bibinfo{title}{{The UKIDSS Galactic Plane Survey},} \mnras, 391, 136, \dodoi{10.1111/j.1365-2966.2008.13924.x}

\bibitem[{G. {Marton} {et~al.}(2016){Marton}, {T{\'o}th}, {Paladini}, {Kun}, {Zahorecz}, {McGehee}, \& {Kiss}}]{Marton2016}
{Marton}, G., {T{\'o}th}, L.~V., {Paladini}, R., {et~al.} 2016, \bibinfo{title}{{An all-sky support vector machine selection of WISE YSO candidates},} \mnras, 458, 3479, \dodoi{10.1093/mnras/stw398}

\bibitem[{M.~A. {Moreno-Corral} {et~al.}(1993){Moreno-Corral}, {C.}, {de La Ra}, \& {Wagner}}]{Moreno-Corral1993}
{Moreno-Corral}, M.~A., {C.}, C.-K., {de La Ra}, E., \& {Wagner}, S. 1993, \bibinfo{title}{{H-alpha interferometric, optical and near IR photometric studies of star forming regions. I. The Cepheus B/Sh2-155/Cepheus OB3 association complex.},} \aap, 273, 619

\bibitem[{S. {Noll} {et~al.}(2009){Noll}, {Burgarella}, {Giovannoli}, {Buat}, {Marcillac}, \& {Mu{\~n}oz-Mateos}}]{Noll2009}
{Noll}, S., {Burgarella}, D., {Giovannoli}, E., {et~al.} 2009, \bibinfo{title}{{Analysis of galaxy spectral energy distributions from far-UV to far-IR with CIGALE: studying a SINGS test sample},} \aap, 507, 1793, \dodoi{10.1051/0004-6361/200912497}

\bibitem[{G.~N. {Ortiz-Le{\'o}n} {et~al.}(2018){Ortiz-Le{\'o}n}, {Loinard}, {Dzib}, {Kounkel}, {Galli}, {Tobin}, {Evans}, {Hartmann}, {Rodr{\'\i}guez}, {Brice{\~n}o}, {Torres}, \& {Mioduszewski}}]{Ortiz-Leon2018}
{Ortiz-Le{\'o}n}, G.~N., {Loinard}, L., {Dzib}, S.~A., {et~al.} 2018, \bibinfo{title}{{Gaia-DR2 Confirms VLBA Parallaxes in Ophiuchus, Serpens, and Aquila},} \apjl, 869, L33, \dodoi{10.3847/2041-8213/aaf6ad}

\bibitem[{V. {Ossenkopf} \& T. {Henning}(1994){Ossenkopf} \& {Henning}}]{Ossenkopf1994}
{Ossenkopf}, V., \& {Henning}, T. 1994, \bibinfo{title}{{Dust opacities for protostellar cores.},} \aap, 291, 943

\bibitem[{A.~K. {Pandey} {et~al.}(2008){Pandey}, {Sharma}, {Ogura}, {Ojha}, {Chen}, {Bhatt}, \& {Ghosh}}]{Pandey2008}
{Pandey}, A.~K., {Sharma}, S., {Ogura}, K., {et~al.} 2008, \bibinfo{title}{{Stellar contents and star formation in the young star cluster Be 59},} \mnras, 383, 1241, \dodoi{10.1111/j.1365-2966.2007.12641.x}

\bibitem[{E. {Peeters} {et~al.}(2002){Peeters}, {Hony}, {Van Kerckhoven}, {Tielens}, {Allamandola}, {Hudgins}, \& {Bauschlicher}}]{Peeters2002}
{Peeters}, E., {Hony}, S., {Van Kerckhoven}, C., {et~al.} 2002, \bibinfo{title}{{The rich 6 to 9 vec mu m spectrum of interstellar PAHs},} \aap, 390, 1089, \dodoi{10.1051/0004-6361:20020773}

\bibitem[{M.~E. {Pelayo-Bald{\'a}rrago} {et~al.}(2023){Pelayo-Bald{\'a}rrago}, {Sicilia-Aguilar}, {Fang}, {Roccatagliata}, {Kim}, \& {Garc{\'\i}a-{\'A}lvarez}}]{Pelayo-Balarrago2023}
{Pelayo-Bald{\'a}rrago}, M.~E., {Sicilia-Aguilar}, A., {Fang}, M., {et~al.} 2023, \bibinfo{title}{{Star formation in IC1396: Kinematics and subcluster structure revealed by Gaia},} \aap, 669, A22, \dodoi{10.1051/0004-6361/202244265}

\bibitem[{M.~W. {Phillips} {et~al.}(2020){Phillips}, {Tremblin}, {Baraffe}, {Chabrier}, {Allard}, {Spiegelman}, {Goyal}, {Drummond}, \& {H{\'e}brard}}]{Phillips2020}
{Phillips}, M.~W., {Tremblin}, P., {Baraffe}, I., {et~al.} 2020, \bibinfo{title}{{A new set of atmosphere and evolution models for cool T-Y brown dwarfs and giant exoplanets},} \aap, 637, A38, \dodoi{10.1051/0004-6361/201937381}

\bibitem[{P. {Pilleri} {et~al.}(2012){Pilleri}, {Montillaud}, {Bern{\'e}}, \& {Joblin}}]{Pilleri2012}
{Pilleri}, P., {Montillaud}, J., {Bern{\'e}}, O., \& {Joblin}, C. 2012, \bibinfo{title}{{Evaporating very small grains as tracers of the UV radiation field in photo-dissociation regions},} \aap, 542, A69, \dodoi{10.1051/0004-6361/201015915}

\bibitem[{R. {Pokhrel} {et~al.}(2020){Pokhrel}, {Gutermuth}, {Betti}, {Offner}, {Myers}, {Megeath}, {Sokol}, {Ali}, {Allen}, {Allen}, {Dunham}, {Fischer}, {Henning}, {Heyer}, {Hora}, {Pipher}, {Tobin}, \& {Wolk}}]{Pokhrel2020}
{Pokhrel}, R., {Gutermuth}, R.~A., {Betti}, S.~K., {et~al.} 2020, \bibinfo{title}{{Star-Gas Surface Density Correlations in 12 Nearby Molecular Clouds. I. Data Collection and Star-sampled Analysis},} \apj, 896, 60, \dodoi{10.3847/1538-4357/ab92a2}

\bibitem[{K.~M. {Pontoppidan} {et~al.}(2022){Pontoppidan}, {Barrientes}, {Blome}, {Braun}, {Brown}, {Carruthers}, {Coe}, {DePasquale}, {Espinoza}, {Marin}, {Gordon}, {Henry}, {Hustak}, {James}, {Jenkins}, {Koekemoer}, {LaMassa}, {Law}, {Lockwood}, {Moro-Martin}, {Mullally}, {Pagan}, {Player}, {Proffitt}, {Pulliam}, {Ramsay}, {Ravindranath}, {Reid}, {Robberto}, {Sabbi}, {Ubeda}, {Balogh}, {Flanagan}, {Gardner}, {Hasan}, {Meinke}, \& {Nota}}]{Pontoppidan2022}
{Pontoppidan}, K.~M., {Barrientes}, J., {Blome}, C., {et~al.} 2022, \bibinfo{title}{{The JWST Early Release Observations},} \apjl, 936, L14, \dodoi{10.17909/67ft-nb86}

\bibitem[{T. {Richardson} {et~al.}(2024){Richardson}, {Ginsburg}, {Indebetouw}, \& {Robitaille}}]{Richardson2024}
{Richardson}, T., {Ginsburg}, A., {Indebetouw}, R., \& {Robitaille}, T.~P. 2024, \bibinfo{title}{{An Updated Modular Set of Synthetic Spectral Energy Distributions for Young Stellar Objects},} \apj, 961, 188, \dodoi{10.3847/1538-4357/ad072d}

\bibitem[{G. {Rieke} {et~al.}(2024){Rieke}, {Alberts}, {Shivaei}, {Lyu}, {Willmer}, {Perez-Gonzalez}, \& {Williams}}]{SMILESRieke2024}
{Rieke}, G., {Alberts}, S., {Shivaei}, I., {et~al.} 2024, \bibinfo{title}{{The SMILES Mid-Infrared Survey},} arXiv e-prints, arXiv:2406.03518, \dodoi{10.48550/arXiv.2406.03518}

\bibitem[{G.~H. {Rieke} \& M.~J. {Lebofsky}(1985){Rieke} \& {Lebofsky}}]{Rieke1985}
{Rieke}, G.~H., \& {Lebofsky}, M.~J. 1985, \bibinfo{title}{{The interstellar extinction law from 1 to 13 microns.},} \apj, 288, 618, \dodoi{10.1086/162827}

\bibitem[{T.~P. {Robitaille}(2017){Robitaille}}]{Robitaille2017}
{Robitaille}, T.~P. 2017, \bibinfo{title}{{A modular set of synthetic spectral energy distributions for young stellar objects},} \aap, 600, A11, \dodoi{10.1051/0004-6361/201425486}

\bibitem[{T.~P. {Robitaille} {et~al.}(2007){Robitaille}, {Whitney}, {Indebetouw}, \& {Wood}}]{Robitaille2007}
{Robitaille}, T.~P., {Whitney}, B.~A., {Indebetouw}, R., \& {Wood}, K. 2007, \bibinfo{title}{{Interpreting Spectral Energy Distributions from Young Stellar Objects. II. Fitting Observed SEDs Using a Large Grid of Precomputed Models},} \apjs, 169, 328, \dodoi{10.1086/512039}

\bibitem[{T.~P. {Robitaille} {et~al.}(2006){Robitaille}, {Whitney}, {Indebetouw}, {Wood}, \& {Denzmore}}]{Robitaille2006}
{Robitaille}, T.~P., {Whitney}, B.~A., {Indebetouw}, R., {Wood}, K., \& {Denzmore}, P. 2006, \bibinfo{title}{{Interpreting Spectral Energy Distributions from Young Stellar Objects. I. A Grid of 200,000 YSO Model SEDs},} \apjs, 167, 256, \dodoi{10.1086/508424}

\bibitem[{M. {Shuntov} {et~al.}(2025){Shuntov}, {Akins}, {Paquereau}, {Casey}, {Ilbert}, {Arango-Toro}, {McCracken}, {Franco}, {Harish}, {Kartaltepe}, {Koekemoer}, {Yang}, {Huertas-Company}, {Berman}, {McCleary}, {Toft}, {Gavazzi}, {Achenbach}, {Bertin}, {Brinch}, {Champagne}, {Chartab}, {Drakos}, {Egami}, {Endsley}, {Faisst}, {Fan}, {Flayhart}, {Hartley}, {Hatamnia}, {Gozaliasl}, {Gentile}, {Jermann}, {Jin}, {Kakiichi}, {Khostovan}, {K{\"u}mmel}, {Laigle}, {Laishram}, {Lambrides}, {Liu}, {Lyu}, {Magdis}, {Mobasher}, {Moutard}, {Renzini}, {Robertson}, {Schefer}, {Scognamiglio}, {Scoville}, {Sattari}, {Sanders}, {Taamoli}, {Trakhtenbrot}, {Valentino}, {Wang}, {Weaver}, \& {Yang}}]{cosmoscat2025}
{Shuntov}, M., {Akins}, H.~B., {Paquereau}, L., {et~al.} 2025, \bibinfo{title}{{COSMOS2025: The COSMOS-Web galaxy catalog of photometry, morphology, redshifts, and physical parameters from JWST, HST, and ground-based imaging},} arXiv e-prints, arXiv:2506.03243, \dodoi{10.48550/arXiv.2506.03243}

\bibitem[{R.~L. {Snell} {et~al.}(1984){Snell}, {Scoville}, {Sanders}, \& {Erickson}}]{Snell1984}
{Snell}, R.~L., {Scoville}, N.~Z., {Sanders}, D.~B., \& {Erickson}, N.~R. 1984, \bibinfo{title}{{High-velocity molecular jets.},} \apj, 284, 176, \dodoi{10.1086/162397}

\bibitem[{V. {Strai{\v{z}}ys} {et~al.}(2014){Strai{\v{z}}ys}, {Maskoli{\={u}}nas}, {Boyle}, {Prada Moroni}, {Tognelli}, {Zdanavi{\v{c}}ius}, {Zdanavi{\v{c}}ius}, {Laugalys}, \& {Kazlauskas}}]{Straizys2014}
{Strai{\v{z}}ys}, V., {Maskoli{\={u}}nas}, M., {Boyle}, R.~P., {et~al.} 2014, \bibinfo{title}{{The distance to the young cluster NGC 7129 and its age},} \mnras, 438, 1848, \dodoi{10.1093/mnras/stt2334}

\bibitem[{M. {Szil{\'a}gyi} {et~al.}(2021){Szil{\'a}gyi}, {Kun}, \& {{\'A}brah{\'a}m}}]{Szilagyi2021}
{Szil{\'a}gyi}, M., {Kun}, M., \& {{\'A}brah{\'a}m}, P. 2021, \bibinfo{title}{{The Gaia view of the Cepheus flare},} \mnras, 505, 5164, \dodoi{10.1093/mnras/stab1496}

\bibitem[{A.~G.~G.~M. {Tielens}(2008){Tielens}}]{Tielens2008}
{Tielens}, A.~G.~G.~M. 2008, \bibinfo{title}{{Interstellar polycyclic aromatic hydrocarbon molecules.},} \araa, 46, 289, \dodoi{10.1146/annurev.astro.46.060407.145211}

\bibitem[{P.-E. {Tremblay} {et~al.}(2024){Tremblay}, {B{\'e}dard}, {O'Brien}, {Munday}, {Elms}, {Gentillo Fusillo}, \& {Sahu}}]{Tremblay2024}
{Tremblay}, P.-E., {B{\'e}dard}, A., {O'Brien}, M.~W., {et~al.} 2024, \bibinfo{title}{{The Gaia white dwarf revolution},} \nar, 99, 101705, \dodoi{10.1016/j.newar.2024.101705}

\bibitem[{S. {Wang} \& X. {Chen}(2019){Wang} \& {Chen}}]{Wang&Chen2019}
{Wang}, S., \& {Chen}, X. 2019, \bibinfo{title}{{The Optical to Mid-infrared Extinction Law Based on the APOGEE, Gaia DR2, Pan-STARRS1, SDSS, APASS, 2MASS, and WISE Surveys},} \apj, 877, 116, \dodoi{10.3847/1538-4357/ab1c61}

\bibitem[{R. {Wen} {et~al.}(2022){Wen}, {An}, {Zheng}, {Shi}, {Qin}, {Gonzalez}, {Bian}, {Xu}, {Pan}, {Tan}, {Liu}, {Fang}, {Ren}, {Zhang}, {Qiao}, \& {Liu}}]{Wen2022}
{Wen}, R., {An}, F., {Zheng}, X.~Z., {et~al.} 2022, \bibinfo{title}{{The Physical Properties of Star-forming Galaxies with Strong [O III] Lines at z = 3.25},} \apj, 933, 50, \dodoi{10.3847/1538-4357/ac7392}

\bibitem[{A. {Whitworth}(2018){Whitworth}}]{Whitworth2018}
{Whitworth}, A. 2018, \bibinfo{title}{{Brown Dwarf Formation: Theory},} arXiv e-prints, arXiv:1811.06833, \dodoi{10.48550/arXiv.1811.06833}

\bibitem[{B.~A. {Wilking} {et~al.}(1989){Wilking}, {Mundy}, {Blackwell}, \& {Howe}}]{Wilking1989}
{Wilking}, B.~A., {Mundy}, L.~G., {Blackwell}, J.~H., \& {Howe}, J.~E. 1989, \bibinfo{title}{{A Millimeter-Wave Spectral Line and Continuum Survey of Cold IRAS Sources},} \apj, 345, 257, \dodoi{10.1086/167901}

\bibitem[{C. {Zucker} {et~al.}(2020){Zucker}, {Speagle}, {Schlafly}, {Green}, {Finkbeiner}, {Goodman}, \& {Alves}}]{Zucker2020}
{Zucker}, C., {Speagle}, J.~S., {Schlafly}, E.~F., {et~al.} 2020, \bibinfo{title}{{A compendium of distances to molecular clouds in the Star Formation Handbook},} \aap, 633, A51, \dodoi{10.1051/0004-6361/201936145}

\end{thebibliography}
